\documentclass[12pt]{article}

\usepackage{amsmath,amsfonts,amssymb,color,graphicx,overpic,}
\usepackage{slashed}
\usepackage{epsfig}
\usepackage[utf8]{inputenc}
\usepackage{amsmath}
\usepackage{ wasysym }
\usepackage{graphicx}
\usepackage[english]{babel}
\usepackage{graphic x}
\usepackage{relsize}
\usepackage{cite}

\begin{document}

\begin{titlepage}
\rightline{April 2018}
\vskip 2.2cm
\centerline{\large \bf
Resolution of the small scale structure issues with}
\vskip 0.24cm
\centerline{\large \bf
dissipative dark matter from multiple Standard Model sectors}

\vskip 2.4cm
\centerline{R. Foot\footnote{
E-mail address: rfoot@unimelb.edu.au}}

\vskip 0.4cm
\centerline{\it ARC Centre of Excellence for Particle Physics at the Terascale,}
\centerline{\it School of Physics, The University of Sydney, NSW 2006, Australia}
\vskip 0.1cm
\centerline{and}
\vskip 0.1cm
\centerline{\it ARC Centre of Excellence for Particle Physics at the Terascale,}
\centerline{\it School of Physics, The University of Melbourne, VIC 3010, Australia
}

\vskip 1.9cm
\noindent
Dissipative dark matter arising from a hidden sector consisting of $N_{\rm sec}$ 
exact copies of the Standard Model is discussed. The particles from each sector 
interact with those from the other sectors by gravity and via the kinetic mixing interaction, 
described by the dimensionless parameter, $\epsilon$. It has been known for some time 
that models of this kind are consistent with large scale structure and the 
cosmic microwave background measurements. Here, we argue that such models can potentially
explain various observations on small scales, including the observed paucity and planar distribution
of satellite galaxies, the flat velocity function of field galaxies, 
and the structure of galaxy halos. The value of the kinetic mixing parameter
is estimated to be $\epsilon \approx 1.2 \times 10^{-10}$ for $N_{\rm sec} = 5$, the example studied in most detail here. 
We also comment on cluster constraints such as those which arise from the Bullet cluster.

\end{titlepage}

\section{Introduction}

The origin of the dark matter in the Universe is perhaps one of the most fascinating scientific puzzles
of modern times. 
From studies of Large Scale Structure (LSS) and the Cosmic Microwave Background (CMB), not only can 
one infer that dark matter exists, but also that it behaves cosmologically the same as if it were 
composed of collisionless cold dark matter particles, e.g. \cite{planck}.
However, it is also known that these large scale probes 
are not very sensitive to the detailed properties of dark matter. Dark matter which has
significant self interactions, is multicomponent, or even comes with a
dark radiation component, is, like collisionless cold dark matter, consistent with LSS and CMB observations, 
e.g. \cite{ber1,IV,ber2,footcmb,late1,late2}.

There are a number of small scale issues that might shed light on the nature of dark matter.
Perhaps the most important of these are:
The paucity cf. \cite{def1,def2,def2b} and distribution \cite{satmw1,satmw2,satmw3,satmw4,sat2,sat2b,sat3,sat3b} of satellite galaxies,
the flat velocity function of field galaxies \cite{def3,def4,def5,zwaan},
and the structure of galaxy halos.
With regard to the latter, a vast amount of information has been collected
from rotation curve measurements, starting in the 1970's, e.g.  \cite{vera1,vera2,bosma,vera3},
and continuing to the present, e.g. \cite{things,littlethings}.
The conclusion is rather surprising: the structure of dark matter halos, the density profile, is in essence,
dictated by the baryonic distribution, e.g. \cite{154,blokrubin,blok0,blok1,blok2,blok3,sal17}. 
This baryon - dark matter connection is encapsulated 
in various scaling relations \cite{tf,btf,btfnew,salucci,KF,DS,Lelli,MC,Lelli2,stacy}. 
This baryon - dark matter connection, along with the
other small scale issues, can potentially illuminate the nature of dark matter.
At least, that is the view adopted here.

The alternative view is that the small scale issues 
can be explained by baryons interacting with dark matter via gravity only.
Baryonic physics is complex, and the early period of galaxy formation is poorly
constrained by observations, so many things are possible, e.g.\cite{cd1,cd2,cd3,whitebook}.
However, in practice the baryonic physics invoked involves many phenomenological
parameters, compromising efforts to test that model on small scales. Moreover,
the baryon - dark matter connection is, at best, accommodated rather than explained. 
The empirical baryon-dark matter scaling relations would seem to be suggestive of a more
direct dark matter connection with the baryons than could be expected in the 
collisionless cold dark matter model. 

Dissipative dark matter appears capable of explaining observations on both large and small
scales.
The possibility that dissipative dark matter might be implicated by the various observations
on small scales has been studied over the past decade or so, especially 
with regard to halo dynamics \cite{sph,foot13,foot14,sunny1,olga,footpaperI,footpaperII},
and the paucity and distribution of satellite galaxies \cite{sunny2,zurab1,lisa1}.
Much of that work considered mirror dark matter,
where dark matter arises from an exact duplicate of the Standard Model (SM) \cite{flv}. 
That model is the
most theoretically constrained nontrivial dark matter model known; all the dark sector parameters
are fixed and the particles from the dark sector interact with the standard particles 
(collectively labelled here as the `baryons')
via gravity and via the kinetic mixing interaction \cite{holdom,he}. 
The latter gives a mechanism for 
baryons to influence dark matter, which, as discussed above, appears necessary to account for the
baryon - dark matter connection.

The purpose of this paper is to explore the implications of extending the mirror dark matter model
to one with $N_{\rm sec}$ exact copies of the SM. 
Such a construction is still theoretically constrained \cite{fff1}, 
but has some flexibility given that $N_{\rm sec}$ is not known and can be treated, effectively, 
as a parameter. In fact, we provide arguments which indicate that a model with 
$N_{\rm sec} \sim 5$ could possibly be preferred over the minimal mirror model case with just one 
additional sector.

The outline of this paper is as follows:
In Section 2 we review the overall picture of how dissipative dark matter can potentially
explain the observations on large and small scales.
In Section 3 we provide some details on the generalized mirror model with $N_{\rm sec}$
dark SM copies. The relevant particle physics is reviewed, along with a brief dissection 
of the early Universe cosmology. 
In Section 4 we examine the matter power suppression of small scale structure and compute the halo mass function
within the Extended Press Schechter formalism.
The velocity function is evaluated, making use of the halo mass - rotational velocity relation suggested
by the dissipative dynamics.
This provides an estimate of the kinetic mixing parameter.
In Section 5 we qualitatively discuss 
dark and baryonic disk formation, which is anticipated to occur during an 
early phase of galaxy formation.
Section 6 deals with the structure of galaxy dark matter halos
in this framework.
These considerations lead to
another estimate for the kinetic mixing parameter, which is compared with the value obtained in Section 4
from the velocity function.  
In Section 7, the steady state solutions, presumed to describe the dark halos around idealized spherically symmetric systems,
are computed for a set of model galaxies for the generalized mirror model with $N_{\rm sec} = 5$. 
In Section 8 we briefly discuss 
the implications of the model for 
dark matter direct detection experiments, and in
Section 9, we conclude.

\section{A review of dissipative mirror dark matter}

Both the large scale and small scale issues can be addressed within the framework of dissipative dark matter.
Here we aim to provide a brief overview of the emerging picture of a 
dissipative dark matter Universe, which has been developed in recent years in 
e.g. \cite{ber1,IV,ber2,sph,footcmb,footreview,sunny1,sunny2,footpaperII}. The subject, though,
has a long history \cite{LY,Okun,pavsic,blinn,hodges,foot04} (see e.g. \cite{footreview} for a more detailed bibliography).
These days, consideration of nonbaryonic dark matter with nontrivial particle
properties has become 
rather
common, and there is, of course, a wider literature
of related models and ideas, e.g. \cite{wider1,wider1b,wide1c,cline,wider2,wider3,wider4,RFAN,wider5,wider6,wider7,wider8}.   

A theoretically constrained dissipative dark matter model arises if one contemplates a hidden sector
exactly isomorphic to the Standard Model. That is, one considers particle physics described by a Lagrangian of the form: 
\begin{eqnarray}
{\cal L} = {\cal L}_{\rm SM}(e,u,d,\gamma,...) + {\cal L}_{\rm SM}(e',u',d',\gamma',...) 
+ {\cal L}_{\rm mix}
\ .
\label{1x1}
\end{eqnarray}
Such a Lagrangian has a discrete $Z_2$ invariance with respect to the interchange of each ordinary particle with its hidden sector partner,
denoted with a prime ($'$) in Eq.(\ref{1x1}).
If left and right chiral fermion fields are swapped in the hidden sector, then this discrete symmetry operation is
an improper Lorentz transformation \cite{flv,flv2}. 
This construction thus has an interesting symmetry motivation,
but also provides specific theoretically constrained dark matter candidate(s) with rather nontrivial particle
properties. 

The hidden sector particles have been called `mirror' particles, the mirror electron ($e'$),
mirror photon ($\gamma'$) etc., and sometimes we will also use more general notation, e.g. `dark electron',
`dark photon' etc.
The mirror particles interact with ordinary particles via gravity and via the kinetic
mixing interaction \cite{holdom,he}, which also leads to photon-mirror photon kinetic mixing:
\begin{eqnarray}
{\cal L}_{\rm mix} 
= \frac{\epsilon}{2} F^{\mu \nu} F^{'}_{\mu \nu}
\label{mix2}
\end{eqnarray}
where $F^{\mu \nu}$ ($F^{'}_{\mu \nu}$) is the field strength tensor for the photon (mirror photon).
This is a renormalizable interaction, consistent with the symmetries of the theory.

For the mirror dark matter model, the broad picture is the following one.
The early Universe can be described by a sector of ordinary particles, including a radiation component 
with temperature $T(t)$ and with the 
matter energy density described in terms of the usual $\Omega_b$ parameter.
In addition, there is also the set of mirror particles, including a dark radiation component with temperature $T'(t)$,
and with matter energy density described by $\Omega_{\rm dark}$.
The Universe is expanding, and the evolution of the temperatures, $T(t)$, $T'(t)$, is described by the Friedmann equation and also
by the second law of thermodynamics.
Since the two sectors are (almost) thermally decoupled, they can have different thermal histories 
so that $T' \neq T$. 
The mirror baryon abundance, like that of the baryons,
is expected to be set by a particle - antiparticle asymmetry, arising at very early times.
The details are model dependent, but
$\Omega_{\rm dark}/\Omega_b$ can naturally be of order one \cite{ber5,fv1,fv2}.

To have a consistent picture we require suitable initial conditions.
In the limit where $T'/T \to 0$, the time when dark hydrogen 
recombination occurs goes to zero; this means that
dark sector particles  take  the form of neutral atoms in the early Universe.
In that circumstance, the dark sector
becomes, cosmologically, indistinguishable from collisionless cold dark matter which is known to lead to
remarkably successful CMB and LSS.
Fits of collisionless cold dark matter to the CMB anisotropy spectrum give $\Omega_{c} \approx 5.4\Omega_b$ \cite{planck}. 
These considerations
suggest that
suitable initial conditions for mirror dark matter are:
\begin{eqnarray}
\frac{T'}{T} \ll 1, \ \ \Omega_{\rm dark} \approx 5.4\Omega_b
\ .
\label{1x}
\end{eqnarray}
Note that the temperature asymmetry is
quite a natural outcome of inflationary models \cite{kolb,hodges,mohap},
and is also  consistent with the symmetric Lagrangian, Eq.(\ref{1x1}). 

If it were to happen that $T'/T \neq 0$ in the early Universe,
then one can talk about an era prior to mirror hydrogen recombination.
In that era, the mirror particles form a tightly coupled plasma, where pressure plays an important role, and
the growth of density perturbations is impacted by dark acoustic oscillations and dark photon diffusion \cite{ber1,IV,ber2,footcmb}.
These effects are analogous to the corresponding processes affecting the baryons prior to hydrogen recombination.
If $T' < T$, then  dark acoustic oscillations and dark photon diffusion occur earlier, 
and suppression of power on small scales is a consequence.
In fact, such a suppression of power is 
desirable because it can explain the paucity 
of satellite galaxies cf. \cite{def1,def2,def2b},
as well as the flat velocity function 
of field galaxies
\cite{def3,def4,def5,zwaan}.


A nonzero value of $T'/T$ necessarily occurs in the early Universe if 
the kinetic mixing interaction exists, as particle processes: $e \bar e \to \bar e' e'$ lead 
to a transfer of entropy  from the
ordinary sector to the mirror sector \cite{cg}.
Such particle processes freeze out at $T \sim m_e$, leading to an asymptotic $T'/T$ value:
\cite{footc,p1}
\begin{eqnarray}
T'/T \simeq 0.31 \sqrt{\epsilon/10^{-9}}
\ ,
\ \ \ {\rm for} \ \epsilon \lesssim 10^{-9}\ .
\label{2x}
\end{eqnarray}
Having $\epsilon \neq 0$, has the effect of generating new physical scales, the Dark Acoustic Oscillation
scale, $L_{\rm DAO} (\epsilon)$ and the Dark Photon Diffusion Scale (Dark Silk Damping), $L_{\rm DSD} (\epsilon)$.
If one attempts to match these  scales to observations, e.g. by evaluating the velocity function,
then an estimate of $\epsilon$ can be made.  A calculation along these lines indicates that $\epsilon \sim 2\times 10^{-10}$ for the 
mirror dark matter case ($N_{\rm sec} = 1$) \cite{sunny2}. 
For such a small value of $\epsilon$, and the associated small value of $T'/T$ [Eq.(\ref{2x})], the model and
collisionless cold dark matter become
indistinguishable as far as
LSS and CMB anisotropies are concerned \cite{footcmb,wall}.
On small scales however, the model is very different, especially with regard to galaxy
formation and evolution.

Dark photon diffusion leads to exponential power suppression, somewhat reminiscent of the situation with warm dark matter, although the physical
origin of the effect is very different.
A consequence of this dramatic matter power suppression is that hierarchical clustering
has a low mass cutoff, $M^*(\epsilon)$; only collapsed structures
with masses greater than $M^*(\epsilon)$ can form from primordial density perturbations.
For $\epsilon \sim 2\times 10^{-10}$, $M^* \sim 2\times 10^9\ m_\odot$. \footnote{One effect
of this low mass cutoff is that there should be relatively few (minor) mergers affecting galaxy evolution.
In fact, this quiet merger history may help account for the observations of bulgeless disk galaxies, e.g. \cite{Kautsch,kom9}.
That is, pure disk galaxies with no evidence for merger-built bulges.
These observations seem to be difficult to reconcile with
hierarchical clustering down to arbitrarily low scales 
cf. \cite{kom1,kom2}. 
}

Consider now a
galaxy-scale perturbation that is sufficiently large ($m > M^*$) so that it is not greatly affected by 
the small scale matter power suppression. 
If one contemplates the evolution of such a galaxy-mass-scale perturbation,
then collapse occurs when the mean overdensity of the perturbation reaches a critical
value, $\delta \sim 1$.  During the nonlinear collapse phase, the dissipative dark matter 
can undergo complex processes, shock heating etc., but is expected to ultimately cool and form a dark disk.
The baryonic matter, which is also collapsing at this time,
undergoes similar processes and forms a baryonic disk. 
Naturally, the dark disk can have a subcomponent of ordinary matter and vice-versa.
The dark matter, having collapsed to from a disk, would be very compact, much more compact than the extended
dark matter halos inferred to exist around disk galaxies today.
The enhanced gravitational field strength affects also
the baryonic disk, which would also be very compact at this time.
Such a dense environment should
be conducive to star formation which could begin even before these disks are fully 
formed.\footnote{ 
The compact baryonic and dark matter distributions would be expected to 
support the rapid growth of supermassive black holes at an early epoch. 
Indeed, a recent study, which invoked only a small subcomponent 
of dissipative dark matter, found that substantial enhancement of black hole growth was possible \cite{panci}.
Such studies could be very important in view of
observations indicating the early formation of supermassive black holes e.g. \cite{quasar1,quasar2,quasar3}.}

In the meantime,  
gravity would tend to make these two disks coalesce on a time scale of only $\sim$ 10 Myr cf. \cite{Fan}.
However, shortly after star formation begins, 
kinetic mixing induced processes in ordinary Type II supernovae (SNe) located in the baryonic disk can lead to 
heating of the dark disk 
(more details of this mechanism will be discussed in a moment).
If dark supernovae occur, corresponding processes can heat the ordinary disk.
It has been suggested \cite{sunny2} that these disk heating effects could possibly
create a pressure force that overwhelms 
gravity. If this does indeed happen, then the dark disk would be expected to evolve until it is
orthogonal to the baryonic disk.
(If the pressure force is weaker than gravity, the disks would evolve until they are aligned in the same plane.)
The heating of the disks will also regulate star formation. 
As they heat up, star formation can slow.
Eventually, the gas component of one disk would become completely disrupted and star formation in that disk effectively ceases.
It is difficult to determine which of these disks will survive and flourish as 
the relative star formation rates are rather 
uncertain.\footnote{
One important factor is the chemical composition:
With $T' \ll T$ in the early Universe, mirror BBN calculations indicate that the dark sector is helium dominated, 
while the baryonic sector is hydrogen dominated \cite{ber1}.
Although it is known that helium dominated stars evolve more rapidly than hydrogen dominated
ones \cite{ber4}, their formation rate is the more relevant consideration here, and is difficult to reliably
estimate. However,
the fact that helium has no low energy excitations is one factor suggesting a reduced mirror
star formation rate.} 
A consistent picture for the observed galaxies requires the dark disk to be the one which is disrupted, so that it can then expand to form
a roughly spherical dark halo.
Naturally, this will require a substantial amount of energy to be transmitted to the halo, which we will 
return to in a moment.

In this picture, the dwarf spheroidal satellite galaxies are observable remnants of the dark disk; 
these galaxies are not primordial, but grew out of perturbations at the edge of this disk, 
either during the disk formation or shortly after. 
They are `top down' forming structures, originating from larger scale density perturbations.
If this formation mechanism is correct, then the satellite galaxies are expected to be 
co-rotating and orbit in the same plane as the dark disk, kinematic features consistent with observations 
of the satellites around the Milky Way \cite{satmw1,satmw2,satmw3,satmw4}, Andromeda \cite{sat2,sat2b}, 
and Centaurus A \cite{sat3,sat3b} (see also
\cite{paww} for earlier historical references).

The kinetic mixing interaction provides the bridge between the ordinary particles and 
their mirror sector partners. This interaction, even if very tiny ($\epsilon \sim 10^{-9}-10^{-10}$),
transforms Type II SNe into powerful heat sources for the dark sector \cite{raffelt,raffelt2,zurab}.
Kinetic mixing induced processes in the SN core generate an expanding energetic plasma,  
initially comprising mainly the light mirror particles, $e', \bar e', \gamma'$, 
with total energy up to around $\sim 10^{53}$ erg for the kinetic mixing strength considered.
Insight into the evolution of this energetic plasma can be obtained from the fireball model of Gamma
Ray Bursts \cite{grb1,grb2,grb3,grb4,fireball}.
The dark plasma quickly evolves into a relativistic fireball and
sweeps up the nearby mirror baryons as it propagates outward from the SN.
The energy of the fireball is anticipated to be transferred to the mirror baryons.
This flow eventually decelerates, and part of this kinetic energy
is converted back into thermal energy which can radiatively cool producing dark radiation.
The end result is that the energy sourced from ordinary Type II SNe can be transmitted to the dark 
matter halo in two distinct ways: via dark photons and also 
via the heating of mirror baryons in the SN vicinity (and beyond via convection/conduction processes).

If the energy transmitted to the mirror baryons is substantial, 
then the mirror gas component 
of the dark disk can be completely disrupted as it
expands in response to the heating.
As the dark matter expands, this influences the baryonic Star Formation Rate (SFR). 
The SFR is known to be strongly correlated with the gas density \cite{ken,Schmidt}, and
this density reduces when the dark matter is expanding due to the weakening gravity. 
The system is a complicated one,
with baryons coupled to the dark matter via gravity, while the dark matter is coupled to the baryons via SN sourced heating.
One envisages a fairly rapid dark matter expansion which ultimately slows due to the decreasing energy input as the SFR
subsides.  
There may of course be various complications during this upheaval, such as baryon and dark
baryon outflows etc. 

This dynamically evolving system can be modelled by fluid equations.
It is assumed that
the system eventually relaxes  to a steady state configuration.
In principle, one needs to model the SFR in this evolving
environment and solve the time-dependent fluid equations to check that this actually happens.
But, if it does, then
this provides for a fairly simple description
of the physical properties of the dark matter halo around galaxies with active star formation.\footnote{
The halos around galaxies without active star formation are expected to have very different physical properties.
In particular the dwarf spheroidal galaxies and elliptical galaxies have relatively little active star
formation; for these galaxies the halo is expected to have cooled and condensed into dark stars, black holes etc.
}
The density and temperature are determined by the steady state
conditions, and in the limit of no bulk halo motion, these conditions are:
\begin{eqnarray}
{\cal H} &=& {\cal C}
\ , \nonumber \\
\bigtriangledown  P &=& - \rho \nabla \phi  
\ .
\label{SSX}
\end{eqnarray}
The first equation equates the local heating and cooling rates of the dark halo plasma, while
the second equation is the balancing of the pressure gradient with the gravitational force (hydrostatic equilibrium).

These steady state equations have been studied in a number of papers, e.g. \cite{foot13,foot14}, with \cite{footpaperI} and \cite{footpaperII}
providing the most detailed description for mirror dark matter.
Some simplifying assumptions were made, including spherically symmetric modelling of both the halo and the baryonic components.
Despite these simplifications, a broadly consistent description 
of the physical properties of the dark halo 
emerged with halo heating sourced by local processes in the vicinity of ordinary SNe.
Assuming a halo with negligible mirror metal content,    
the halo properties depend only
on the amount of energy transmitted to the halo per (average) SN, denoted by
the parameter $L_{\rm SN}^{e'}$.
If $L_{\rm SN}^{e'}$ is approximately constant for all galaxies, 
then this dynamics becomes quite predictive. 
In fact, realistic rotation curves for the modelled spirals were obtained for 
$L_{\rm SN}^{e'} \approx 10^{53}$ erg \cite{footpaperII}.
This value is close to the maximum available energy from a Type II SN, and a tad higher
than expectations for $\epsilon \sim 2 \times 10^{-10}$,
although there are considerable uncertainties and
a precise calculation of $L_{\rm SN}^{e'}$ in terms of the kinetic mixing parameter
is not yet available.

Realistic rotation curves result largely because 
the dark matter distribution is strongly influenced by the 
baryons (given that Type II SNe are the primary heat source).
Not only does this lead to an approximately cored halo density profile, but the core radius, $r_0$,
correlates with the baryonic scale length, $r_D$. 
These properties lead to a linearly rising rotational velocity in the inner region
$r \lesssim r_D$, consistent with observations, e.g. \cite{154,blokrubin,blok0,DS,littlethings,stacy}.
Another consequence of this dynamics is that the 
normalization of halo rotational velocities
follows Tully Fisher - type relations.
In fact, the entire halo dark matter distribution was found \cite{footpaperII} to be dictated by the 
baryonic distribution in a manner broadly consistent
with observations.
It appears that the baryon - dark matter connection can be
explained in this dynamics. 

The picture sketched above assumed
mirror dark matter,  but a more generic dissipative model is, of course, possible.
Mirror dark matter though, is rather special in that it is strongly theoretically constrained.
There is only one new fundamental parameter, the kinetic mixing strength $\epsilon$.
Actually, the generalization of mirror dark matter to a model with several additional isomorphic sectors,
is also strongly theoretically constrained, although obviously less minimal. Here, we denote by $N_{\rm sec}$ 
the number of additional sectors, so that the mirror dark matter case corresponds to having $N_{\rm sec} = 1$.
The aim of this paper is to explore the generalized mirror dark matter model where there are multiple additional 
sectors $N_{\rm sec} \ge 1$. 
We are interested in the case where $N_{\rm sec}$ is not so large, $N_{\rm sec} \lesssim 10$, so that
the above broad picture can result from such models.


\section{Generalized Mirror Dark Matter}

\subsection{The particle physics}

We consider the generalized mirror model, which consists of $N_{\rm sec}$ identical dark sectors that are copies 
of the SM \cite{fff1}. That is, the 
Lagrangian describing fundamental physics is
\begin{eqnarray}
{\cal L} = {\cal L}_{\rm SM}(e,u,d,\gamma,...) + \sum_{i=1}^{N_{\rm sec}} {\cal L}_{\rm SM}(e_i,u_i,d_i,\gamma_i,...) 
+ {\cal L}_{\rm mix}
\ .
\label{yyy6}
\end{eqnarray}
These dark sectors shall be assumed to be exact copies of the SM, that is, the chirality of the dark sector fermions is not flipped.
In this case the Lagrangian
possesses a $C_{N_{\rm sec}+1}$ discrete symmetry, the permutation symmetry of $N_{\rm sec}+1$ objects.
With this discrete symmetry, the 
kinetic mixing interaction is described by one dimensionless parameter, $\epsilon'$, and takes the form
\begin{eqnarray}
{\cal L}_{\rm mix} = \frac{\epsilon'}{4} \sum_{i,j} F^{i\mu \nu} F^{j}_{\mu \nu}
\label{mix1}
\end{eqnarray}
where $F^{i\mu \nu}$ is the field strength tensor of the $i^{th}$ sector $U(1)_Y$ gauge boson.
Here, the sum runs over $i,j = 0,...,N_{\rm sec}$, with $j \neq i$, where $i=0$ is taken as the SM 
sector.\footnote{In general, one can consider the case of $p$ ordinary sectors and $q$ chirality flipped
sectors. In such case the structure of the discrete symmetry is different and the kinetic mixing
interaction involves several parameters \cite{fff1}.}
As in the mirror model case, kinetic mixing induces tiny ordinary electric charges for the charged
dark sector particles. The induced electric charges for the $i^{th}$ sector proton ($p_i$) and electron ($e_i$) are
$\pm \epsilon e$, where $\epsilon$ can be defined in terms of $\epsilon'$ and the weak mixing angle, $\theta_w$:
$\epsilon = \epsilon' \cos^2\theta_w/[1 + \epsilon'(N_{\rm sec} -1)]$.

Each sector has an associated Higgs doublet,  $\phi_i$, and in general there can also be 
a quartic Higgs mixing term: ${\cal L}_{\rm mix} = \lambda' \sum_{i,j} \phi_i^\dagger \phi_i \phi_j^\dagger \phi_j$. 
Analysis of the Higgs potential, including the quartic mixing term,
finds that the $C_{N_{\rm sec}+1}$ symmetric vacuum, in which the Vacuum Expectation 
Values (VEV) for the Higgs doublets are identical with
each other, results for a large range of parameters \cite{fff1,fff2}.  The point where $\lambda' = 0$
is also contained within this parameter space, which is fortunate as there are
fairly strong contraints on a nonzero $\lambda'$ from collider experiments \cite{vol97,footreview} and
generally tighter constraints from early Universe cosmology \cite{vol98,vol99}.
Such a symmetric vacuum state will be assumed, which means that the
symmetry between each sector, including the sector describing the ordinary particles,
is exact and unbroken. It follows that the particle masses of the corresponding particles in each
sector are identical, so for example, $m_e = m_{e_1} = ...= m_{e_{N_{\rm sec}}}$.

In addition to dark matter,
neutrino masses and the baryon number asymmetry of the Universe
can be accommodated in this framework, but there are many possibilities.
Typically this involves replacing ${\cal L}_{\rm SM}$ in Eq.(\ref{yyy6})
with an extension containing new particles and interactions (e.g. by including right handed neutrinos
to allow for neutrino masses).
Finally, the construction Eq.(\ref{yyy6}) might have some unexpected connection to gravity.
A curious feature
of the SM Higgs effective potential is the possibility that it contains two degenerate
vacua; the usual one at the weak scale and
another one near the Planck scale, e.g. \cite{xyz,xyx1,xyx2}. 
This could perhaps be taken as a hint that a more fundamental theory
might involve multiple SM sectors; one (or more) of these sectors gains the Planck scale
VEV and the remaining sectors gain the electroweak scale VEV. 
If these scalars couple to the curvature, $R$, in a (minimal) scale invariant manner, with coupling
constant of order unity, then the Einstein-Hilbert action results after spontaneous symmetry breaking.
The classical scale invariance can be extended to the full Lagrangian and 
a fairly constrained model can be constructed which reduces
to Eq.(\ref{yyy6}) in the low energy limit \cite{archie}. 

\subsection{Dark matter from multiple SM sectors}

In this type of model the dark baryons can form the dark matter. 
The existence of dark baryons, like ordinary baryons, requires a particle-antiparticle
asymmetry to be generated in the very early Universe.
The origins of such particle asymmetries are unknown; many possibilities
have been proposed in the context of mirror dark matter \cite{ber5,fv1,fv2}, 
related models \cite{Melb1,Melb2}, and more generally e.g. \cite{kali,zurek}.
Such a framework has the potential to explain why the mass densities of the dark matter
and ordinary matter are similar, $\Omega_{\rm dark} \sim \Omega_b$, 
as the ordinary and dark sector asymmetries can be interconnected.
Mirror dark matter, and indeed dark matter arising from multiple copies 
of the Standard Model,
are perhaps of particular interest in this regard as such models require the mass scale of the baryons
and dark sector baryons to be identical. Understanding why the dark and ordinary sectors have a similar
mass scale is, of course, required for $\Omega_{\rm dark} \sim \Omega_b$ to be fully explained. 

With multiple SM copies the possibility that 
each sector has an identical matter content arises, i.e. 
\begin{eqnarray}
\Omega_1 = ....= \Omega_{N_{\rm sec}}
\ .
\label{fl}
\end{eqnarray}
If $\Omega_1 = \Omega_b$, then
$N_{\rm sec} = 5$ is preferred, although the best estimates of $\Omega_{\rm dark}/\Omega_b$ available 
disfavour this quantity being an integer, $\Omega_{\rm dark}/\Omega_b = 5.40 \pm 0.11$ \cite{planck}.

In addition to the matter content, each sector has a radiation component which can be probed via the CMB anisotropies and 
also via Big Bang Nucleosythesis (BBN).
These considerations can be used to place limits on the energy density of the dark sector particles during the radiation
dominated era.   
For example, Planck observations combined with
other astrophysical data give $N_{\rm eff} = 3.15 \pm 0.23$ for the effective number of relativistic degrees of freedom \cite{planck},
which is consistent with the SM value of $N_{\rm eff} = 3.046$. This gives a CMB limit of around $\delta N_{\rm eff}^{\rm CMB} \le 0.56$ at
$95  \%$ C.L.
Since the dark sectors are thermally decoupled from each other, each sector can be described by a distinct
temperature $T_i$,
that is also distinct from the baryonic sector temperature, $T$. 
Clearly, the case where $T_i = T$ is excluded,
and the energy density requirements can be satisfied if there is a temperature asymmetry present prior to
hydrogen recombination:
\begin{eqnarray}
\frac{T_i}{T} \lesssim \frac{0.5}{(N_{\rm sec})^{1/4}}\ , \  
    \   \  \ i=1,...,N_{\rm sec}
\ .
\label{asymm}
\end{eqnarray} 
This temperature asymmetry also needs to be present prior to the BBN epoch if the primordial
abundances of light elements are to be explained in the usual way. 

Inflationary models which give $T_i/T \ll 1$ have been constructed within the context of mirror dark matter
and related models \cite{kolb,hodges,mohap}, 
and this motivates an effective initial condition for the 
temperature asymmetry:
\begin{eqnarray}
\frac{T_i}{T} \ll 1\ , \    \   \  \ i=1,...,N_{\rm sec}
\ .
\label{yyy7}
\end{eqnarray}
It should perhaps be emphasized that, within the context of inflationary models, this
asymmetric effective initial condition is not in any formal conflict with having a symmetric Lagrangian 
[Eq.(\ref{yyy6})] with symmetric initial conditions prior to inflation.
Also, having asymmetries of the form Eq.(\ref{fl}), with $\Omega_1 \simeq \Omega_b$,  
is similarly  not inconsistent with having asymmetric temperatures, Eq.(\ref{yyy7}). In fact, simple asymmetry generating mirror models
have been constructed \cite{ber5} that are consistent with the conditions 
Eq.(\ref{fl}) (with $\Omega_1 \simeq \Omega_b$) and Eq.(\ref{yyy7}).
In general, $\Omega_1$ need not be exactly $\Omega_b$ in such models, it is dependent
on the details of the specific asymmetry generating mechanism and associated thermal history.

In the presence of kinetic mixing, with the effective initial conditions, Eq.(\ref{fl}), entropy
is transferred from the ordinary particles to the dark sector particles, and $T_i/T$ becomes nonzero.
The fundamental process driving this entropy transfer is $e \bar e \to e_i \bar e_i$ cf. \cite{cg}. 
Since $n_e \simeq n_{\bar e} \propto T^3$,
$\sigma \propto 1/T^2$, we have $\Gamma \propto T$ 
and $\Gamma/H \propto 1/T$.\footnote{
Natural units with $\hbar = c = k_B=1$ are used throughout, unless otherwise indicated.}
It follows that this process is most important 
at low temperature, $T \lesssim 100$ MeV, where
$T_i/T \propto 1/T^{1/4}$ \cite{p1},
and increases until $T \sim m_e$. After this time the number of $e, \bar e$ becomes Boltzmann 
suppressed and $T_i/T$ asymptotes to: \cite{footc}
\begin{eqnarray}
T_i/T \simeq 0.31 \sqrt{\epsilon/10^{-9}}\ .
\label{goodeq}
\end{eqnarray}
The above result assumed that
the dark sector has energy density much less than the ordinary sector, an assumption valid
for $\epsilon$ values satisfying the CMB constraint, Eq.(\ref{asymm}). 
Notice that the asymptotic value of $T_i/T$ is independent of $N_{\rm sec}$. 

As with the mirror dark matter 
case, reviewed in Sec. 2, the nonzero $T_i/T$ means that there was an era prior to dark
hydrogen recombination, where the dark sector particles formed a plasma.
This will have important ramifications for small scale structure (discussed in the following section).
The composition of the dark plasma is also potentially quite important.
In addition to dark electrons, dark protons and dark photons,
the plasma contained 
dark helium, formed somewhat earlier during dark BBN. In fact, the dark sectors are helium dominated
cf. \cite{ber1},
with mass fraction around $Y'_p \approx 0.95$ for $\epsilon \sim 2\times 10^{-10}$ \cite{p2,footreview}. 

\section{Power suppression on small scales}

\subsection{The power spectrum}

Structure in the Universe is seeded from tiny perturbations, $\delta \sim 10^{-5}$,
at very early times. These perturbations evolve according to a set of 
coupled Boltzmann and Einstein equations. For the
dark matter model under discussion, with $N_{\rm sec}$  exact copies of the SM, this evolution
can be separated into two regimes; the period prior to dark hydrogen recombination,
where the dark matter consists of a plasma containing dark electrons, dark protons and
dark photons, and the period after, where the dark photons decouple from the dark
matter. Prior to recombination the dark matter plasma can be modelled as a tightly
coupled fluid which evolves under the influence of gravity and pressure. Dark acoustic
oscillations and dark photon diffusion are important processes in this era. 
These processes also arise for baryons prior to hydrogen recombination, but since
hydrogen recombination happens at a later time (given $T_i < T$) the baryon
acoustic oscillations can occur on much larger scales. 
After dark hydrogen recombination,
the dark matter is influenced only by gravity, and evolves in the linear regime in 
the same way as collisionless cold dark matter.

With the assumed effective initial conditions, Eq.(\ref{yyy7}),
the time of dark hydrogen recombination, when $T_i \approx 0.25$ eV, is controlled by
the kinetic mixing parameter, $\epsilon$, via Eq.(\ref{goodeq}). As $\epsilon \to 0$, the time
when dark hydrogen recombination occurs also goes towards zero, and it follows that in the linear
regime
generalized mirror dark matter would evolve (in that limit) in the same way as collisionless cold dark matter on all scales.
For $\epsilon \neq 0$, departures occur, which only 
occur on small scales, smaller than the sound horizon at dark recombination. 

The evolution of perturbations in the early Universe are described by  
Boltzmann and Einstein equations which can be solved when  the
relevant perturbations are less than unity (linear regime). 
Since the dark sector consists of copies of
the Standard Model, the relevant equations are analogous to the equations 
governing baryons, e.g. \cite{dodelson}, and were given explicitly in \cite{footcmb} for the mirror dark matter case
($N_{\rm sec} = 1$). These equations can easily be adapted to the more general case of $N_{\rm sec}$ 
exact SM copies, and will be solved here to obtain the matter power spectrum.
In the numerical work the cosmological parameters, 
in the usual notation, e.g. \cite{dodelson}, were taken as follows:
$\Omega_m h^2 = 0.142$, $\Omega_b h^2 = 0.022$, $\Omega_\Lambda = 1 - \Omega_m$, $n_s = 0.97$, $h = 0.7$ 
($\Omega_m = \Omega_{\rm dark} + \Omega_b$).

\begin{figure}[t]
  \begin{minipage}[b]{0.5\linewidth}
    \centering
    \includegraphics[width=0.7\linewidth,angle=270]{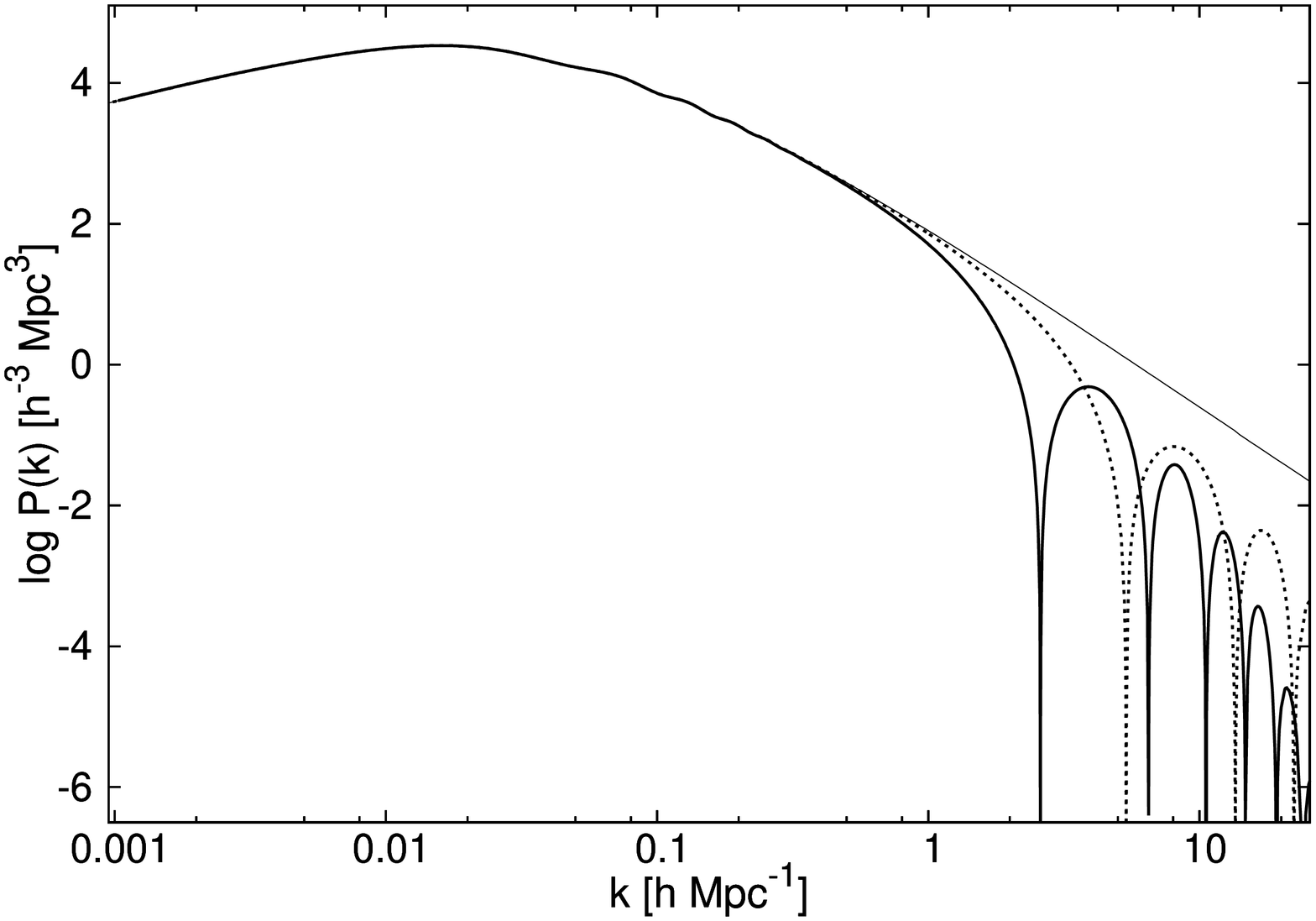}
     (a)
    \vspace{4ex}
  \end{minipage}
  \begin{minipage}[b]{0.5\linewidth}
    \centering
    \includegraphics[width=0.7\linewidth,angle=270]{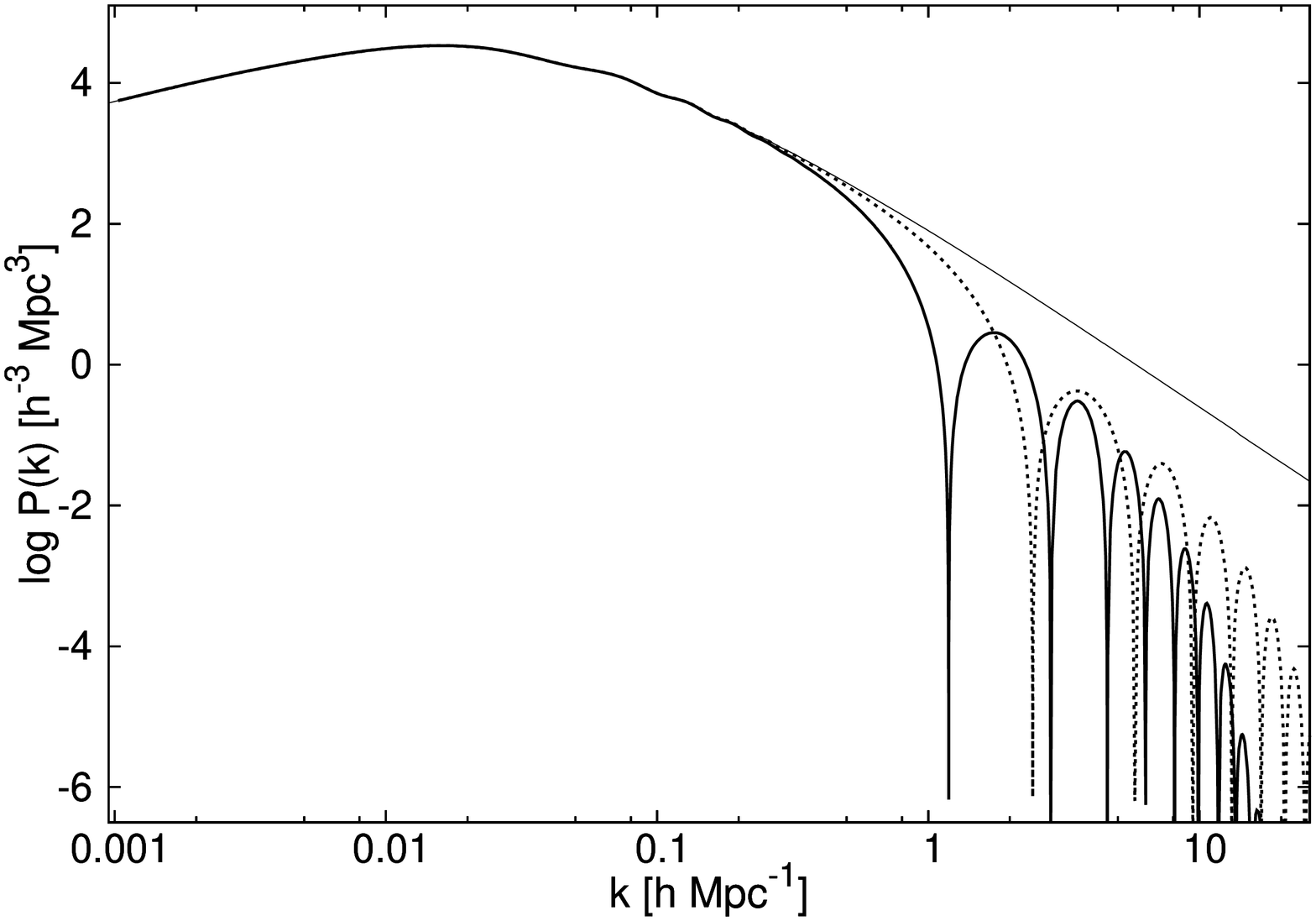}
    (b)
    \vspace{4ex}
  \end{minipage}
\vskip -1.0cm
\caption{
\small
Generalized mirror dark matter power spectrum for (a) $\epsilon = 10^{-10}$, with $N_{\rm sec} = 5$  (thick solid line),
and $N_{\rm sec} = 1$ (dashed line). The thin upper line is for $\epsilon = 0$, cosmologically equivalent to collisionless
cold dark matter. (b) same as (a) but for $\epsilon = 2 \times 10^{-10}$.
} 
\end{figure}

The matter power spectrum is defined in terms of the Fourier transform of the fractional 
overdensity,
$\delta (\mathbf{k})$, by:
\begin{eqnarray}
\langle \delta (\mathbf{k}) \delta ^{\star} (\mathbf{k'}) \rangle = (2\pi)^3P(k)\delta ^3 (\mathbf{k}-\mathbf{k'}) 
\end{eqnarray}
where the brackets indicate an ensemble average.
In Figure 1 we show the computation of this quantity, at redshift $z=0$, in the generalized mirror model for several values of 
$\epsilon,\ N_{\rm sec}$. The case of $\epsilon = 0$, where the equations are equivalent 
to those of the collisionless cold dark matter 
model, is also shown for comparison.

As discussed above, the suppression of power on small scales can be understood from
the effects of dark acoustic oscillations and dark photon diffusion.
These scales
can be estimated analytically from linear perturbation theory, and can be extracted from the calculation given
in the appendix of \cite{sunny2}:
\begin{eqnarray}
L_{\rm DAO}^i &=& 8.6 \left(\frac{\Omega_{\rm dark}}{\Omega_i}\right)^{1/2}
\left(\frac{\epsilon}{10^{-9}}\right)^{5/4} \ h^{-1} \ {\rm Mpc}
\ ,
\nonumber \\
L_{\rm DSD}^i &=& \left(\frac{\Omega_{\rm dark}}{\Omega_i}\right)^{1/2}
 \left(\frac{\epsilon}{10^{-9}}\right)^{3/4} \ h^{-1}\ {\rm Mpc} 
\ .
\label{s8}
\end{eqnarray}
This estimate for $L_{\rm DSD}^i$ also takes into acount the effects of the dark helium 
mass fraction, i.e.  $L_{\rm DSD}^i \propto (1-Y'_p/2)^{-1/2}$ from \cite{dodelson},
with $Y'_p \approx 0.95$ used.  
For each of these length scales the
associated wavenumber scale is roughly $k \approx \pi/L$.
For identical matter content in each dark sector, Eq.(\ref{fl}), it follows that $\Omega_{\rm dark}/\Omega_i = N_{\rm sec}$.
Dark acoustic oscillations provide only a moderate suppression of power, while dark diffusion damping
provides a much sharper exponential decline. 
Since $L_{\rm DAO}$ and $L_{\rm DSD}$ are proportional to $\sqrt{N_{\rm sec}}$, the small scale power 
suppression will encroach larger scales for models with higher values of 
$N_{\rm sec}$ (unless compensated by a smaller $\epsilon$ value).  

The mass scales corresponding to these length scales are: 
$M_{\rm DAO}  \sim \pi \rho_{\rm crit} \Omega_m L^3_{\rm DAO}/6$, 
$M_{\rm DSD}  \sim \pi \rho_{\rm crit} \Omega_m L^3_{\rm DSD}/6$, where 
$\rho_{\rm crit} \simeq 1.8788 \times 10^{-29}\ h^2 \ {\rm g/cm^3}$ is the critical density.
The paucity of satellite galaxies suggests that $M_{\rm DSD} \sim 10^9 - 10^{10} \ m_\odot$, which
provides a rough estimate of $\epsilon \sim 2.5 \times 10^{-10}/N_{\rm sec}^{2/3}$. 
For such values of kinetic mixing strength, we have $M_{\rm DAO} \sim 2\times 10^{11}  m_\odot/N_{\rm sec}$. 
The moderate
suppression of power due to dark acoustic oscillations can help explain the 
rather flat velocity function measured for field galaxies \cite{def3,def4,def5,zwaan},
which can also be used to extract another 
estimate of $\epsilon$ [to be looked at in Sec. 4.3].

\subsection{The halo mass function}

Naturally, we would like to know the number and distribution of galaxies in the Universe.
To this end, it is useful to calculate the halo mass function. This is the number density
of dark halos per mass interval.
The halo mass function will be here computed at redshift $z=0$ using the
Extended Press Schechter (EPS) formalism \cite{ps,bond}.
The EPS method assumes linear growth of perturbations until the
mass overdensity reaches a certain critical threshold, at which point the halo is assumed to immediately collapse.
Here we give only the equations to be solved. For a description of the EPS formalism, along with more
extensive bibliography, see e.g. \cite{psreview}.

The halo mass function depends on the choice of filter function.
Here, we make use of the
sharp-$k$ filter, a top-hat function in Fourier space: $W(k;R) \equiv \Theta (1 - kR)$,  
where $\Theta$ denotes the Heaviside step function. 
As discussed in \cite{benson,schneider}, the more commonly used filter functions, including the
top-hat function in real space, fail
when applied to theories with exponentially suppressed 
power on small scales.

In the EPS formalism the
halo mass function is:
\begin{eqnarray}
\frac{dn}{d\log m_{\text{halo}}} = 
\frac{\bar{\rho}}{m_{\text{halo}}} \nu f(\nu)
\frac{d\ln \nu}{d\log m_{\text{halo}}}
\label{hmfnew}
\end{eqnarray}
where $\bar \rho = \Omega_m \rho_{\rm crit}$ is the average density in the Universe.
Here, $f(\nu)$ is the first-crossing distribution. In the case of
ellipsoidal collapse \cite{smt}:
\begin{eqnarray}
f(\nu) = A\sqrt{\frac{2}{\pi}} \left [ 1 + (\nu) ^{-2p} \right ] e ^{-\frac{\nu^2}{2}} 
\label{firstc}
\end{eqnarray}
where $A = 0.3222$ and $p = 0.3$.
Also, $\nu$ is defined as:
\begin{eqnarray}
\nu \equiv \frac{\delta_c}{\sigma(R)} 
\label{peakheight}
\end{eqnarray}
where the critical overdensity for collapse at redshift $z=0$ is $\delta_c \simeq 1.686$ and 
$\sigma(R)$ is the variance:
\begin{eqnarray}
\sigma^2(R)  = \frac{1}{2\pi ^2}\int  dk \ k^2 P(k)|W(k;R)|^2 \, .
\label{variance}
\end{eqnarray}
Here, $P(k)$ is the linear matter power spectrum at redshift $z=0$. 
The mapping between the filter scale $R$ and the mass scale is
$m_{\text{halo}} = 4\pi \bar{\rho}(cR)^3/3$, where $c = 2.5$ \cite{benson,s13,schneider}.


\begin{figure}[t]
  \begin{minipage}[b]{0.5\linewidth}
    \centering
    \includegraphics[width=0.7\linewidth,angle=270]{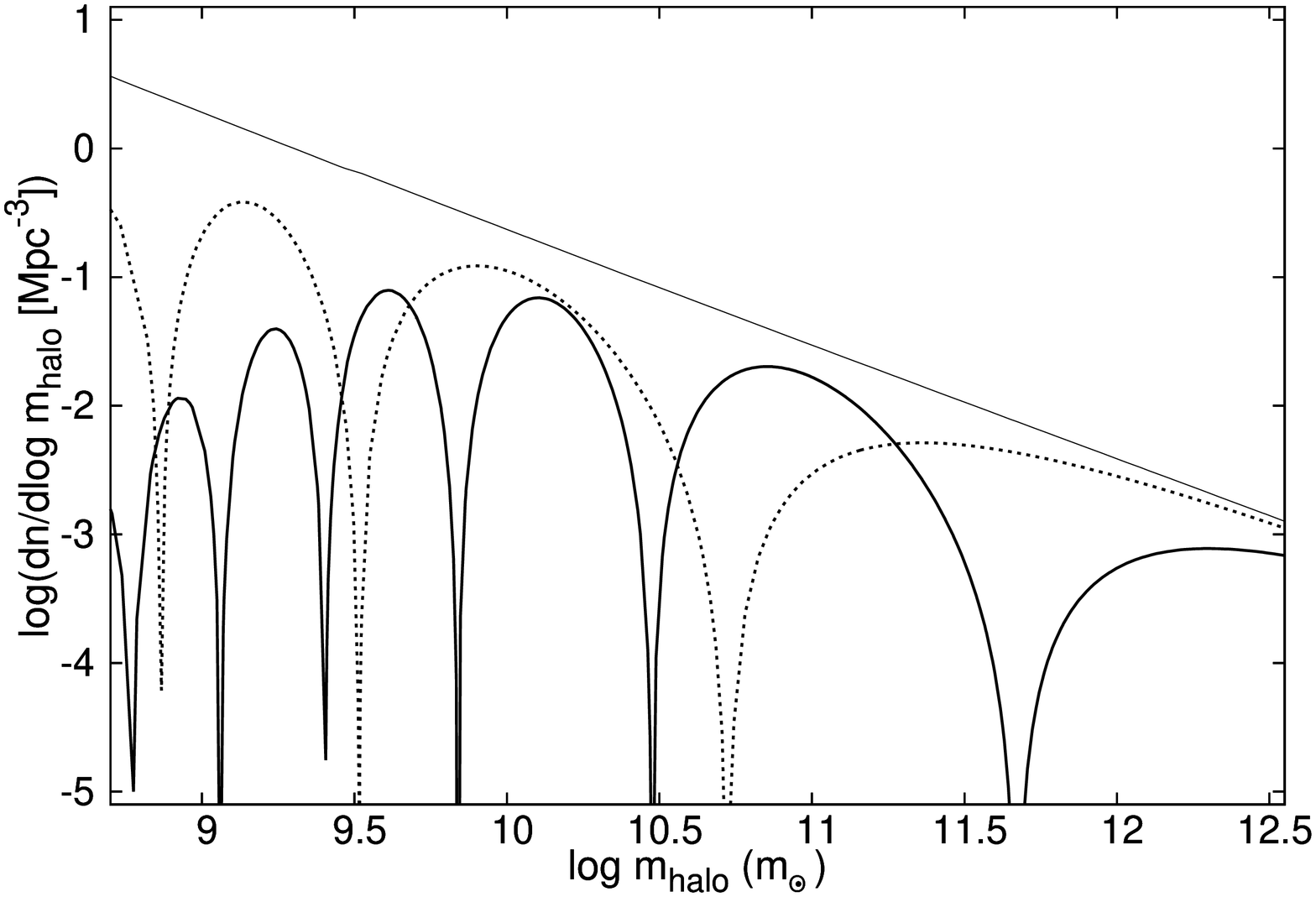}
     (a)
    \vspace{4ex}
  \end{minipage}
  \begin{minipage}[b]{0.5\linewidth}
    \centering
    \includegraphics[width=0.7\linewidth,angle=270]{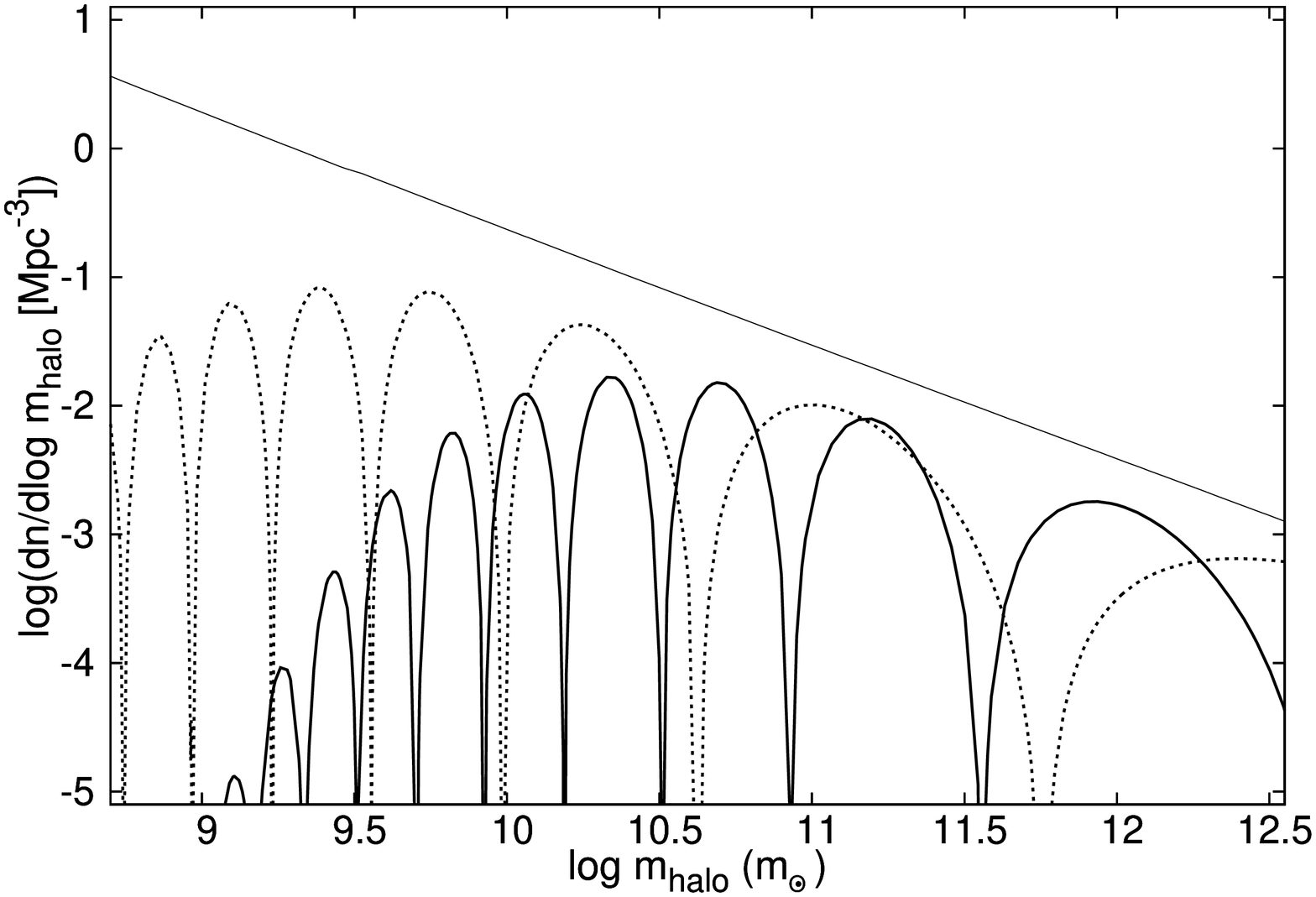}
    (b)
    \vspace{4ex}
  \end{minipage}
\vskip -1.0cm
\caption{
\small
Halo mass function calculated within the EPS formalism for the
generalized mirror dark matter model. Shown are the results for 
(a) $\epsilon = 10^{-10}$, with $N_{\rm sec} = 5$  (thick solid line),
and $N_{\rm sec} = 1$ (dashed line). The thin upper line is for $\epsilon = 0$, cosmologically equivalent to collisionless
cold dark matter. (b) same as (a) but for $\epsilon = 2\times 10^{-10}$.
} 
\end{figure}

The halo mass function was calculated by numerically solving the above set of equations
for a generic dissipative dark matter model in \cite{sunny2}.
Here, we repeat this exercise for the generalized mirror dark matter model.
The results for several examples 
are shown in Figure 2 (the parameters chosen are the same as Figure 1). 
The oscillations
in the figure are due to the dark acoustic oscillations, which are not diminished by the use
of the sharp-$k$ filter. These oscillations are expected to be much less pronounced if 
a more physical filter could be found. In any case, they would be  
smoothed out by galaxy evolutionary processes.

\subsection{The velocity function}

To compare with observations it is necessary to relate the halo mass to something more
easily observable; the baryon mass, or more ideally, the velocity function. 
Dissipative halo dynamics gives such a relation.
As will be discussed in more detail in Sec. 6, 
galaxy halos are assumed to have evolved into a steady 
state configuration,  and as a consequence, the halo density 
can be related to the Type II SN distribution in galaxies.
This allows one to conect the
halo mass to the maximum halo rotational velocity, $v_{\rm halo}^{\rm max}$.
For an exponential SN spatial distribution, the halo mass relation that follows from such considerations is:
\begin{eqnarray}
m_{\rm halo} &\simeq & \frac{8.54}{G_N} \left[ v_{\rm halo}^{\rm max} \right]^2 r_D \nonumber \\
& \simeq & 4.0\times 10^{11} m_\odot \ \left[ \frac{v_{\rm halo}^{\rm max} }{200 \ {\rm km/s}}\right]^2 \left[ \frac{r_D}{5 \ {\rm kpc}}\right]
\label{13xx}
\end{eqnarray}
where $G_N$ is Newton's constant.
Here, $r_D$ is the baryonic scale length, which is a length scale associated with the distribution of Type II supernova,
and can be approximated by the stellar disk scale length.
The above relation, which is independent of $N_{\rm sec}$, has also been obtained in the context of the mirror model \cite{footpaperII}.

\begin{figure}[t]
  \begin{minipage}[b]{0.5\linewidth}
    \centering
    \includegraphics[width=0.7\linewidth,angle=270]{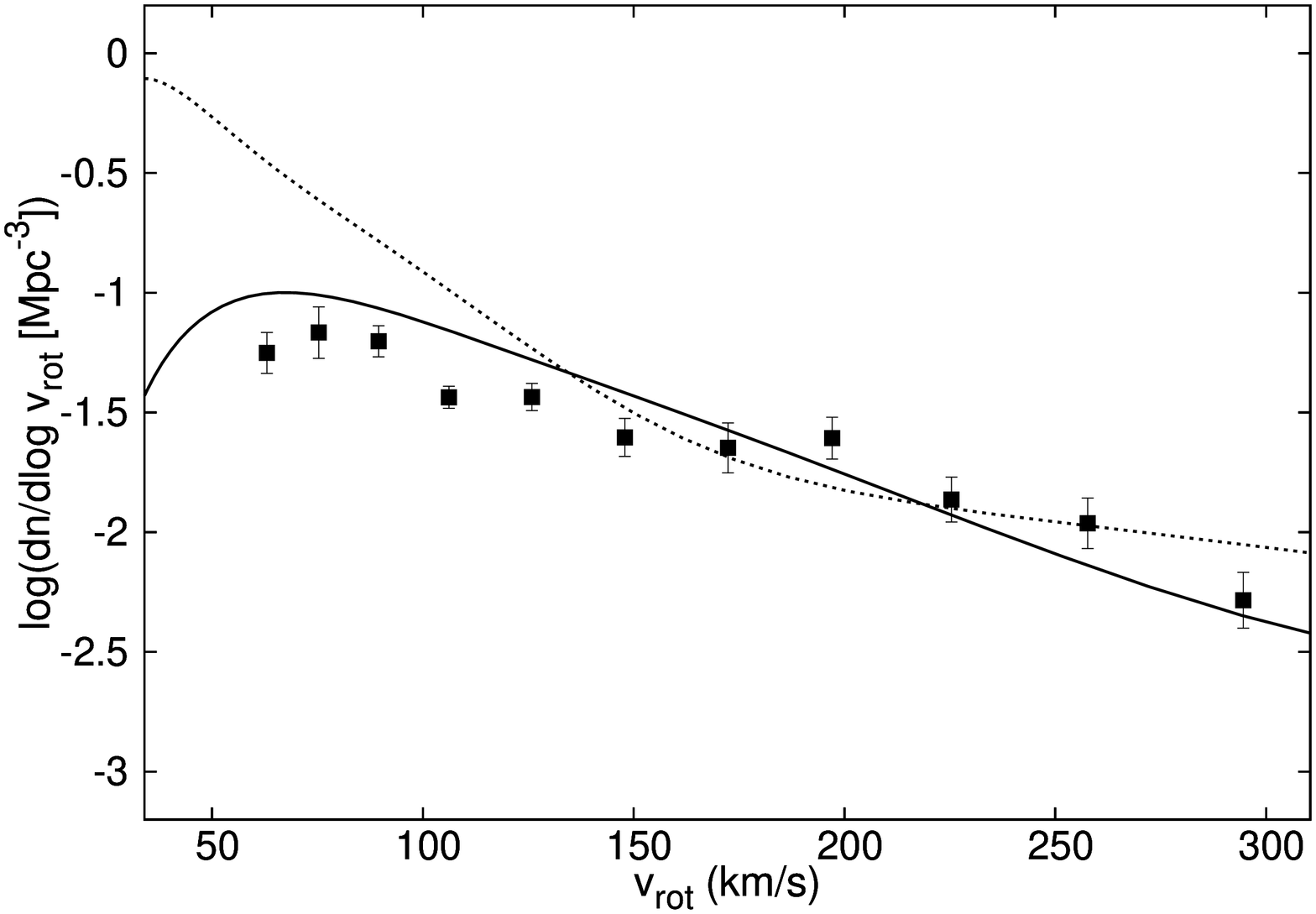}
     (a)
    \vspace{4ex}
  \end{minipage}
  \begin{minipage}[b]{0.5\linewidth}
    \centering
    \includegraphics[width=0.7\linewidth,angle=270]{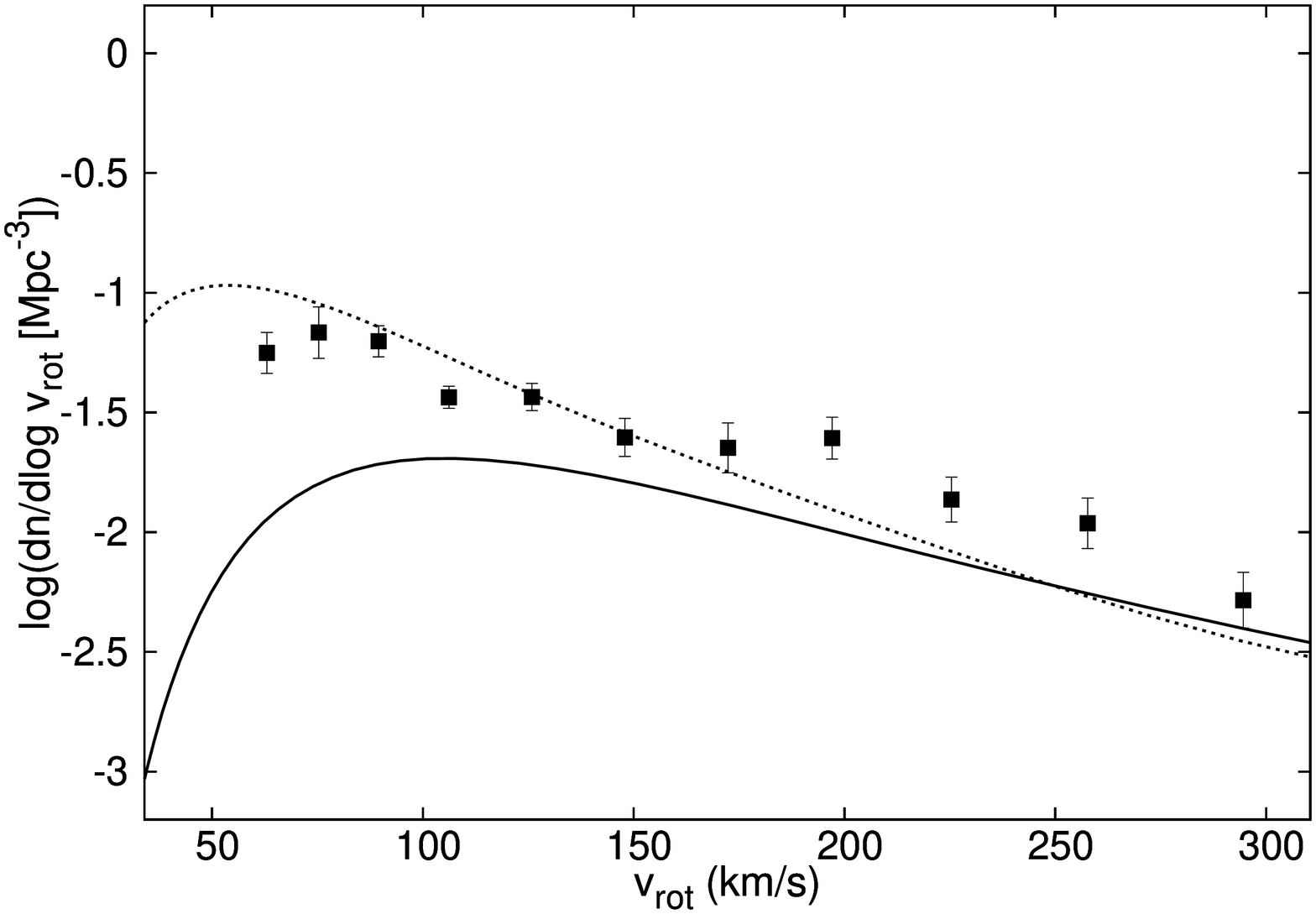}
    (b)
    \vspace{4ex}
  \end{minipage}
\vskip -1.0cm
\caption{
\small
Velocity function calculated within the EPS formalism. 
Shown are the results for generalized mirror dark matter with 
(a) $\epsilon = 10^{-10}$ with $N_{\rm sec} = 5$  (thick solid line),
and $N_{\rm sec} = 1$ (dashed line). 
(b) same as (a) but for $\epsilon = 2\times 10^{-10}$.
Data is obtained from \cite{zwaan}.
} 
\end{figure}

There are some important caveats with regard to the use of the halo mass relation, Eq.(\ref{13xx}).
Firstly, Eq.(\ref{13xx}) refers to the halo mass after a period of evolution leading to the steady state configuration. 
If there were large dark baryon outflows 
due to galaxy evolutionary processes,
then the halo mass relation would not be so useful for our purposes here, as 
such processes are not incorporated into the Press Schechter formalism.
On the other hand,
the halo mass relation would still be useful if
baryon outflows occur, as it relates the halo mass directly to
the halo rotational velocity. 
It doesn't matter that baryon outflows are not incorporated in the EPS formalism as baryons
constitute only a subcomponent of the mass.
Also,
the halo mass relation would be
invalid if there are (currently) a significant proportion of galaxies with halos not in the steady state configuration,
or without dissipative halos. In particular, it is not expected to be valid for elliptical galaxies and
some dwarf irregular galaxies  (although it may still approximate these objects at some level). 
Given these significant caveats, only a rough estimate for the velocity function
is possible.

To make contact with observations,
we shall make the simple assumption that $v_{\rm halo}^{\rm max}$ in Eq.(\ref{13xx}) can 
be identified, approximately, with the measured
asymptotic rotational velocity, $v_{\rm rot}$.
Furthermore, there are baryonic scaling relations which, as will be shown in a moment, imply a tight correlation 
between $r_D$ and $v_{\rm rot}$: 
\begin{eqnarray}
\frac{r_D}{{\rm kpc}}  \approx 3.9 \ \left[ \frac{v_{\rm rot}}{200 \ {\rm km/s}}\right]^\delta
\ . 
\label{rdr}
\end{eqnarray}
This relation, together with Eq.(\ref{13xx}) implies:
\begin{eqnarray}
\frac{m_{\rm halo}}{10^{12}\ m_\odot} \approx 0.31 \left( \frac{v_{\rm rot}}{200\ {\rm km/s}}\right)^{2 + \delta}
\ .
\label{13xxq}
\end{eqnarray}
The index $\delta$ can be estimated in the following way.
The stellar disk scale lengths are observed to be tightly correlated with baryon mass via:
\begin{eqnarray}
\log (r_D/{\rm kpc})
= s_1 [\log (m_{\rm baryon}/m_\odot)  - 10] + B
\label{s1soc}
\end{eqnarray}
where $s_1  = 0.385^{+0.008}_{-0.013}$, $B = 0.281^{+0.010}_{-0.009}$ \cite{diskrD}.
In addition, there is the baryonic Tully Fisher relation, 
\begin{eqnarray}
\log (m_{\rm baryon}/m_\odot)  =  s_2 \log(v_{\rm rot}/{\rm km/s}) + A
\label{s2soc}
\end{eqnarray}
where $s_2 = 3.75 \pm 0.11$, $A = 2.18\pm 0.23$  \cite{btfnew}.
Combining these two scaling relations gives Eq.(\ref{rdr}),
with
$\delta = s_1 s_2  = 1.44 \pm 0.06$.

The velocity function can now be determined by combining the halo mass function, computed in Sec.4.2, with Eq.(\ref{13xxq}),
which gives the $m_{\rm halo}-v_{\rm rot}$ connection.
As already mentioned, the oscillations in the halo mass function are expected to be smoothed via various processes.
To take into account this averaging, the velocity function can be convolved with a Gaussian
\begin{eqnarray}
\langle \frac{dn}{d\log v}\rangle =
\frac{1 }{\sqrt{2\pi} \sigma} 
\int e^{-\frac{1}{2}\left[ (\log \stackrel{\sim}{v} - \log v)/\sigma \right]^2 } \ \frac{dn}{d \log \stackrel{\sim}{v}} d\log \stackrel{\sim}{v}
\ .
\end{eqnarray}
In Figure 3
we plot this smoothed velocity function, with variance $\sigma = 0.15$,
for the same parameters as Figures 1,2.
Also shown in the figure is the measured velocity function from \cite{zwaan}.
That study used data from the 
Calar Alto Legacy Integral Field Area survey
(CALIFA) \cite{califa} for large galaxies,
combined with the velocity function 
measured from HI Parkes All Sky Survey
HIPASS \cite{def3} for the smaller galaxies.

\begin{figure}[t]
    \centering
    \includegraphics[width=0.48\linewidth,angle=270]{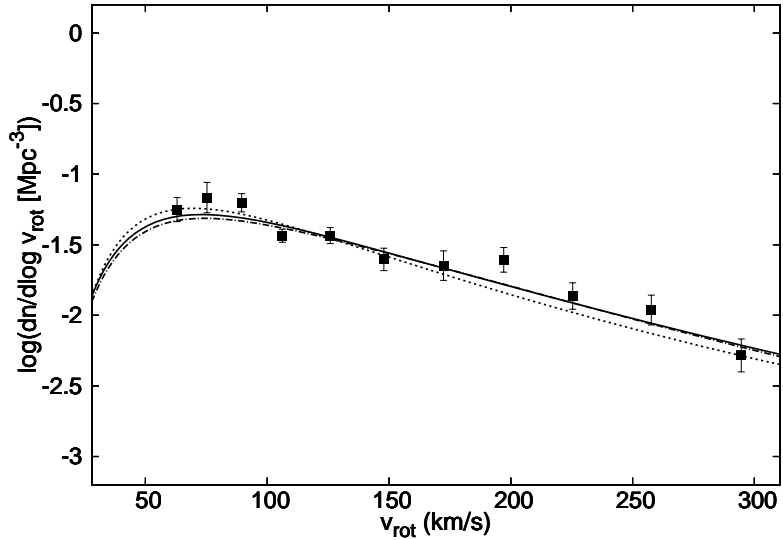}
\caption{
\small
Velocity function calculated within the EPS formalism.  Data is obtained from \cite{zwaan}.
An example near the
best fit is shown for $N_{\rm sec} = 1$ (dashed line), 
$N_{\rm sec} = 5$ (solid line), $N_{\rm sec} = 10$ (dashed-dotted line).
The $\epsilon/10^{-10}$ values are $2.4$, $1.2$ and $0.85$ respectively.
}
\end{figure}

To make a rough estimate for the region of parameter space of interest we have 
performed a simple $\chi^2$ analysis.
Specifically, we have calculated the $\chi^2$ function:
\begin{eqnarray}
\chi^2(\epsilon, N_{\rm sec})  = \sum_j \left[ \frac{ \langle \frac{dn}{d\log v} \rangle - {\rm data}[j]}{\delta {\rm data}[j]}\right]^2
\end{eqnarray}
for the data (from \cite{zwaan}) over the range $60$ km/s $\lesssim v_{\rm rot} \lesssim$ 230 km/s.
This corresponds to 9 data points in the sum. 
This range was chosen as the halo mass relation, Eq.(\ref{13xx}),
is possibly unreliable for the largest galaxies where ellipticals can dominate. 
[Although, using the entire data range from \cite{zwaan}, $60$ km/s $\lesssim v_{\rm rot} \lesssim 300$ km/s,
did not significantly alter the results obtained.]

For the purposes of this simple $\chi^2$ analysis, for each $\epsilon, \ N_{\rm sec}$,
we have minimized $\chi^2$ over variations of (a) the variance $\sigma$ ($\sigma \le 0.2$),
(b) $2+\delta = 3.44 \pm 0.20$, and (c) allowed for an overall normalization uncertainty
of $\pm 30\%$ for the predicted velocity function (due to various uncertainties inherent in the
Press Schechter approach adopted).
The resultant $\chi^2$ function was then minimized with respect to variations in $\epsilon$.
The result of this procedure was a  $\chi^2_{\rm min}/d.o.f \simeq 0.9$, which was 
roughly independent of $N_{\rm sec}$ in the range considered ($1 \le N_{\rm sec} \le 12$).
The best fit $\epsilon$ values follow 
$\epsilon \simeq 2.4\times 10^{-10}/N_{\rm sec}^{2/5}$,
a behaviour which can be understood from $L_{\rm DAO}^i$ in Eq.(\ref{s8}).
The velocity function that  results from a
kinetic mixing value near the best fit for $N_{\rm sec} = 1, 5, 10$
is shown in Figure 4.
We have also made an estimate of the 
 $\epsilon,\ N_{\rm sec}$ values  preferred by the measured 
velocity function, shown in Figure 5.
This region was obtained
by requiring  $\chi^2 \le \chi^2_{\rm min} + 25$ (roughly a $5\sigma$ allowed region).

\begin{figure}[t]
\centering
\includegraphics[width=0.48\linewidth,angle=270]{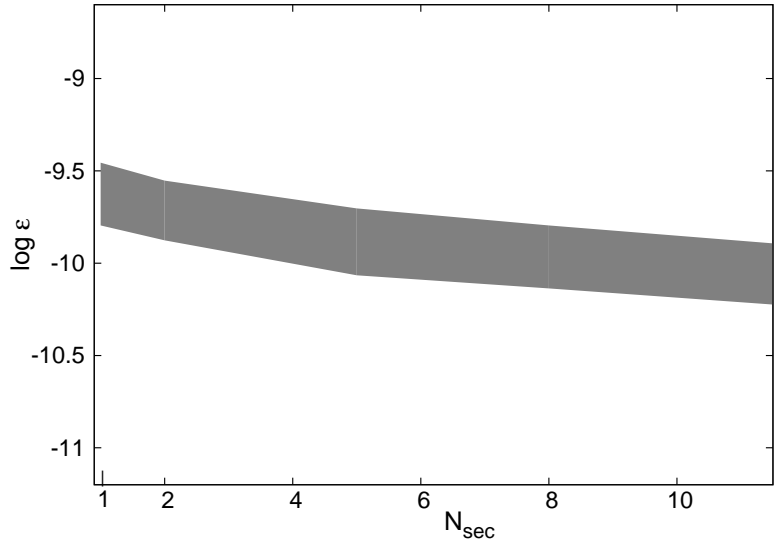}
\caption{
\small
Allowed region in the $N_{\rm sec},\ \epsilon$ plane, from a fit of the predicted velocity function
in the generalized mirror model to the data obtained from \cite{zwaan}.
}
\end{figure}

The allowed region obtained in Figure 5 is compatible with the earlier estimate, 
$\epsilon \sim 2.5 \times 10^{-10}/N_{\rm sec}^{2/3}$, 
inspired by the paucity of satellite galaxies, so that a self consistent picture emerges.
The measured flat velocity function is the result of dark acoustic oscillations,
while at smaller scales, exponential suppression from dark photon diffusion provides a sharp
cutoff. 
This cutoff is responsible for the observed deficit of small satellite galaxies. Actually,
the number of satellites
observed around the Milky Way and Andromeda
is more than would be expected, so these satellite galaxies presumably originated
out of the collapse of larger scale density perturbations, i.e. in a top-down fashion,
as will be discussed in  more detail in the following section.

\section{Origin of satellite planes}


Observations of the satellite galaxies orbiting around the two largest galaxies in
the local group, the Milky Way and Andromeda, 
indicate that their distribution is anisotropic.
In each of these galaxies, many of the satellites have been found to orbit in a co-rotating planar
distribution. 
Considering Andromeda, it was found \cite{sat2,sat2b} that about half of the
satellites belong to a vast extremely thin planar structure,
about 400 kpc in diameter but only about 14 kpc in thickness. 
Of the 15 satellites in this thin plane, 13 of
these were found to be co-rotating sharing the same direction of angular momentum.
Simulations with collisionless dark matter indicate that the probability of such an occurrence 
is only around 0.1\% \cite{1799}. 
A qualitatively similar planar distribution has been found to exist around the Milky Way \cite{satmw1,satmw2,satmw3,satmw4}, 
and also, Centaurus A \cite{sat3,sat3b}. See \cite{paww} for a review and more extensive bibliography.

Exponential power suppression on small scales suggests that the satellite galaxies could not have
arisen from the growth of tiny perturbations seeded in the Early Universe.
Presumably, the satellite galaxies formed top-down, i.e. out of larger mass scale perturbations.
As reviewed in Sec.2, and discussed in \cite{sunny2} (see also \cite{zurab1,lisa1}), 
it is envisaged that the satellite galaxies arose as a consequence
of dark disk formation; the satellite galaxies formed from perturbations at the
edge of the dark disk during (or shortly after) the complicated nonlinear dissipative collapse.
The satellites should be very old, forming around the same time as the oldest baryonic stars.
Despite the inherent complexities in such a formation scenario,
this picture has the potential to provide a very simple explanation for the observed planar distribution of the
satellite galaxies around the Milky Way, Andromeda and Centaurus A.
 It seems particularly important, therefore, to try to understand as much
as possible about this early phase of galaxy formation and evolution.   

According to the standard picture of galaxy formation, a galaxy mass scale perturbation evolves 
linearly initially, governed by Boltzmann and Einstein equations. Eventually such a perturbation
reaches a critical overdensity, $\delta \sim 1$, and gravitational collapse begins.
During the nonlinear collapse, the baryonic and dark matter initially free fall, with the kinetic 
energy of the infalling matter converted to thermal energy via shock heating.
Since both baryonic matter and dark matter are dissipative, cooling processes can occur in each sector,
and it is anticipated that baryonic and dark disks can form.  
Naturally, these disks could only have been formed if the collapsing cloud had some initial
angular momentum. The standard theory
describing angular momentum generated via tidal torques, e.g. \cite{hoyle,peebles,peebles2,peebles3,peebles4},
could be applied to this dissipative dark matter model. 
Some 
important 
modifications are likely, e.g. the angular
momentum losses due to dynamical friction can be mitigated due to the suppression of
small scale structure cf.\cite{dolgov}.

If dark and baryonic disks do indeed form at this early stage of galaxy formation, then their evolution
is rather important.
A pertinent discussion about the formation and evolution of dark disks has been given in a somewhat related kind of dissipative 
dark matter model in \cite{Fan}. 
There are important differences: The dissipative dark matter 
was assumed in that reference to be a subcomponent of the dark matter and there was no kinetic mixing induced heating.
Under those conditions, the authors of \cite{Fan} argued that the baryonic and dark disks would form and evolve under gravity
until the disk planes merged. 
If the satellite galaxies were indeed formed out of perturbations at the edge of the dark disk,
then the satellite plane should be aligned with the plane of the baryonic disk; this would be at odds with the
orthogonal orientation of the planar distribution of Milky Way satellites.
It would also be at odds with the orientation of the satellite plane around Andromeda, which is observed to be inclined to the baryonic disk by 
$\theta_{\rm disk} \approx 51^\circ$ \cite{sat2b}.

\begin{figure}[t]
\centering
\includegraphics[width=0.48\linewidth,angle=0]{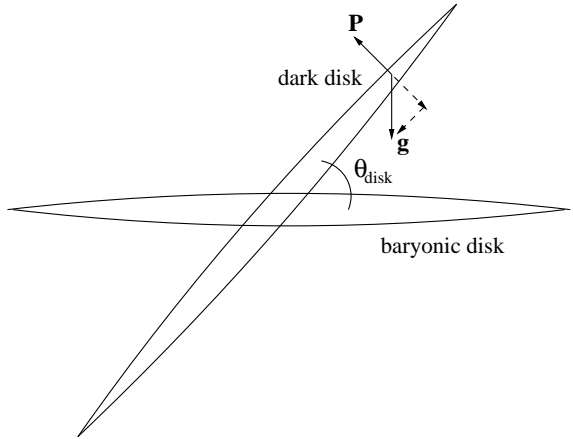}
\caption{
\small
The geometry: edge on view of the dark and baryonic disks. 
}
\end{figure}

Gravity is not the only force that can act on the two disks.
A pressure force could arise
if either (or both) of the disks were heated from heat sources in the other disk. 
As will be discussed in more detail in Sec.6, kinetic mixing induced processes in
ordinary (baryonic) SNe will produce a huge dark energy release
heating dark baryons directly and via dark radiation.
The disks have finite thickness 
and the two sides of the disk would be heated
unevenly (assuming that $\theta_{\rm disk} \neq \{0,\ \pi/2\}$), that is the side of the 
disk facing the other disk would be the hot side.  This uneven heating would create a 
temperature gradient across the disk thickness, $\partial T/\partial z$, 
and also a pressure force, 
$\partial P/\partial z \sim n\partial T/\partial z$. 
(See the appendix for a crude estimate of the pressure force.)
In fact, this force will be in the direction perpendicular to the disk plane opposing gravity (Figure 6). 
Baryonic SN can also produce a substantial dark baryonic wind, and conversely, dark SN
can produce a baryonic wind. These winds could  also contribute to the pressure, and may well
dominate over the other effects.


If the pressure force overwhelms gravity 
then the dark and baryonic disks would evolve until they are orthogonal, 
$\theta_{\rm disk} = \pi/2$.  
The expected time scale is short, $t  \sim 10$ Myr cf.\cite{Fan}.
Satellites which formed out of material at the edges of the dark disk could
presumably share the same plane as the dark disk given the short time scale of the disk evolution.
This seems to provide a possible explanation for the orthogonal disk of satellites observed around the Milky Way.
\footnote{In the absence of a pressure force, a consistent picture for the Milky Way satellites might still arise if the 
satellites  
originated as tidal dwarf galaxies formed during an ancient merger
event \cite{pav1,pav2}. In that scenario they can be still be dark matter dominated because they formed from 
material of the merged dark and baryonic disks \cite{zurab1,lisa1}.}
In the case of Andromeda, the satellite plane is not orthogonal, $\theta_{\rm disk} \approx 51^\circ$.
There appears to be a readily available explanation though, namely that
this offset is the result of a major merger event. 
In fact, there is evidence that Andromeda underwent a 
major merger just 2-3 Gyr ago \cite{hammersave}. 
The Milky Way, on the other hand, is known to have a 
relatively quiet merger history, e.g. \cite{hammerq}, which could explain why its 
satellite plane remains relatively unperturbed with an approximately polar orientation.
This all seems very convenient. Naturally, one might need to check that 
such a violent incident in Andromeda's past could 
disrupt the baryonic disk, affecting its orientation, but leaving the plane of satellites largely undisturbed.

The stability of satellite planar systems, especially if formed very early ($\sim 10$ Gyr ago) 
as in the scenario advocated here,
is a potentially serious issue. However, the polar alignment with respect
to the baryonic disk  is one of the few configurations where a thin satellite disk can
persist over a cosmological time scale \cite{bowden,fernandoI}.
Also, destabilizing effects due to dark subhalos, which can be quite debilitating \cite{fernando},  
are expected to be greatly mitigated in the
dissipative dark matter model due to the small scale power suppression effects discussed in Sec.4.
Of, course, at $\theta_{\rm disk} \sim 51^\circ$, the plane of satellites around Andromeda would not be expected to survive very long,
so the relatively recent nature of the major merger event, possibly just $2$ Gyr ago \cite{hammersave},
appears to be an important consideration. 

The particular scenario advocated here suggests that satellite planes around host galaxies should be common
in the Universe. The orientation of this plane should be preferentially polar,
especially around host galaxies with quiet merger histories.
The dwarf satellite galaxies are interesting in their own right, especially as they originate (in this picture) in a way which is quite different to other galaxies. 
The satellites 
are not primordial, but are `top down' forming structures, which grew out of perturbations at the edge of this disk, 
either during the disk formation or shortly after. 
They can be strongly dark matter dominated, and have internal properties quiet different to isolated field galaxies.

To summarize, galaxy evolution is envisaged to proceed in two stages.
In the first stage, dark and baryonic disks formed and evolved. In the second stage the dark gas 
component expanded to form an extended distribution due to heating 
from Type II SNe once substantial star formation began. 
In this section we have qualitatively discussed the early stage of evolution where dark and baryonic disks formed.
The disk forming nonlinear collapse process is not expected to be uniform, and perturbations leading to
small satellite galaxies could thereby have arisen.
If this formation mechanism is correct, then the satellite galaxies are anticipated to be 
co-rotating and orbit in the same plane as the dark disk, kinematic features consistent with observations 
of the satellites around the Milky Way \cite{satmw1,satmw2,satmw3,satmw4}, Andromeda \cite{sat2,sat2b} and Centaurus A \cite{sat3,sat3b}.
Moreover, the planar distribution of satellite galaxies with respect to the baryonic disk can be preferentially polar,
an orientation which appears to be necessary to explain the observations.



\section{Halo dynamics: Heating, cooling, and analytic considerations}
\subsection{Some preliminaries}

In the previous section the first stage of galaxy evolution was discussed, where dark and baryonic disks formed and evolved.
In the second stage of galaxy evolution, the dark gas component of the dark disk is assumed to be heated by ordinary Type II SNe to form a roughly spherical
distribution.
The evolution of such a system is quite complex, with the dark matter gas component governed by fluid equations, influenced by both heating
and cooling processes.
In general, one would expect three possible outcomes of this evolution:
\vskip 0.3cm
\noindent  
(i) There is insufficient heating of the halo, so that a long lived spherical halo does not arise. In such a situation the dark matter would presumably evolve
into compact dark stellar mass objects (dark stars).
\vskip 0.3cm
\noindent  
(ii) There is excessive heating of the halo, so much so that runaway expansion occurs. This would leave the galaxy largely devoid
of dark matter.
\vskip 0.3cm
\noindent  
(iii) The halo evolves into a long lived steady state configuration. In such a configuration, heating and cooling rates locally
balance and the dark plasma is in hydrostatic equilibrium.
\vskip 0.3cm
\noindent  
Halo evolution to the steady state configuration (iii), it turns out, leads
to various scaling relations which closely resemble empirical observations
(see \cite{footpaperII} and references therein).
This result has lead to the conjecture that
the dark halo around (typical) disk galaxies
is in a steady state configuration.
Even if the halo around most disk galaxies
is currently in a steady state configuration,
all three types of systems listed above should exist in the Universe, and specific candidate examples will be discussed.


For a steady state solution to arise from this dynamics appears at first sight to be quite nontrivial.
To describe the qualitative picture evisaged, consider the heating and cooling rates
suitably averaged over the halo, $\langle {\cal H} \rangle$, $\langle {\cal C} \rangle$.
The idea is that a roughly spherical halo  
is formed after the onset of star formation, due to SN sourced halo heating, with $\langle {\cal H} \rangle \gg \langle {\cal C} \rangle$
initially.
Once the halo is heated to the point where it is roughly spherical, further expansion would
be expected to reduce the cooling rate
$\langle {\cal C} \rangle$. This is expected because of the dependence of ${\cal C}$ on the halo density, $\rho$, which reduces as the halo expands 
[${\cal C} \propto \rho^2$, given that radiative cooling results from two body collisional processes].
One might imagine that runaway expansion is the inevitable outcome.
However, as the halo expands the heating rate is also modified. The expanding halo will result in a more diffuse baryonic
distribution in the weakening gravity, which can lead to a reducing star formation rate.
Thus, it is possible for $\langle {\cal H} \rangle - \langle {\cal C} \rangle$ to decrease in an expanding halo.
That is, a spherical halo with  $\langle {\cal H} \rangle > \langle {\cal C} \rangle$ can expand, but can evolve dynamically to a point
where $\langle {\cal H} \rangle = \langle {\cal C} \rangle$.
In such a situation, one expects the system to relax into a true steady state configuration where the local
heating and cooling rates balance, ${\cal H} = {\cal C}$.




The above picture was developed within the framework of the mirror model, i.e.  the minimal $N_{\rm sec} = 1$ case.
The models with $N_{\rm sec} > 1$ have important differences.
Indeed, for $N_{\rm sec} > 1$, the halo is composed of several fluids weakly coupled to each other via gravity, but strongly coupled to baryons via the SN sourced heating.
It is quite unclear whether such a multifluid halo could reach a steady state configuration which is dynamically stable.
In particular, it seems much more nontrivial than the $N_{\rm sec} = 1$ model.
In the remainder of this paper, though,
we shall study the steady state solution, and leave open the important question as to whether such a solution is dynamically stable.


\subsection{The fluid equations}



The dark halo gas component is governed by fluid equations, and if
dark magnetic fields can be neglected, then these equations take the form:
\begin{gather}
\frac{\partial \rho}{\partial t} + \nabla \cdot (\rho \mathbf{v}) = 0
\nonumber \ , \\
\frac{\partial \mathbf{v}}{\partial t} + (\mathbf{v} \cdot
\nabla)\mathbf{v} = -\left ( \nabla \phi + \frac{\nabla P}{\rho} \right)
\nonumber \ , \\
\frac{\partial}{\partial t} \left [\rho \left ( \frac{\mathbf{v}^2}{2} +
{\cal E} \right ) \right ] + \nabla \cdot \left [\rho \left (
\frac{\mathbf{v}^2}{2} + \frac{P}{\rho} + {\cal E} \right ) \mathbf{v}
\right ] - \rho \mathbf{v} \cdot \nabla \phi = {\cal H} - {\cal C}
\ .
\label{euler}
\end{gather}
Here $P$, $\rho$ and $\mathbf{v}$, denote the pressure, mass density and
velocity of the fluid, and $\phi$ is the gravitational potential. 
${\cal E}$ is the internal energy per unit mass
of the fluid, so that $\rho \left (\mathbf{v}^2/2 + {\cal E} \right )$
is the energy per unit volume. Finally, ${\cal H}$ and ${\cal C}$ 
are the local heating and cooling rates per unit volume. 

In general, the system is a complicated one, as the dark matter is coupled to the baryons via SN
sourced heating, and the baryons are coupled to the dark matter via gravity.
If the system is able to evolve to a steady state configuration (which, as mentioned earlier, is far from
certain in the $N_{\rm sec} > 1$ case), then significant simplifications arise.
Assuming there is no steady state velocity flow, the system is 
governed by just two time-independent equations, Eq.(\ref{SSX}).
These equations have been numerically solved for the mirror dark matter case in \cite{footpaperI,footpaperII},
and we aim to extend this work here by considering the case with $N_{\rm sec}$ dark SM sectors.
To solve these equations we will first need to model the local heating and cooling rates.
Since many details are similar to the mirror dark matter case, 
in the ensuing discussion we mainly focus on the details which
depend on the number of dark sectors, $N_{\rm sec}$.

\subsection{Heating}

In galaxies with active star formation, including spirals and
gas rich dwarf irregular galaxies, type II supernovae can provide the primary halo heat source \cite{sph}.
Type II supernovae are important because the core temperature is high enough for the production
of dark sector electrons and positrons, and these particles can be copiously produced
if the kinetic mixing interaction exists. In fact, for $\epsilon \sim 10^{-9}$ the amount
of energy transferred to dark sector particles in an ordinary SN can compete with the energy carried away by
the neutrinos. 

The amount of energy transferred to dark sector particles from Type II supernovae can be estimated 
from \cite{raffelt,raffelt2}. 
The effect of kinetic mixing is to embellish the dark sector charged particles with a tiny electric charge, so that
the $i^{th}$ sector dark electrons have electric charge $-\epsilon e$.
Dark electrons and positrons are the only dark sector charged particles that can be produced in a SN given
the typical core collapse temperature of $T_{\rm SN} \sim 30$ MeV.
According to \cite{raffelt,raffelt2}, 
the dominant production mechanism is via plasmon decay, $\gamma^* \to \bar e_i e_i$, and the energy loss rate
for each sector 
is estimated to be:\footnote{
There are additional contributions to $\bar e_i e_i$ production from nucleon-nucleon bremsstrahlung, potentially
comparable to the plasmon decay process \cite{mohap9}, but are uncertain due to nonperturbative effects.
If these processes are important then the rate, Eq.(\ref{raflx}), can be enhanced, and in any case, their
possible significance increases the uncertainty in the estimate, Eq.(\ref{11yyy}).
}
\begin{eqnarray}
Q_P = \frac{8 \zeta_3}{9 \pi^3}  \epsilon^2 \alpha^2 
\left( \mu^2_e + \frac{\pi^2 T_{\rm SN}^2}{3}\right) T_{\rm SN}^3 Q_1
\label{raflx}
\end{eqnarray}
where $\zeta_3 \simeq 1.202$ is the Riemann zeta function,
$Q_1$ is a factor of order unity and $\mu_e$ is the electron chemical potential.

Putting in typical SN parameters, 
the amount of energy transferred to dark sector particles in the core of a Type II supernova is estimated to be:
\begin{eqnarray}
L_{\rm SN}^{\rm dark} \approx N_{\rm sec} \left( \frac{\epsilon}{10^{-10}}\right)^2 \ 10^{51} \ \ {\rm erg}
\ .
\label{11yyy}
\end{eqnarray}
This calculation is valid provided that $L_{\rm SN}^{\rm dark} \lesssim 10^{53}$ erg, so that the cooling time
of the SN core is not greatly affected, consistent with 
the observations of around a dozen neutrino events associated with SN1987A \cite{snref1,snref2}.
The requirement that
$L_{\rm SN}^{\rm dark} \lesssim 10^{53}$ erg,
yields
\begin{eqnarray}
N_{\rm sec} (\epsilon/10^{-10})^2 \lesssim 10^2 \ .
\label{ul}
\end{eqnarray}
For $N_{\rm sec} = 1$ this constraint gives the well-known SN bound of $\epsilon \lesssim 10^{-9}$ \cite{raffelt2}.

The kinetic mixing induced processes in the Type II SN core generate an expanding energetic plasma,  
initially comprising mainly the light dark sector particles, $e_i, \bar e_i, \gamma_i$, 
with total energy of at least $10^{51}$ erg for the kinetic mixing strength range considered.
By analogy with the fireball model of Gamma Ray Bursts \cite{grb1,grb2,grb3,grb4,fireball},
it is expected that the dark plasma evolves into a relativistic fireball and
sweeps up the nearby dark baryons as it propagates away from the SN.
The energy of the fireball is anticipated to be transferred to the dark baryons.
This flow eventually decelerates, and part of this kinetic energy
is converted back into thermal energy which can radiatively cool producing dark radiation.
The end result is that the energy sourced from ordinary Type II SNe is transmitted to the halo 
in two distinct ways: via dark photons and also 
via the heating of dark baryons in the SN vicinity.

In the fireball model many details, including the proportion of energy transferred into radiation 
versus local heating of the swept up baryons, remains uncertain.
In this paper we follow \cite{footpaperII} and assume that the local heating of the dark 
baryons in the SN vicinity dominates the halo heating.
With this assumption, the (average) halo heating rate per galaxy is given by:
\begin{eqnarray} 
\kappa = R_{\rm SN} L_{\rm SN}^{\rm dark}
\end{eqnarray}
where $R_{SN}$ is the rate of Type II SNe in the galaxy under consideration.
Using Eq.(\ref{11yyy}), we then have:
\begin{eqnarray}
\kappa \approx N_{\rm sec} \left( \frac{\epsilon}{10^{-10}}\right)^2 
\left[ \frac{R_{\rm SN}}{0.1\ {\rm yr}^{-1}}\right]  \ 3 \times 10^{42} \ {\rm erg/s}
 \ .
\label{st2}
\end{eqnarray}
If the dark sector consists of $N_{\rm sec}$ copies of the Standard Model, 
then the symmetric kinetic mixing interaction, Eq.(\ref{mix1}),
indicates that each copy would receive the same energy, so that
\begin{eqnarray}
\kappa_1 = \kappa_2 = ...= \kappa_{N_{\rm sec}} = \kappa/N_{\rm sec}
\ .
\label{kap1}
\end{eqnarray}
These equations give the rate of SN sourced heating per galaxy. To proceed, we need to model
how this heating is distributed within the galaxy, since the primary item of interest for the fluid
equations is the {\it local} halo heating, ${\cal H}$.

As in earlier work, we consider idealized spherically symmetric galaxies. 
That is, for a given disk galaxy we construct a spherically symmetric analogue system where both
the baryons and dark matter have spherically symmetric distributions.
A spherically symmetric baryon stellar distribution can be obtained  by requiring
$\int \rho dV = \int \Sigma dA$ which implies $\rho = \Sigma/2r$,
where $\Sigma$ is the stellar disk surface density.
Taking the Freeman surface density: $\Sigma = e^{-r/r_D}/(2\pi r_D^2)$, 
which is known to be a reasonable description for the star 
forming region of many spirals and dwarfs \cite{freeman}, we have:
\begin{eqnarray}
\rho_{\rm baryon}^{*} (r) = m_{*} \ \frac{e^{-r/r_D}}{4\pi r_D^2 r}
\ .
\label{doc}
\end{eqnarray}
Here, $m_*$ is the stellar mass parameterized in terms of a stellar
mass fraction: $m_* = f_s m_{\rm baryon}$.
In addition to stars, there is also a baryonic gas component,
which we model with a spatially more extended distribution of the form
Eq.(\ref{doc}), but with
$r_D^{\rm gas} = 3r_D$ and total mass $m_{\rm gas} = (1-f_s)m_{\rm baryon}$ cf. \cite{sal17}.
Taking ${\cal H}_i \propto \rho_{\rm baryon}^{*}$, and normalizing the distribution
so that $\int {\cal H}_i dV = \kappa_i$ gives 
\begin{eqnarray}
{\cal H}_i = \frac{\kappa_i e^{-r/r_D}}{4\pi r_D^2 r}
\ .
\label{h1}
\end{eqnarray}
This is the SN sourced halo heating component, which is the primary halo heat source. 
In addition, one can have secondary heating from halo re-absorption of cooling radiation,
and conduction/convection processes can also, in general, contribute to the local heating.

For the purposes of this paper, we shall only consider generic spherical galaxies with the exponential distribution, Eq.(\ref{h1}). 
Nevertheless, we digress here to mention that for a specific galaxy, 
a better representation for ${\cal H}_i$ would be given in terms of the luminosity density of the UV sources, ${\cal L}_{\rm UV}$,
as these should more accurately map the distribution of large stars that are the progenitors of Type II SNe.
Since we are considering idealized spherically symmetric systems, which are supposed to represent actual galaxies at some level,
the luminosity density of UV sources could be modelled from the UV surface brightness of the disk via: ${\cal L}_{\rm UV} = \Sigma_{\rm UV}/2r$.
In this case, the SN sourced halo heating component takes the form:
\begin{eqnarray}
{\cal H}_i = \frac{\kappa_i \Sigma_{\rm UV}}{2r L_{\rm UV}}
\label{h4}
\end{eqnarray}
where $L_{\rm UV} = \int {\cal L}_{\rm UV} dV = \int \Sigma_{\rm UV}  dA$ is the galaxy's UV luminosity.
Furthermore, the rate of Type II SN is expected to be proportional to the UV luminosity of a galaxy,
so that $\kappa_i/L_{\rm UV}$ is anticipated to be a galaxy-independent constant.
Note that the UV surface brightness can be directly measured, e.g. \cite{galex,evergreen}, so that in principle, 
Eq.(\ref{h4}), provides a fairly constrained description of the SN heat sources.

\subsection{Cooling}

The radiative cooling processes, bremsstrahlung, line emission, and dark electron capture, each involve two particle
collisions and thus have a cooling rate proportional to the density squared: ${\cal C}_i = \Lambda \rho^2_i $.
If each of the dark sector components satisfies the steady state condition, Eq.(\ref{SSX}), then ${\cal H}_i = {\cal C}_i$.
Since each sector is heated the same [Eq.(\ref{kap1})], and each sector has identical particle properties, we expect the
density of each component to be equal: 
\begin{eqnarray}
\rho_1 = \rho_2 = ....=\rho_{N_{\rm sec}} = \rho/N_{\rm sec}
\end{eqnarray}
where $\rho = \sum \rho_i$ is the total dark sector mass density.
The radiative cooling rate in each sector therefore takes the form:
\begin{eqnarray}
{\cal C}_i = \Lambda  \rho^2/N_{\rm sec}^2
\label{15x}
\end{eqnarray}
and the total radiative cooling rate is ${\cal C} = N_{\rm sec} {\cal C}_i$, i.e.
\begin{eqnarray}
{\cal C} = \Lambda \rho^2/N_{\rm sec}
\ .
\label{15xb}
\end{eqnarray}
The cooling rate rapidly goes down as we consider models with a larger number of isomorphic sectors, $N_{\rm sec} \gg 1$.
In general, $\Lambda$ depends on the chemical composition, ionization state and temperature.
The ionization state can be determined in the steady state limit by balancing dark electron capture with the
ionization processes, and in the low density optically thin limit, the ionization state depends on temperature only.
The technical details, including the cross sections used, are given in \cite{footpaperI}.

Consider now the chemical composition.
Dark helium is synthesized in the early Universe, much like ordinary helium, but with one major difference:
Dark helium synthesis occurs significantly earlier given that $T_i < T$.
In fact as $\epsilon \to 0$, the freeze-out temperature of 
$n' \leftrightarrow p'$ weak interactions
becomes much greater than the neutron-proton mass difference,
so that $n_{p'} \simeq n_{n'}$ results.
That is, in this $\epsilon \to 0$ limit 
the primordial helium mass fraction, $Y'_p \equiv \rho_{\rm He'}/(\rho_{\rm H'} + \rho_{\rm He'}) \to 1$.
Of course, the kinetic mixing parameter of interest is not zero, 
and it is found that for $\epsilon \sim 2 \times 10^{-10}$
the primordial dark helium mass fraction is around 0.95 \cite{p2,footreview}. 
Those calculations were done in the $N_{\rm sec} = 1$ model, but since
Eq.(\ref{2x}) is valid for each sector for $N_{\rm sec} \ge 1$ 
[Eq.(\ref{goodeq})],
this result also follows for each sector in the general case.

\begin{figure}[t]
\centering
\includegraphics[width=0.48\linewidth,angle=270]{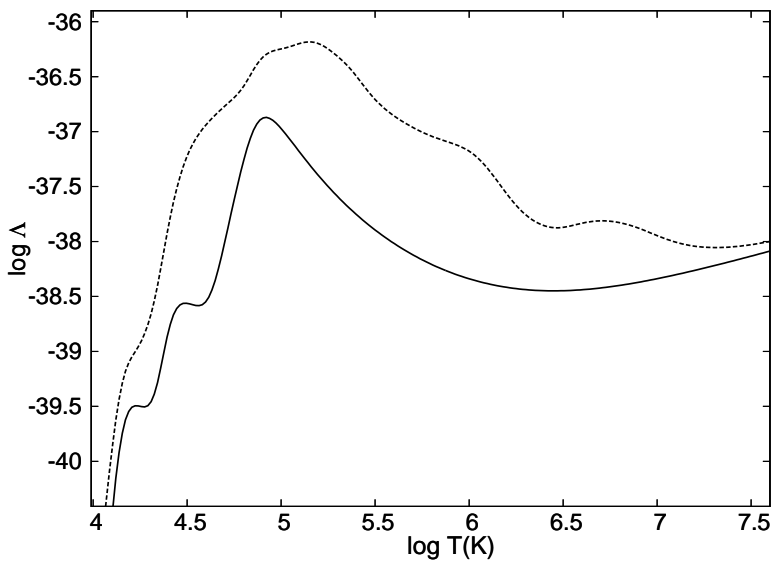}
\caption{
\small
Optically thin cooling function, $\Lambda$ [${\rm erg/cm}^3/{\rm s} ][({\rm kpc}^6/m_\odot^2]$,
defined in Eq.(\ref{15x}), computed for the generalized mirror dark matter model.
The chemical composition of each of the $N_{\rm sec}$ dark sectors is assumed identical, with
${\rm log} [n_{\rm He'}/n_{\rm H'}] = 0.68$, and a dark metal mass fraction, $f_{\rm metal}$ = 0 (solid line),
$f_{\rm metal} = 0.02$ (dashed line). See text for further details.
}
\end{figure}

A dark metal component might also exist, generated in dark stars at an early epoch.
Naturally, this metal component is uncertain both in composition and overall abundance.
In each sector,
we consider a metal component consisting of elements ${\rm C'}, \ {\rm O'}, 
\ {\rm Ne'}, \ {\rm Si'},\ {\rm Fe'}$, with relative proportion: $n_{\rm A'}/n_{\rm H'} = n_{\rm A}/n_{\rm H}$ (solar abundance).
The overall abundance can be parameterized by introducing the dark metal mass fraction: 
$f_{{\rm metal}} = (\rho - \rho_{\rm H'} - \rho_{\rm He'})/\rho$.
[Here $\rho_{\rm H'}$, $\rho_{\rm He'}$ are the mass density of all dark sector ${\rm H'}$ and ${\rm He'}$ components.]
Having defined the chemical composition,
the cooling function can now be computed
using the code developed in \cite{footpaperI}.
The results are shown in Figure 7
for $f_{\rm metal} = 0$ and $f_{\rm metal} = 0.02$. 

In the absence of heat sources the dark plasma would cool.
The cooling time scale for a dark plasma of mass density $\rho$ and temperature $T$ is estimated to be:
\begin{eqnarray}
t_{\rm cool} \sim \frac{3\sum n_A T/2}{{\cal C}} 
             = \frac{3T N_{\rm sec}}{2\bar m \Lambda \rho}
\label{cooly}
\end{eqnarray}
where $\bar m$ is the mean mass parameter:
\begin{eqnarray}
\bar m \equiv \frac{\rho}{\sum n_A} = \frac{\sum n_A m_A}{\sum n_A}
\ .
\label{sum}
\end{eqnarray}
Here, the sum runs over all matter particles in each sector, the dark ions and free dark electrons.
For a fully ionized dark sector plasma, with a dark helium dominated composition ($Y'_p = 0.95$),
we have $\bar m \simeq 1.16$ GeV. Considering typical parameters for the halo of a Milky Way scale galaxy yields:
\begin{eqnarray}
t_{\rm cool} \sim N_{\rm sec}  
\left( \frac{T}{0.5\ {\rm keV}}\right)
\left( \frac{0.3 \ {\rm GeV/cm^3}}{\rho}\right) 
\left( \frac{10^{-38.5}\ 
[{\rm erg/cm^3/s}][{\rm kpc^6/m_\odot^2}]}{\Lambda} 
\right) \ 50\ {\rm Myr}\ .
\notag \\
\label{cts}
\end{eqnarray}
Observe that $t_{\rm cool} \propto N_{\rm sec}$, and can extend to several hundred million years for the $N_{\rm sec} = 5$ case.
In principle, the halo heating would need to be averaged over this time scale, so any `graininess' effects arising from the discrete nature
of SN, in both time and space, is not expected to be important.

\subsection{Analytic halo density}

If the halo is optically thin, and 
if the SN sourced heating component dominates the halo heating, then
a useful analytic estimate for the dark halo mass density
arises by matching this heat component [Eq.(\ref{h1})] with the cooling rate [Eq.(\ref{15x})].
That is, the steady state condition ${\cal H} = {\cal C}$ can be solved to give
the halo density:
\begin{eqnarray}
\rho(r) =   \frac{\sqrt{N_{\rm sec} \kappa}  }{\sqrt{4\pi \Lambda}\ r_D^{3/2}}  \frac{e^{-r/2r_D}}{\sqrt{r/r_D}} 
\label{r5x}
\end{eqnarray}
where $\kappa \equiv \sum_i \kappa_i$.
This equation generalizes the result obtained for mirror dark matter ($N_{\rm sec} = 1$) given in \cite{footpaperII}.
For an isothermal halo, $\Lambda$ is approximately a spatial constant.

The halo rotation
curve which results from the analytic density profile, Eq.(\ref{r5x}), with spatially constant $\Lambda$, follows from Newton's law and is given by
\begin{eqnarray}
v_{\rm halo}^2 =
\frac{G_N}{y} \sqrt{ \frac{4\pi N_{\rm sec} \kappa r_D}{\Lambda} }
\left\{ 3\sqrt{2\pi} \ {\rm erf}(\sqrt{y/2}) - 2e^{-y/2} \sqrt{y} (y+3) \right\}
\label{bobm}
\end{eqnarray}
where $y \equiv r/r_D$.
The halo rotation curve has a maximum, which
arises at $r \simeq 5.21 r_D$, and is given by:
\begin{eqnarray}
\left[ v_{\rm halo}^{\rm max} \right]^2 \simeq 3.12 \ G_{N}  \sqrt{ \frac{N_{\rm sec} \kappa  r_D}{\Lambda}}
\ .
\label{ss1}
\end{eqnarray}
Also, the density [Eq.(\ref{r5x})] can be integrated to obtain the halo mass, which is finite, and using Eq.(\ref{ss1}) simplifies to Eq.(\ref{13xx}), 
given earlier.

For a Milky Way scale galaxy,
an estimate of $\kappa$ can be obtained from Eq.(\ref{ss1}). 
Setting $v_{\rm halo}^{\rm max} = 200$ km/s and $r_D = 4.6$ kpc, we have:
\begin{eqnarray}
 \kappa_{\rm MW} N_{\rm sec} = 1.8 \times 10^{44}  \left(\frac{\Lambda}{10^{-38.5} \ 
[{\rm erg}/{\rm cm^3}/{\rm s}] \ [{\rm kpc^6}/m_\odot^2]}\right)
\ \ {\rm erg/s}
\ .
\label{st}
\end{eqnarray}
The halo temperature is not expected to be sensitive to $N_{\rm sec}$, 
and estimates from earlier work with $N_{\rm sec} = 1$
imply that $\langle T \rangle \approx 10^{6.5}$ K for a Milky Way scale galaxy halo \cite{footpaperII}. 
From Figure 7 we see that $\Lambda \approx 10^{-38.5}$ 
[${\rm erg}/{\rm cm^3}/{\rm s}] \ [{\rm kpc^6}/m_\odot^2]$
($\Lambda \approx 1.3 10^{-38}$
\ [${\rm erg}/{\rm cm^3}/{\rm s}]  \ [{\rm kpc^6}/m_\odot^2$]) for 
$f_{\rm metal} = 0 $ ($f_{\rm metal} = 0.02$).

\begin{figure}[t]
  \begin{minipage}[b]{0.5\linewidth}
    \centering
    \includegraphics[width=0.7\linewidth,angle=270]{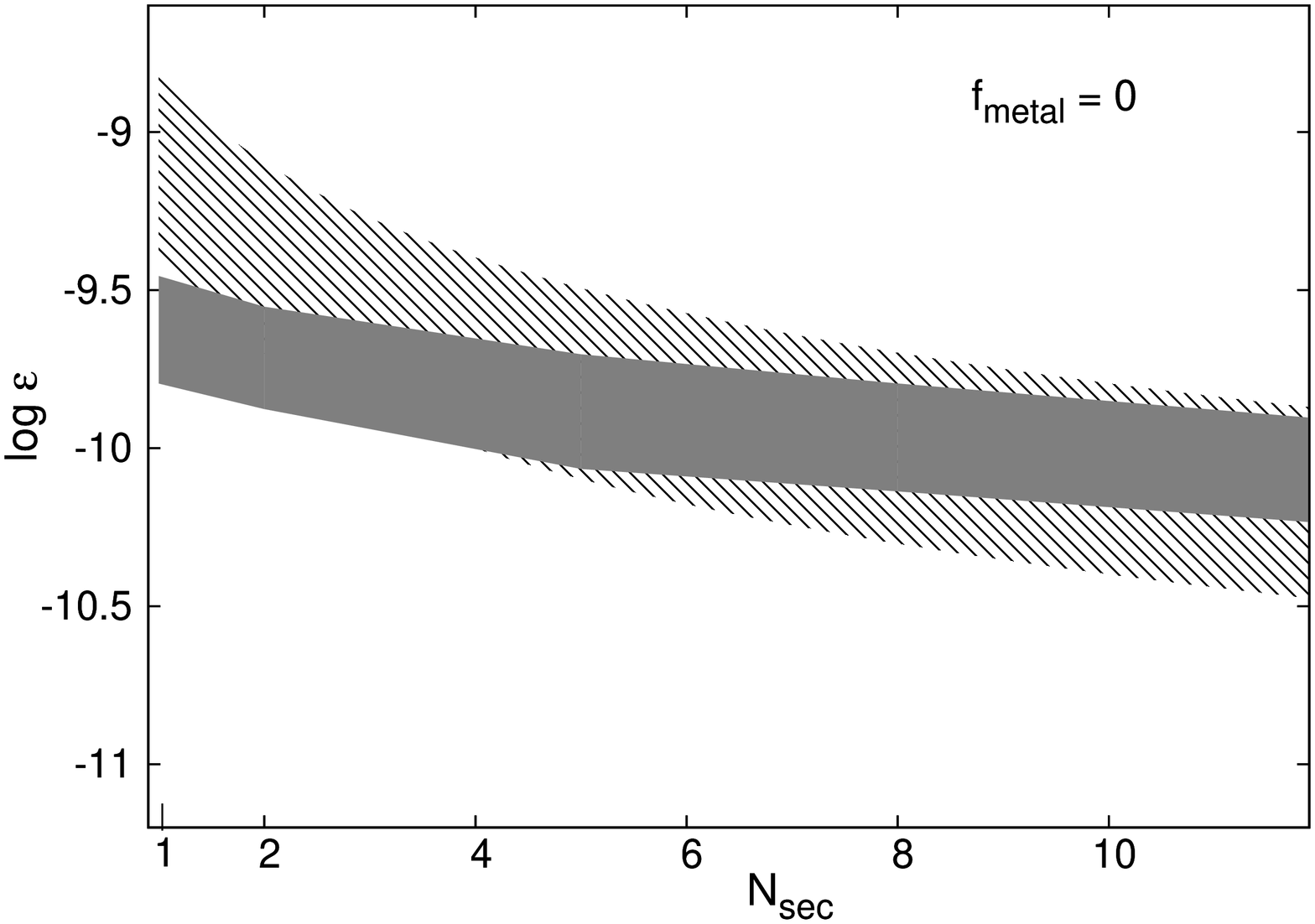}
     (a)
    \vspace{4ex}
  \end{minipage}
  \begin{minipage}[b]{0.5\linewidth}
    \centering
    \includegraphics[width=0.7\linewidth,angle=270]{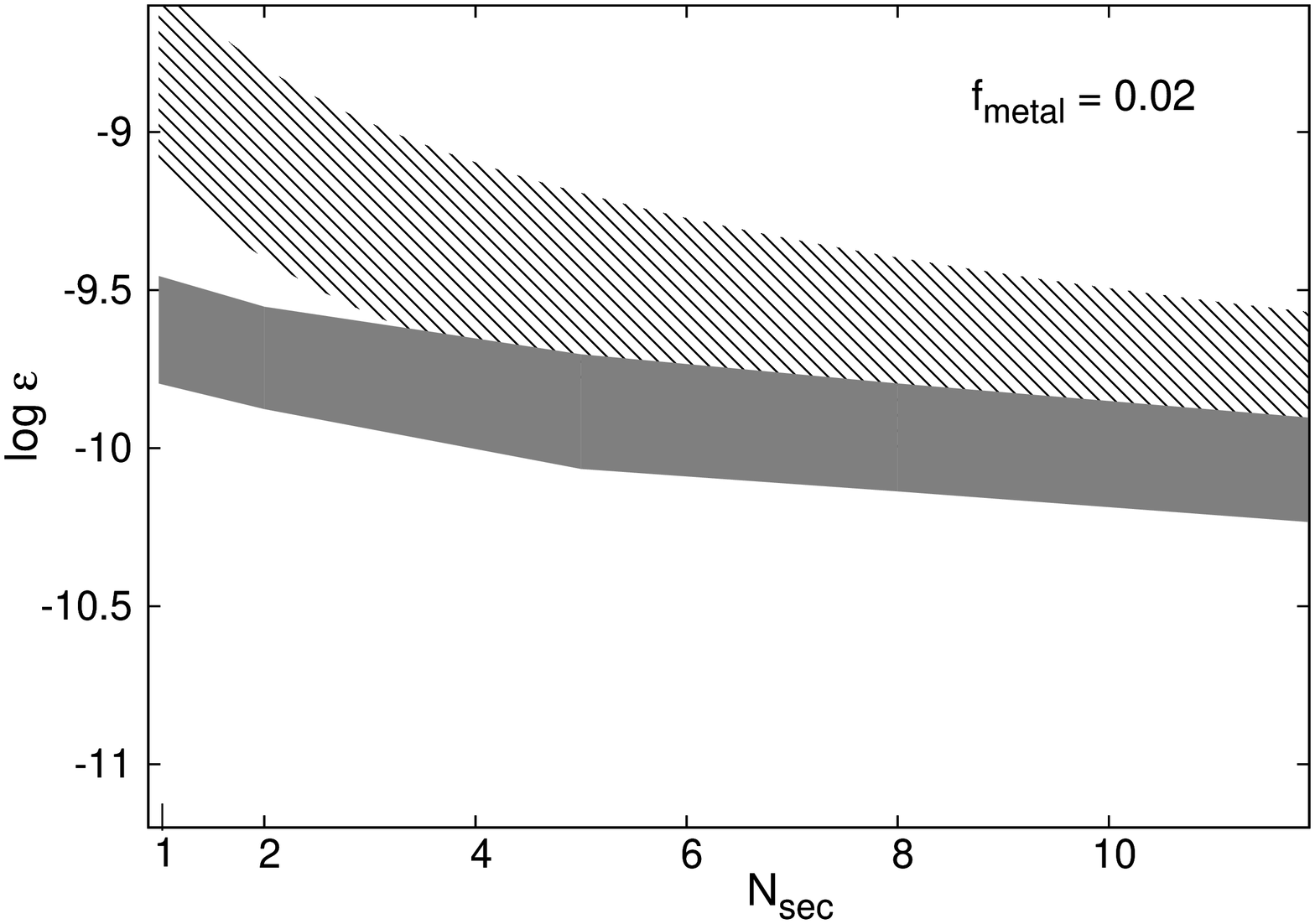}
    (b)
    \vspace{4ex}
  \end{minipage}
\vskip -1.0cm
\caption{
\small
Preferred regions in the $N_{\rm sec}, \ \epsilon$ plane.
The shaded region is the kinetic mixing estimated from 
the velocity function discussed in Sec. 4.3, 
while the hatched region is the region derived
from galaxy halo dynamics [Eq.(\ref{23q})].
The halo dynamics estimate depends on the dark metal mass fraction for spirals,
and (a) is for $f_{\rm metal} = 0$, while (b) is for $f_{\rm metal} = 0.02$.
}
\end{figure}

We now have two equations for the halo heating rate, $\kappa$. 
One relates $\kappa$ to the kinetic mixing parameter, Eq.(\ref{st2}), 
the other estimates the $\kappa$ value required for galaxies to have evolved with realistic rotational velocity, Eq.(\ref{st}). 
Comparison of these two equations, setting $R_{SN} = 0.1 \ {\rm yr}^{-1}$  for
a Milky Way scale galaxy, provides
an estimate for $\epsilon$ of:
\begin{eqnarray}
\epsilon = \begin{cases}
8\times 10^{-10}/N_{\rm sec} \  & {\rm for} \ f_{\rm metal} = 0  \ , \\
1.6 \times 10^{-9}/N_{\rm sec}  \ & {\rm for} \ f_{\rm metal} = 0.02  \ .
\label{23q}
\end{cases}
\end{eqnarray}
Of course, there are various uncertainties, so one must keep in mind that this is only a rough estimate.
These  uncertainties arise from various sources, including in the SN estimate, Eq.(\ref{11yyy}), and also the 
SN rate. There are also uncertainties in some of the modelling assumptions, notably the
use of spherically symmetry and neglect of dark magnetic fields. We have also implicitly assumed that 
the contribution of the halo rotational velocity due to dark stars is negligible relative to the dark fluid (plasma) component. 
[If there were a significant dark stellar component, this would lower the plasma density, and lead to a lower
$\epsilon$ estimate in Eq.(\ref{23q}).]
It is unclear how to quantify these uncertainties, so for illustration, 
we consider $\epsilon$ values within a factor of two [i.e. $\delta \log \epsilon = \pm \log(2)$].
In Figure 8 we compare this band of values centered around Eq.(\ref{23q}), with the preferred region
of kinetic mixing values given earlier in Figure 5 obtained from quite different considerations.
Figure 8 suggests that a self consistent picture is possible
for all values of $N_{\rm sec}$ considered if the dark metal mass fraction is low, while if it is significant, then
$N_{\rm sec} > 1$ might be needed.

\section{Halo dynamics with five SM copies}

\subsection{Steady state solutions}

Dissipative halos around sufficiently isolated and unperturbed galaxies can, in principle, evolve towards the steady
state configuration, where the local heating and cooling rates balance and the halo is in hydrostatic equilibrium.
As discussed, this evolutionary outcome may not always arise, but the assumption of a dark halo in the steady state configuration appears 
to be consistent with 
the properties of most of the observed disk galaxies with active star formation.
Here, we shall proceed to numerically solve the steady state
conditions for the $N_{\rm sec} = 5$  case, and thereby obtain the physical properties of the halo, 
including the dark matter density profile.
As in previous work, we consider an  idealized spherically symmetric system
and do not consider the possible effects of dark magnetic fields. 

The steady state conditions, Eq.(\ref{SSX}), can be solved by the same method used 
in \cite{footpaperII,footpaperI} for the $N_{\rm sec} = 1$ case.
One considers a test density function, calculates the temperature profile from the hydrostatic equilibrium
condition, and then computes the local heating and cooling rates, taking into account halo reabsorption of
cooling radiation, as the halo is not always optically thin at all wavelengths. The ionization
state is also evaluated at each location in the halo by balancing dark electron capture with ionization processes. 
This procedure must be iterated a number of times due to the interdependence
of these quantities.
This whole  process is repeated, modifying the input test density function, until the resulting local heating and cooling
rates match. The end result is a density and temperature profile for which the steady state conditions
are satisfied. 

In practice, a suitably chosen test density function with a finite number of parameters is considered. 
These parameters are varied, and for each parameter choice, ${\cal H}(r)$, ${\cal C}(r)$, $T(r)$ and the ionization
state are computed following the above iterative procedure. To determine 
how well the heating and cooling rates match, we evaluate the functional:
\begin{eqnarray}
\Delta \equiv \frac{1}{R_2-R_1} \int^{R_2}_{R_1}
\frac{|{\cal H}(r') - {\cal C}(r')|}{{\cal H}(r')+{\cal C}(r')} \ dr'
\label{d1}
\end{eqnarray}
and we take $R_1 = 0.15r_D$, $R_2 = 6.4r_D$ in the numerical analysis.
The value of the parameters defining the density which minimize $\Delta$ describe the steady state density solution.
This is, of course, approximate, and could only give a reliable solution  if
the minimum of $\Delta$ is sufficiently small.
The obvious choice for the test density function is motivated by the analytic density formula, Eq.(\ref{r5x}):
\begin{eqnarray}
\rho(r) = {\lambda} \ \frac{e^{-r/2r_D}}{\sqrt{r/r_D}} 
\ .
\label{r1}
\end{eqnarray}
To be a completely general function, we allow for a 
spatially dependent $\lambda$:
\begin{eqnarray}
\lambda \to \lambda \left[1 \ + \ \sum_{n=1}^{N} a_n \left(\frac{r}{r_D}\right)^n 
 +  b_n  \left(\frac{r_D}{r}\right)^n\right]
\ .
\label{2grl}
\
\end{eqnarray}
For the examples considered in this paper,
only the first few terms, $N=2$, are needed to achieve a reasonable minimum: $\Delta_{min} \lesssim 0.02$.

In addition to the baryonic parameters, the dynamics depends on two additional parameters:
the SN  halo heating rate, $\kappa$ (discussed in Sec. 6.3) and 
the dark metal mass fraction of the halo, $f_{\rm metal}$.
Naturally, these quantities will depend on the galaxy under consideration.
As in previous work, we shall assume that $\kappa$
scales with the SFR, so that
\begin{eqnarray}
\kappa  \approx \kappa_{\rm MW} \ \frac{L_{\rm FUV}}{L_{\rm FUV}^{\rm MW}} =  
\kappa_{\rm MW}  \ \frac{10^{-0.4M_{\rm FUV}}}{10^{-0.4M_{\rm FUV}^{\rm MW}}}
\label{sc99}
\end{eqnarray}
where $M_{\rm FUV}^{\rm MW} \approx -18.4$ is the FUV absolute magnitude for a Milky Way scale galaxy.

With the scaling, Eq.(\ref{sc99}), and assuming $f_{\rm metal} = 0$, ref.\cite{footpaperII} found that
a Tully Fisher type relation for the normalization of the halo rotational velocity results.
This scaling was broadly compatible with the sample of galaxies
considered (from THINGS \cite{things} and LITTLE THINGS \cite{littlethings} data), 
however there was some tension between the dwarfs and the spirals.
The dwarfs preferred a larger value of $\kappa_{\rm MW}$ than
the $\kappa_{\rm MW}$ set by the normalization of the spirals.
Another aspect of this tension is that the Tully Fisher relation that results is one involving luminosity
rather than baryon mass. This is largely a consequence of Eq.(\ref{sc99}).
While observations are consistent with a Tully Fisher relation involving luminosity for spirals, e.g. \cite{tf,rachel},
the relationship breaks down when dwarfs are included, e.g. \cite{btf}.
The existence of a hard break in the Tully Fisher luminosity relation suggests that
the dwarfs are characteristically different to that of spirals in a way that
impacts the heating or cooling rates.

One possible resolution to this conundrum is that spirals and dwarfs have a different halo dark sector metal content.
It is known that the dwarfs have a much lower baryonic metal mass fraction than the spirals, and in general,
the baryonic metallicity is observed to sharply decrease for smaller galaxies.
If such a situation were to arise also for the dark sector metal content then the resultant discontinuity in cooling
would lead to a discontinuity in halo density $\rho$, so that the cooling rate can
continue to match the heating rate.
Of course, this may not be the only possible solution, and there may be more than one factor at play,
nevertheless we shall focus on this particular direction as it appears to provide
a simple consistent picture.

\begin{figure}[t]
\centering
\includegraphics[width=0.48\linewidth,angle=270]{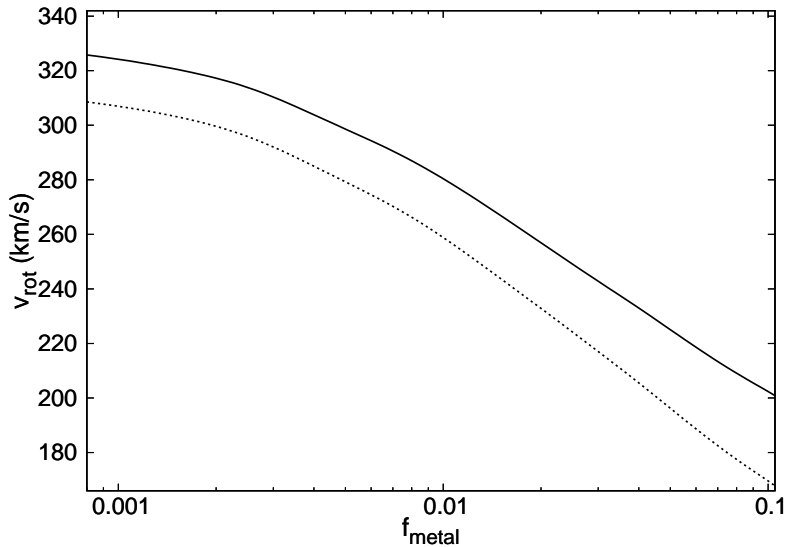}
\caption{
\small
Asymptotic rotational velocity, $v_{rot} (r=6.4r_D)$ (solid line)
and maximum halo velocity, $v^{\rm max}_{\rm halo}$ (dashed line) 
derived from the steady state solutions for a Milky Way scale galaxy 
($m_{\rm baryon} = 10^{11} \ m_\odot$,  $M_{\rm FUV} = -18.4$, $r_D = 4.63$ kpc)
with the dark metal mass fraction, $f_{\rm metal}$, varied.
The halo heating rate parameter was set to $\kappa_{\rm MW} = 2.5\times 10^{44}$ erg/s. 
}
\end{figure}

To be specific, we consider the $N_{\rm sec} = 5$ model
where the metal component of the dwarfs is negligible, $f_{\rm metal} = 0$, while that of the spirals
is nonzero.
With $f_{\rm metal} = 0$, previous work with $N_{\rm sec} = 1$ \cite{footpaperII} 
found that 
$\kappa_{\rm MW} \to 10^{0.8} 2\times 10^{44}$ erg/s was needed to give realistic 
asymptotic halo velocity for the dwarfs. 
As discussed in Sec. 6.4,
the cooling rate goes down rapidly for increasing $N_{\rm sec} > 1$.
Indeed,
if the SN sourced heating is to match cooling, then Eq.(\ref{15xb})
suggests a value of 
$\kappa_{\rm MW} \approx  2 \times 10^{44.8}/N_{\rm sec}$ erg/s, which for $N_{\rm sec} = 5$ gives
$\kappa_{\rm MW} \approx 2.5\times 10^{44} $ erg/s.
With this value of $\kappa_{\rm MW}$ 
we have computed steady state solutions for a Milky Way scale galaxy for a variety of $f_{\rm metal}$ values.
The circular rotational velocity can be determined from the halo density 
profile arising in this numerical solution. In Figure 9 we show the maximum halo
rotational velocity, together with the asymptotic rotational velocity. The latter quantity
includes the baryon contribution, and is defined as $v_{rot}(r=6.4r_D)$.
This figure indicates that for $\kappa_{MW} = 2.5\times 10^{44}$ erg/s, a metal mass fraction of
$f_{\rm metal} \approx 0.02-0.06$ is required to give a realistic asymptotic halo rotational velocity
for a Milky Way scale galaxy.

\begin{table}[t]
\centering
\begin{tabular}{l c c c c}
\hline\hline
$m_{\rm baryon} (m_\odot)$ & $r_D$ (kpc) & $M_{\rm FUV}$ & $f_s$
& $f_{\rm metal}$
  \\
\hline
\rule{0pt}{3ex}
${\rm (i)}\ \ \ 10^{11}$  & 4.63 & -18.4  &     0.8  & 0.02 \\
$\ {\rm (ii)}\ \ 10^{10.5}$  & 2.98  &  -17.9 &      0.8   & 0.02 \\
$\ {\rm (iii)}\ 10^{10}$  & 1.91 &   -17.4 &     0.8  & 0.02 \\
\hline
\rule{0pt}{3ex}
${\rm (iv)}\ 5 10^{8}$  & 0.60  &  -14.6 &      0.2   & 0\\
$\ {\rm (v)}\ 10^{8}$  & 0.50  &  -13.5 &      0.2   & 0 \\
\hline\hline
\end{tabular}
\caption{\small Baryonic properties: baryon mass, baryonic scale length, FUV
absolute magnitude, stellar mass fraction, and dark metal mass fraction
for the canonical model galaxies considered.
}
\end{table}

\begin{figure}[t]
  \begin{minipage}[b]{0.5\linewidth}
    \centering
    \includegraphics[width=0.7\linewidth,angle=270]{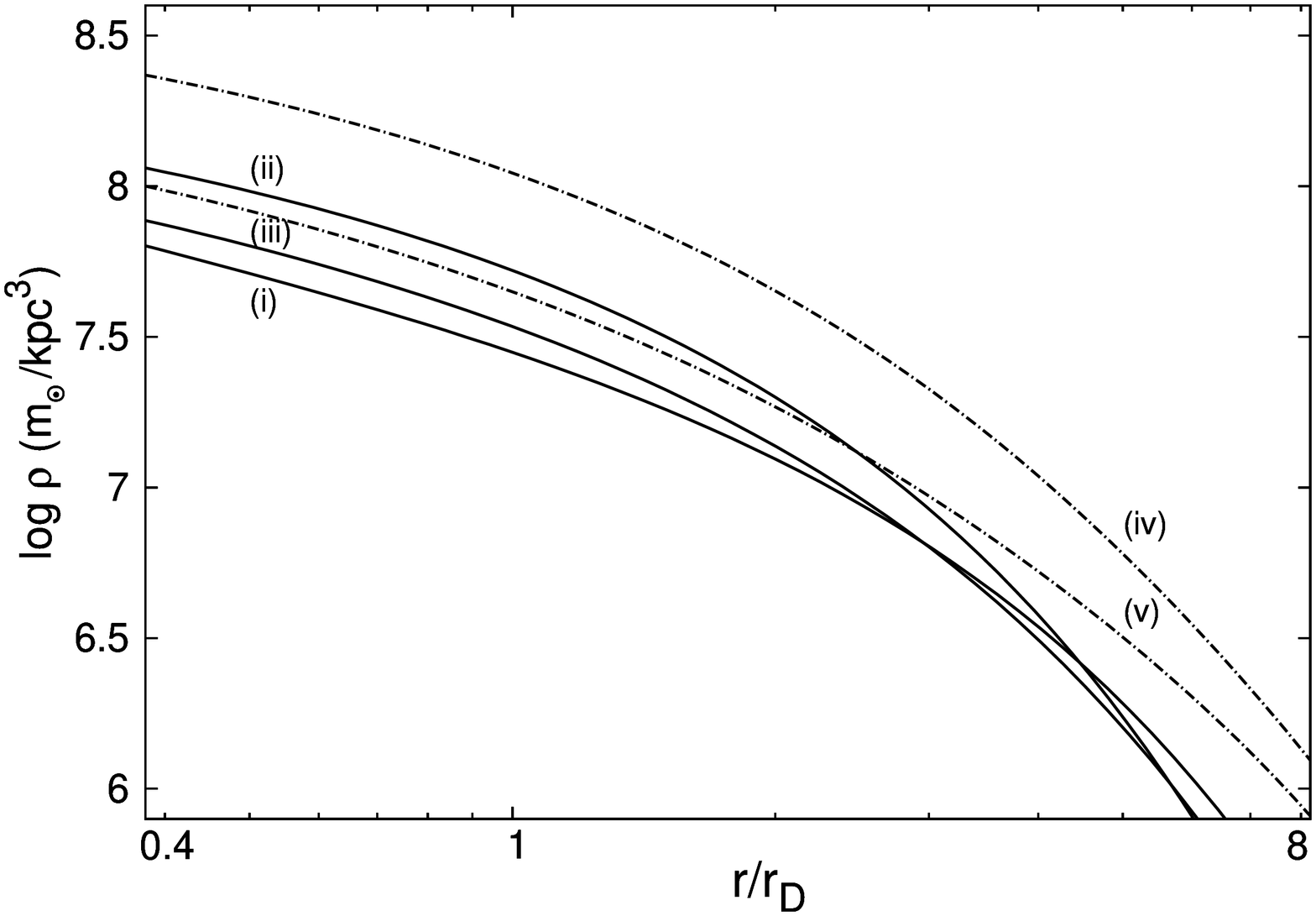}
     (a)
    \vspace{4ex}
  \end{minipage}
  \begin{minipage}[b]{0.5\linewidth}
    \centering
    \includegraphics[width=0.7\linewidth,angle=270]{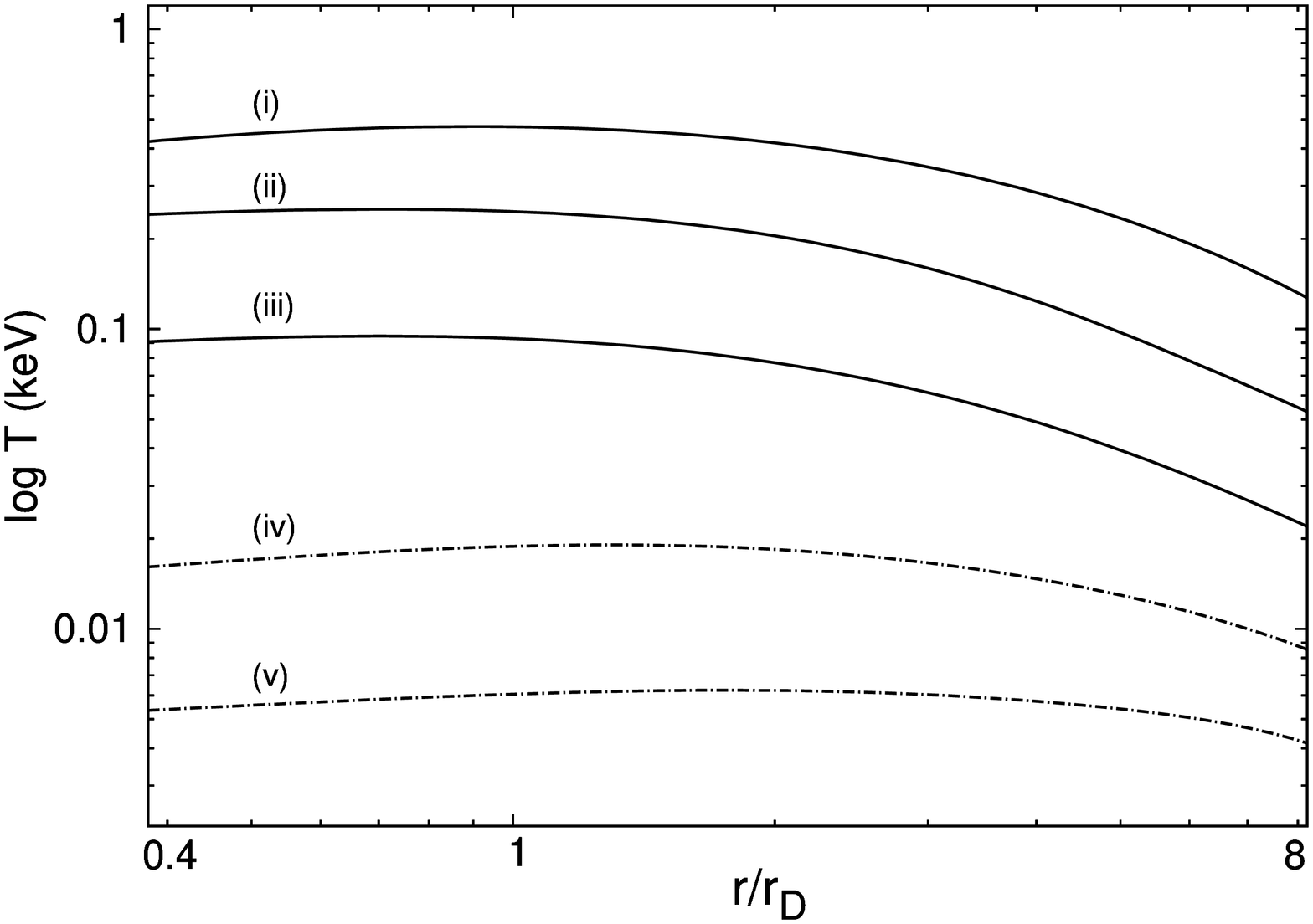}
    (b)
    \vspace{4ex}
  \end{minipage}
  \begin{minipage}[b]{0.5\linewidth}
    \centering
    \includegraphics[width=0.7\linewidth,angle=270]{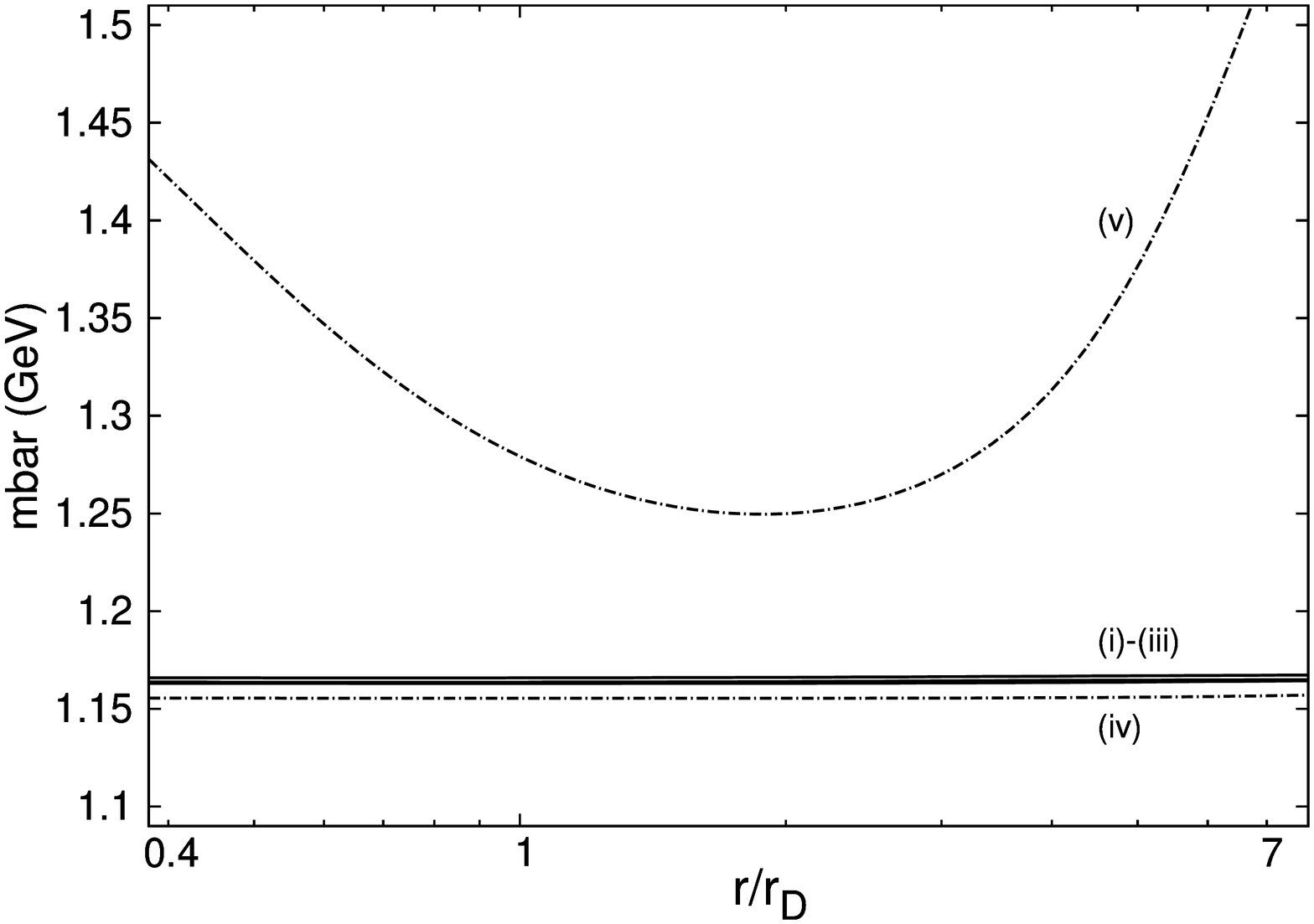}
     (c)
    \vspace{4ex}
  \end{minipage}
  \begin{minipage}[b]{0.5\linewidth}
    \centering
    \includegraphics[width=0.7\linewidth,angle=270]{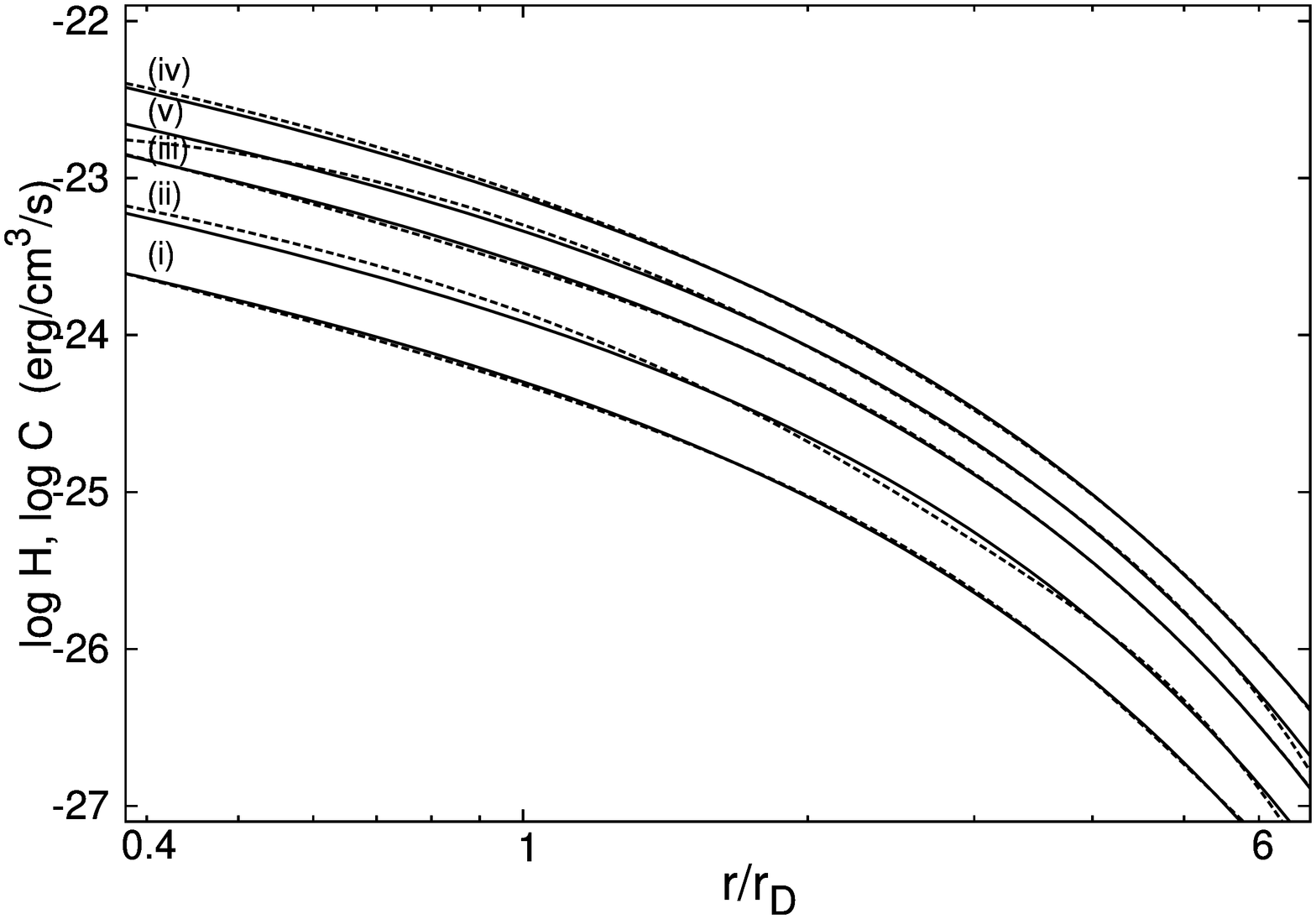}
    (d)
    \vspace{4ex}
  \end{minipage}
\vskip -1.0cm
\caption{\small Properties of the  steady state solutions  computed for
the set of model galaxies considered.
The baryon mass ranges from
$10^8 \ m_\odot$ to $10^{11} \ m_\odot$,
see Table 1 for the other baryonic parameters chosen.
Shown are (a) the
halo density, (b) the halo temperature,
(c) mean mass parameter, $\bar m$,
and (d) the heating, cooling rates [${\cal H}, \ {\cal C}$]
(solid, dashed line).
In Figs. (a),(b),(c) the dwarfs
are distinguished with dashed-dotted lines.
}
\end{figure}

Adopting a lower value in this range, $f_{\rm metal} = 0.02$, 
we calculated steady state solutions
for a variety of model galaxies, with baryonic masses typical of spiral and dwarf irregular galaxies.
The baryonic scale length values considered are consistent with the stellar disk values from
the empirical scaling relation, Eq.(\ref{s1soc}).
The Far Ultraviolet (FUV) magnitudes were taken consistent with GALEX measurements \cite{galex}. 
The baryonic properties adopted for the modelled galaxies are summarized in Table 1.

The $\Delta_{min}$ values for the solutions, (i),(iii),(iv),(v) were 
low, $\Delta_{min} \lesssim 0.02$, while that of (ii) was $\Delta_{min} \approx 0.04$.
Although still reasonable,
the $\Delta_{min}$ values for the modelled spirals are a little higher than for the $N_{\rm sec} = 1$,
$f_{\rm metal} = 0$ case studied in \cite{footpaperII}.
The difference is likely due to the 
abrupt changes in shape of the cooling function, as illustrated already in Figure 7.
Indeed, for the model galaxy (ii), with $m_{\rm baryon} = 10^{10.5}\ m_\odot$, the average
halo temperature is around $\langle T \rangle \approx 0.2$ keV, i.e. $2\times 10^6$ K, which, 
as Figure 7 indicates, corresponds
to a particularly steep section of the cooling function.
This rough terrain will make the density a little more sensitive to the modest variations
in the halo temperature, and a bit less smooth.

In Figure 10, we give the halo density, temperature, mean mass, and local heating and cooling rates 
obtained from the numerical
steady state solutions.
The density closely resembles the phenomenological cored profiles extensively
discussed in the literature, including the pseudo isothermal profile \cite{kent}, and the  Burkert profile \cite{burkert}:
\begin{eqnarray}
\rho (r) = \begin{cases}
\ \ \ \ \frac{\rho_0 r_0^2}{r^2 + r_0^2} \   \ \ \  \ \ \  {\rm Pseudo\ Isothermal} \ , \\
 \frac{\rho_0 r_0^3}{(r + r_0)(r^2 + r_0^2)} \ \ \ {\rm Burkert} \ .
\label{bur}
 \end{cases}
\end{eqnarray}
These profiles depend on two parameters,
the central density ($\rho_0$) and core radius ($r_0$).
Of these two profiles, the Burkert profile provides a marginally better approximation to the
computed density.
A fit of the numerical steady state solution,  shown in Fig.10a,
indicates that the fitted $\rho_0, r_0$ parameters 
are consistent with the empirical scaling relations \cite{KF,DS,SAL}:
$r_0 \propto r_D$ and constant halo surface density: $\rho_0 r_0$ is a constant.
[The fitted parameters are given in Table 2.]
However, the $r_0/r_D$ value is around a factor of two
below the measured value, which might possibly be an artifact of the spherically symmetric modelling.
Quite similar results were obtained for the $N_{\rm sec} = 1, \ f_{\rm metal} = 0$ 
case given in \cite{footpaperII}.

\begin{table}[t]
\centering
\begin{tabular}{c c c c c}
\hline\hline
$m_{\rm baryon} (m_\odot)$ & $r_D$ (kpc) & $r_0/r_D$ & $\rho_0 [10^7 m_\odot/{\rm kpc^3}]$ & ${\rm log}(\rho_0r_0\ [m_\odot/{\rm pc^2}])$
\\
\hline
\rule{0pt}{3ex}
${\rm (i)}\ \ 10^{11}$    & 4.63 & 1.71  & 6.69 & 2.72  \\
$\ {\rm (ii)}\ \ 10^{10.5}$ & 2.98 & 1.25  & 16.1 & 2.78   \\
${\rm (iii)}\ 10^{10}$  & 1.91 & 1.44  &  9.33  & 2.41 \\
${\rm (iv)}\ 5 10^{8}$  & 0.60 & 1.56  &  27.7 & 2.41 \\
${\rm (v)}\ \ 10^{8}$  & 0.50 & 1.70  &  10.6 &  1.95  \\
\hline\hline
\end{tabular}
\caption{\small
Fit of the steady state density solutions for the canonical galaxies of Table 1
in terms of the Burkert profile [Eq.(\ref{bur})].
}
\end{table}

The halo temperature profiles are not far from isothermal with some softening in the inner and outer regions
of the halo. The mean halo temperature scales with rotational velocity: $\langle T \rangle \propto \bar m v_{\rm rot}^2$,
and the halo is hot enough so that it has a high degree of ionization. This means that
the mean mass parameter [Eq.(\ref{sum})], which is a measure of the ionization state of the plasma, 
is approximately constant: $\bar m \simeq 1.16$ GeV, except for the smallest galaxy considered
which has a significant proportion of He II.


\subsection{Rotation curves}

Historically, 
the measured flat rotation curves out to the observational edge of
spiral galaxies  
provided the first clear indication that galaxies
were surrounded by a halo of dark matter \cite{vera1,vera2, bosma, vera3}
(see also \cite{rubin} for a review and more extensive bibliography).
Since those pioneering studies, rotation curve measurements have
continued to supply remarkable information on the dark halo properties,
and on the nature of dark matter itself.
As mentioned in the introduction, these measurements 
indicate that the structure of the dark halo
appears to be dictated by the baryonic properties of the galaxy.
From the previous discussion and earlier work,
dissipative dark matter has the potential to address this issue as the halo density
is strongly influenced by the SN sourced heating.

In Figure 11 we show 
the rotation curves derived from Newton's law using the steady state density  for each of the modelled galaxies.
The halo rotation curves feature an approximate linear rise in the inner region ($r \lesssim r_D$),
which transitions to a roughly flat curve for the observable region of interest.
The transition radius scales with the baryonic scale length, $r_D$. 
As in the $N_{\rm sec} = 1$, $f_{\rm metal} = 0$ case studied in \cite{footpaperII}, the
analytic density profile, Eq.(\ref{r5x}), with spatially constant coefficient (equivalent
to Eq.(\ref{r1}) with spatially constant $\lambda$) is a reasonable first order approximation, and can
be used to understand the origin of these features in this framework.

\begin{figure}[t]
  \begin{minipage}[b]{0.5\linewidth}
    \centering
    \includegraphics[width=0.7\linewidth,angle=270]{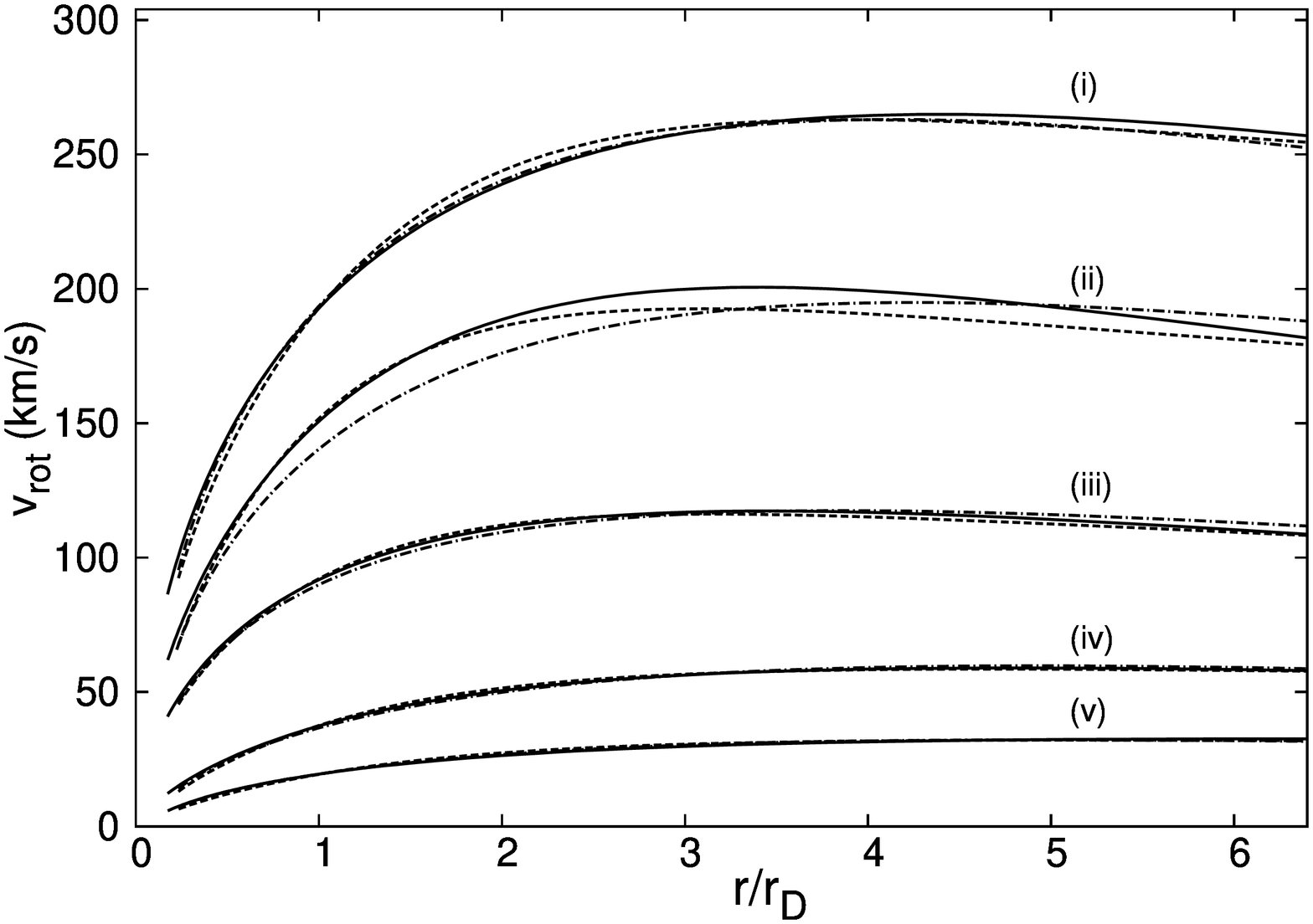}
     (a)
    \vspace{4ex}
  \end{minipage}
  \begin{minipage}[b]{0.5\linewidth}
    \centering
    \includegraphics[width=0.7\linewidth,angle=270]{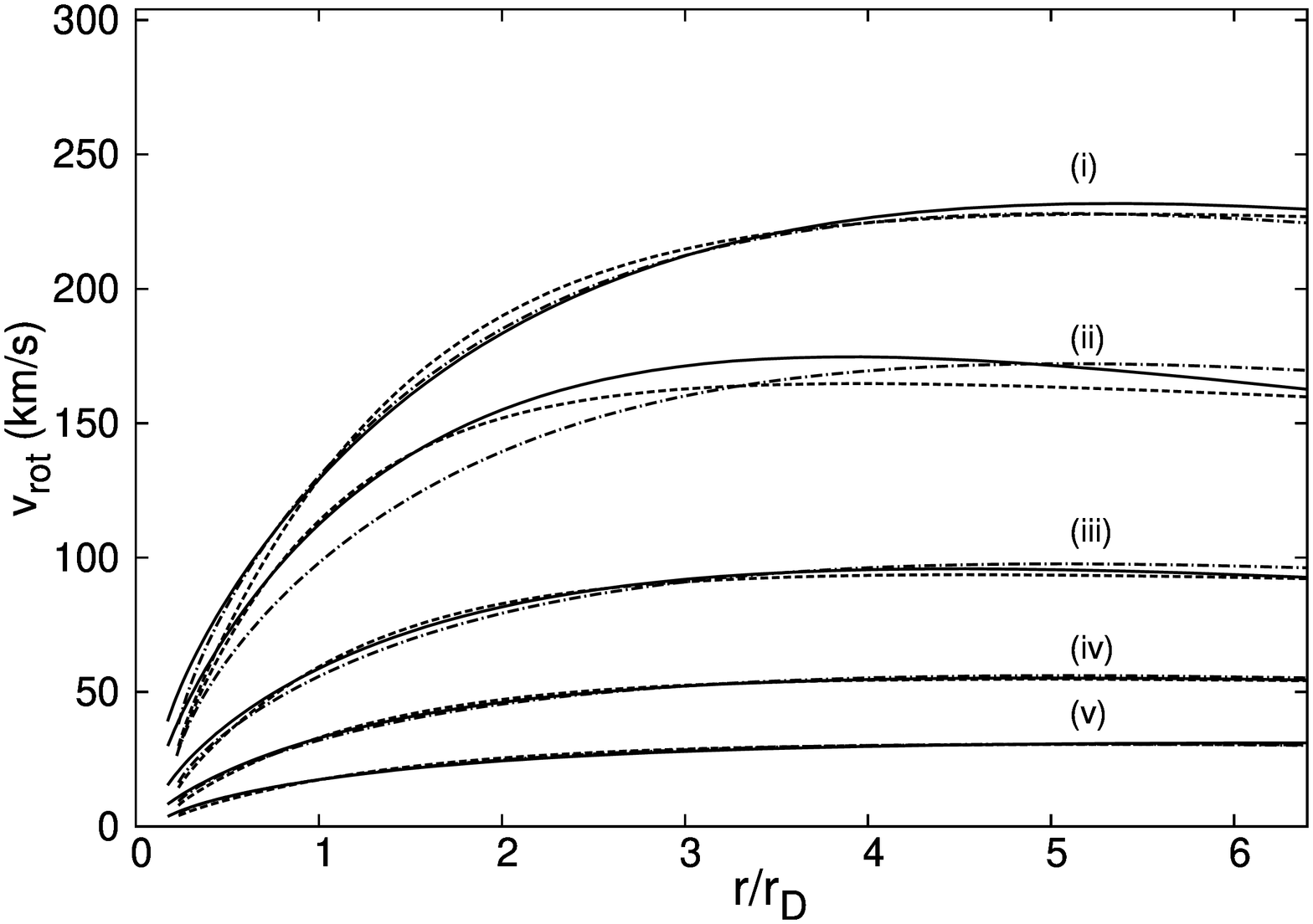}
    (b)
    \vspace{4ex}
  \end{minipage}
\vskip -1.0cm
\caption{
\small
(a) The rotation curves (halo + baryons) [solid line]
derived from the computed steady state solutions for the canonical baryonic parameters of Table 1 . 
(b) The corresponding halo rotation curves (halo contribution only). 
The dotted line shows the rotation curve that results from the 
Burkert profile [Eq.(\ref{bur})] with parameters
adjusted to approximate the steady state density solution.
The dashed-dotted line shows 
the rotation curve that results from
the analytic density profile, Eq.(\ref{r5x}), with spatially constant coefficient.
}
\end{figure}

\begin{figure}[t]
  \begin{minipage}[b]{0.5\linewidth}
    \centering
    \includegraphics[width=0.7\linewidth,angle=270]{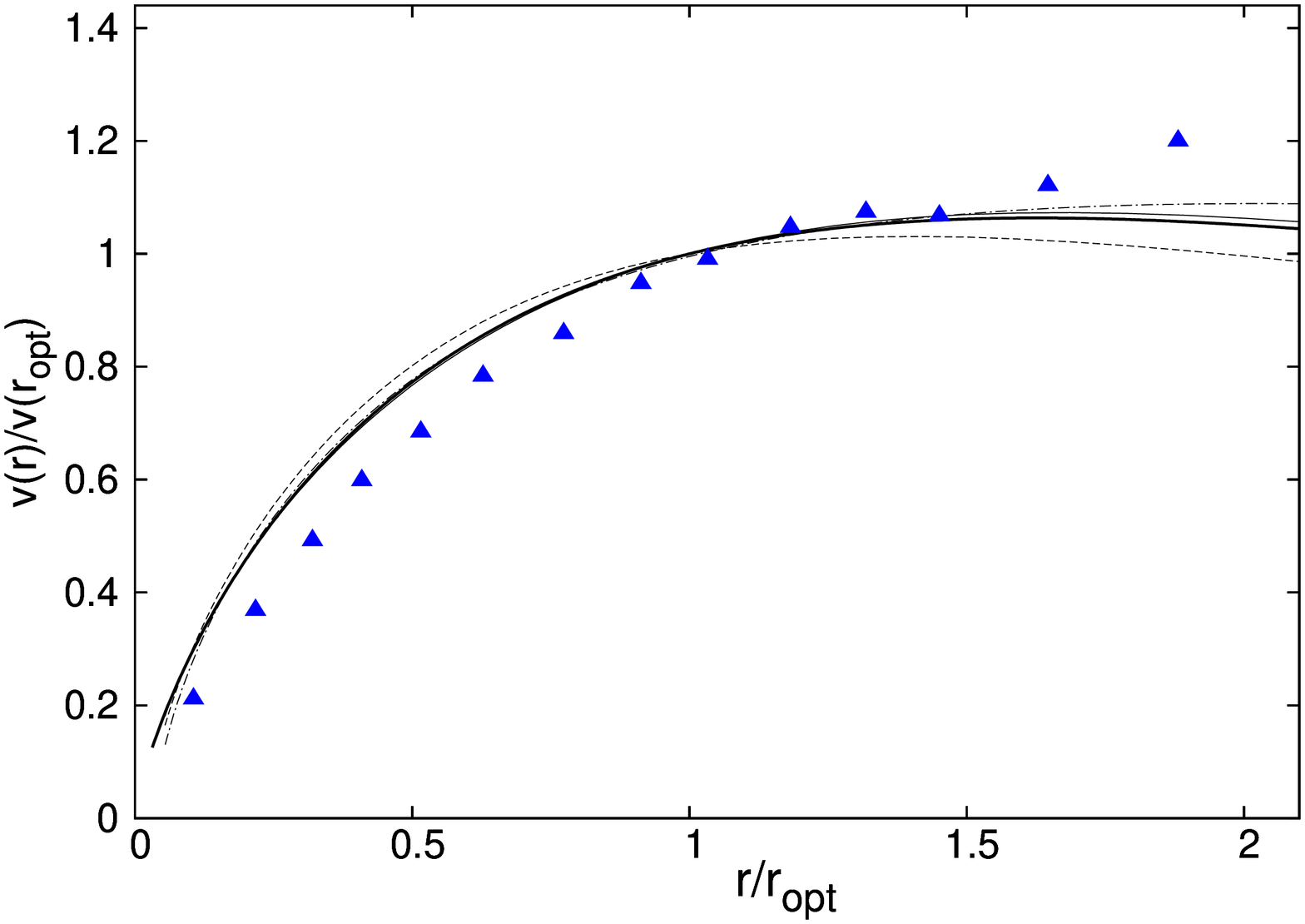}
     (a)
    \vspace{4ex}
  \end{minipage}
  \begin{minipage}[b]{0.5\linewidth}
    \centering
    \includegraphics[width=0.7\linewidth,angle=270]{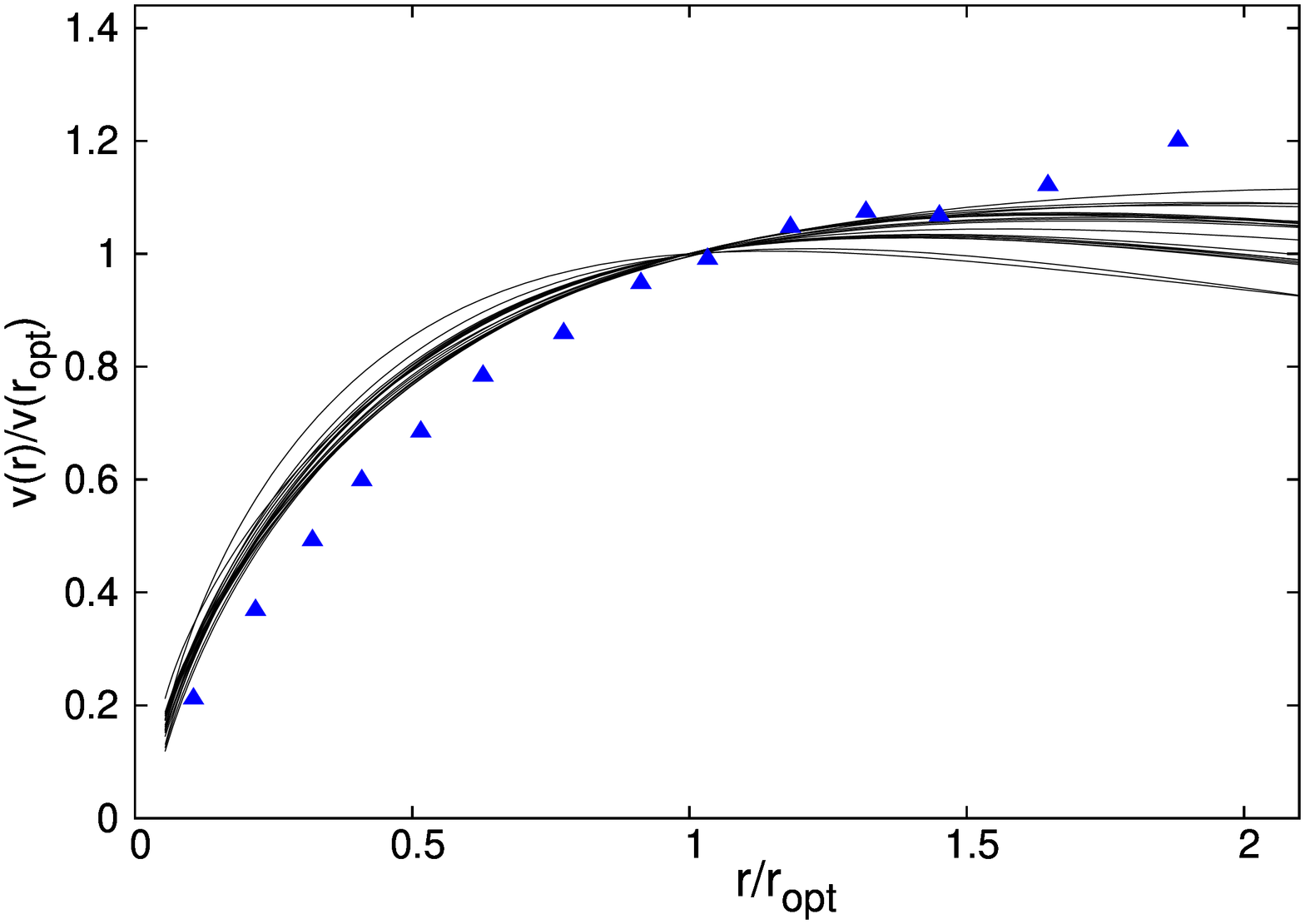}
    (b)
    \vspace{4ex}
  \end{minipage}
\vskip -1.0cm
\caption{
\small
(a) The normalized halo rotational velocity: $v_{\rm halo}(r)/v_{\rm halo} (r_{\rm opt})$ 
resulting from the steady state solutions
for canonical galaxies from Table 1: (i) [solid line], (iii) [dotted line] and (v) [dashed dotted line].
The thick solid line is the analytic normalized halo rotational curve resulting from Eq.(\ref{r5x}).
(b) The normalized halo rotation velocity resulting from the steady state solutions 
for all five galaxies from Table 1, along with
$M_{\rm FUV} = M_{\rm FUV} \pm 0.6$, $r_D \to 2r_D$ parameter variations.
Triangles are the synthetic rotation curve derived from dwarf galaxies \cite{sal17}.
}
\end{figure}

These rotation curves have an approximately universal shape, which we demonstrate in Figure 12 by
considering the normalized halo rotational velocity: $v_{\rm halo}(r)/v_{\rm halo} (r_{\rm opt})$ (with
$r_{\rm opt} = 3.2r_D$ is the so-called optical radius).
In Figure 12b results for a wide variety of baryonic parameters are considered: 
In addition to the canonical parameters we 
consider the variations: $M_{\rm FUV} = M_{\rm FUV} \pm 0.6$ and $r_D \to 2r_D$.
This provides a sample of 20 modelled galaxies covering a wide range of parameters.
Figure 12 indicates that the normalized
rotation curves associated with the steady state solutions are remarkably similar to each other when plotted in terms of the dimensionless variable $r/r_D$. 
These curves are also similar to
the corresponding results obtained for the $N_{\rm sec} = 1$, $f_{\rm metal} = 0$ case given in \cite{footpaperII}.
A universal curve rotation results from Eq.(\ref{bobm}), obtained from
the analytic density formula, Eq.(\ref{r5x}), so these numerical results are 
consistent with the analytical considerations of Sec. 6.5.

To summarize,
the rotation curves derived from the steady state solutions have a universal character,
determined (approximately) from the galaxy's baryonic properties.
Furthermore, these curves are broadly consistent with observations; the mysterious baryon - dark matter connection
discussed in the literature, e.g. \cite{salucci,DS,Lelli2,stacy}, appears to be rather neatly explained by the
SN sourced heating. 
The rotation curves are a reflection
of the halo density, which, in this framework, is strongly influenced by the baryons as Type II SN are the 
heat source shaping the dark halo.

Consider now the normalization of the rotational velocity. With the scaling: $\kappa \propto L_{\rm FUV}$ [Eq.(\ref{sc99})], 
we expect from the analytic result [Eq.(\ref{ss1})] that the maximum
of the halo rotational velocity for the modelled galaxies will approximately satisfy:
\begin{eqnarray}
\Lambda \propto \frac{r_D \ L_{\rm FUV}}{[v_{\rm halo}^{\rm max}]^4 }
\ .
\end{eqnarray}
The cooling function depends on the halo temperature 
and metal mass fraction (Fig.7), and the
temperature scale is set by the rotational velocity (Fig.10b).
To make contact with observable quantities, it is convenient to consider:
\begin{eqnarray}
\stackrel{\sim}{\Lambda} \ \equiv \ \frac{r_D \ 10^{-0.4M_{\rm FUV}}  }{[v^{\rm max}_{\rm halo}]^4}
\label{10x}
\ 
\end{eqnarray}
where $M_{\rm FUV}$ is the (extinction corrected) far UV absolute magnitude.
This is the same quantity introduced in \cite{footpaperII} in the context of the $N_{\rm sec} = 1$ case.
With the halo dark sector metal mass fraction fixed in the considered model,
$\stackrel{\sim}{\Lambda}$ is expected to depend, approximately, only on the asymptotic rotational velocity,
given that it is, essentially, a representation of the cooling function suitably averaged over the halo. 

\begin{figure}[t]
\centering
\includegraphics[width=0.48\linewidth,angle=270]{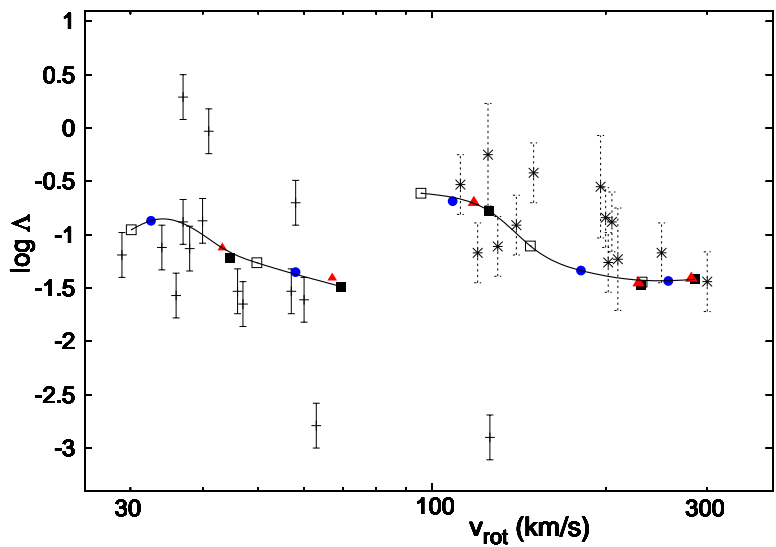}
\caption{
\small
$\log \stackrel{\sim}{\Lambda} \equiv -4\log[v_{\rm halo}^{\rm max} {\rm (km/s)}] +
\log [r_D ({\rm kpc})] - 0.4M_{\rm FUV}$
versus the asymptotic rotational velocity
computed from the steady state solutions found.
Circles denote the baryonic parameters
of Table 1, triangles a $r_D \to 2r_D$ variation, and filled (unfilled) squares for a
$M_{\rm FUV} \to M_{\rm FUV} - 0.6$  ($M_{\rm FUV} \to M_{\rm FUV} + 0.6$) variation.
The solid line is an extrapolation of the computed solutions.
Also shown are the
$\log \stackrel{\sim}{\Lambda}$  values from THINGS spirals  \cite{things}
and LITTLE THINGS  dwarfs \cite{littlethings}.
}
\end{figure}

For each set of galaxy parameters considered, $\stackrel{\sim}{\Lambda}$
can be determined from the computed steady state solutions.
In Figure 13, we give the
results for $\stackrel{\sim}{\Lambda}$ for the five galaxies of Table 1, along with
the luminosity and baryonic scale length variations: $M_{\rm FUV} \to M_{\rm FUV} \pm 0.6$ and $r_D \to 2r_D$.
(These parameter variations are made while keeping the other baryonic parameters fixed.)
Also shown in Figure 13 are the measured $\stackrel{\sim}{\Lambda}$ values from THINGS spirals \cite{things}
and a sample of LITTLE THINGS dwarfs \cite{littlethings}.

\begin{figure}[t]
  \begin{minipage}[b]{0.5\linewidth}
    \centering
    \includegraphics[width=0.7\linewidth,angle=270]{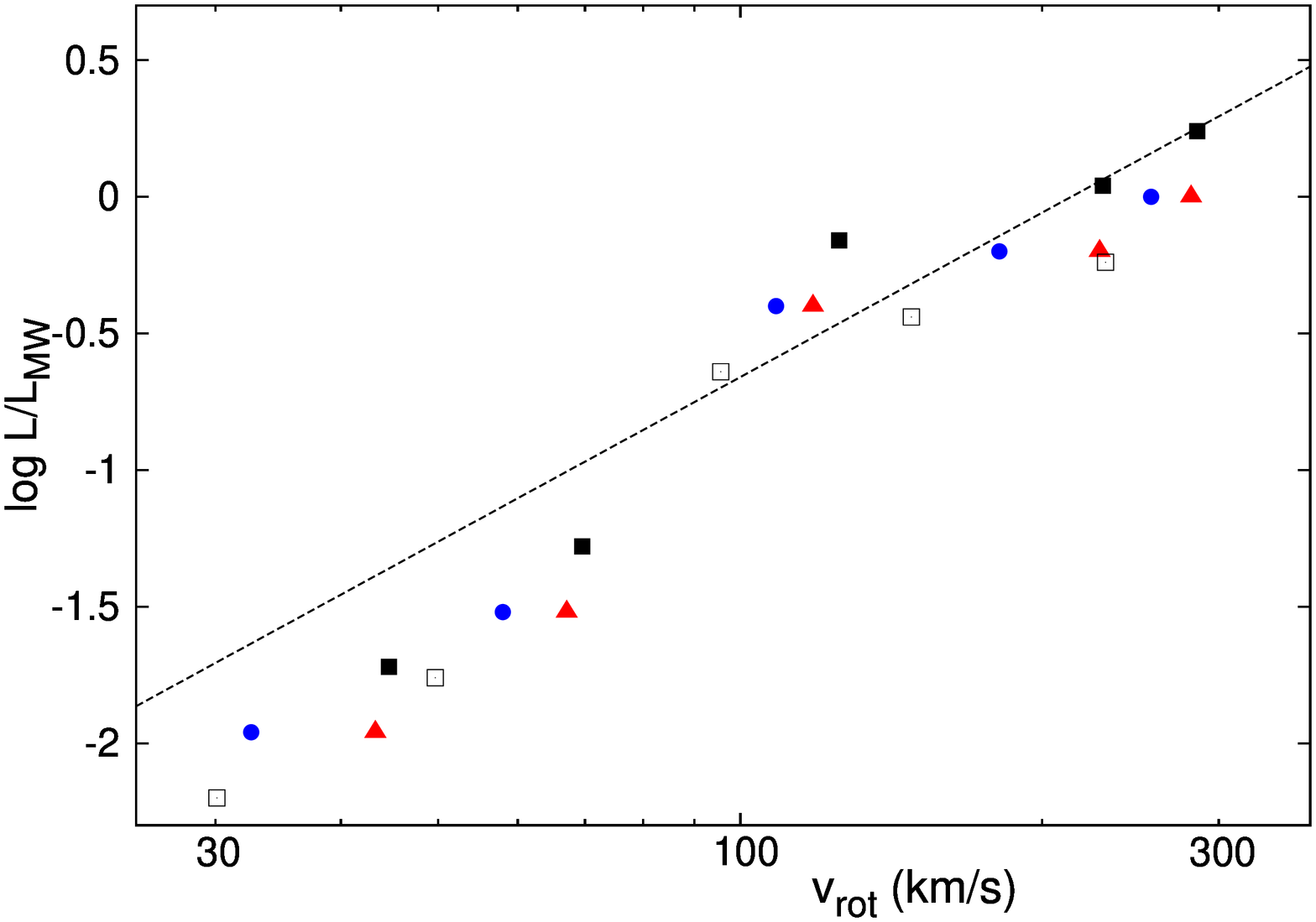}
     (a)
    \vspace{4ex}
  \end{minipage}
  \begin{minipage}[b]{0.5\linewidth}
    \centering
    \includegraphics[width=0.7\linewidth,angle=270]{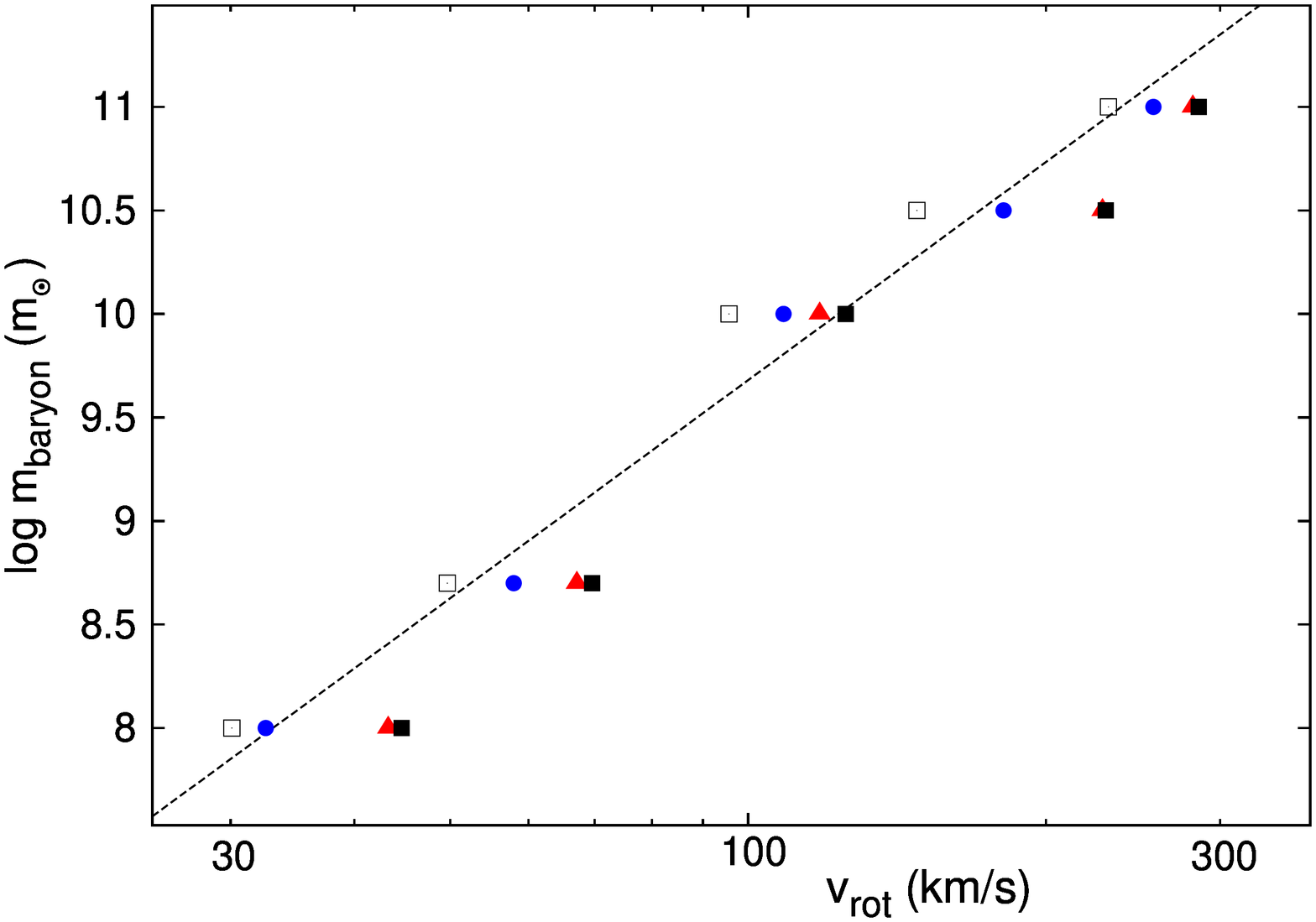}
    (b)
    \vspace{4ex}
  \end{minipage}
\vskip -1.0cm
\caption{
\small
(a)  $L_{\rm FUV}/L^{\rm MW}_{\rm FUV}$ and (b) $m_{\rm baryon}$
versus the asymptotic rotational velocity
computed from the steady state solutions found.
Circles  are the results for the canonical galaxy set of
Table 1, triangles for a $r_D \to 2r_D$ parameter variation, and filled (unfilled) squares  for a
$M_{\rm FUV} \to M_{\rm FUV} - 0.6$ ($M_{\rm FUV} \to M_{\rm FUV} + 0.6$)  variation.
The dashed lines are the power laws:  $L_{\rm FUV} \propto [v_{\rm rot}^{\rm asym}]^{2.0}$ and
$m_{\rm baryon} \propto [v_{\rm rot}^{\rm asym}]^{3.5}$.
}
\end{figure}

In Figure 13 only the `classical' dwarfs are shown, these are the ones where
the rotation curve has an approximately classical shape.
This comprises roughly 2/3 of the LITTLE THINGS sample, the remaining 1/3 are irregularly shaped.
The diversity of rotation curves suggests that many of the dwarfs are not in a steady
state configuration; they are being perturbed in some way, or their star formation rate is unstable
relative to the cooling time scale, $t_{\rm cool}$ [Eq.(\ref{cooly})].
Dwarf galaxies in a starburst phase can have dark halos deviating substantially from
the steady state configuration, i.e.  dark halos undergoing large scale radial expansion and contraction
that is in phase and, possibly, driving the oscillating star formation rate.
The three dwarfs with the highest $\stackrel{\sim}{\Lambda}$ values shown in Figure 13
(DDO50, NGC2366, NGC1569) 
are all known to be starburst galaxies \cite{quinn}, and their relatively high $\stackrel{\sim}{\Lambda}$ values
can thereby be accounted for.
Some of the LITTLE THINGS dwarfs appear to have a collapsed halo, that is,
the halo is no longer dissipative but in the form of dark stars; this would result in
a much lower value of $\stackrel{\sim}{\Lambda}$.
The two dwarfs with the lowest values of $\stackrel{\sim}{\Lambda}$ in Figure 13, DDO101, NGC3738,  are candidates for
such objects.

In this dynamics,
the $\stackrel{\sim}{\Lambda} - v_{\rm rot}^{\rm asym}$ relation emerges as the fundamental relation
governing the normalization of the rotational velocity.  The empirical Tully Fisher relation \cite{tf} and baryonic
Tully Fisher relation \cite{btf} can be viewed as approximate relations stemming from
this fundamental relation along with  (rough) purely baryonic relation between luminosity,  
baryon mass and disk scale length. These Tully Fisher relations are considered in Figure 14.
Figure 14b indicates that there is an approximately linear mass relation 
$\log m_{\rm baryon} \propto 3.5 \log v_{\rm rot}$,
which is roughly consistent with the empirical baryonic Tully Fisher relation \cite{btf,btfnew}.
Having distinct $f_{\rm metal}$ values for the modelled spirals and dwarfs plays an important role here.
If instead we had set $f_{\rm metal} = 0$ for both dwarfs and spirals, then the Tully Fisher luminosity relation 
would be approximately linear, while the Tully Fisher mass relation would have a break between the
modelled dwarfs and spirals, cf. \cite{footpaperII}; behaviour at odds with observations of actual galaxies.
The assumption of 
a different metal mass fraction for the dwarfs and spirals rectifies the situation, 
although, of course, there may be other ways to achieve this.

As briefly discussed in Sec. 6.1, the stability of the steady state solution is an important but quite nontrivial 
issue. It involves the feedback of expanding or contacting halos on the star formation rate, 
the change in cooling due to the change in cooling function, $\partial \Lambda/\partial T$,
and the change in cooling due to the changing density.
Some qualitative discussion was given in \cite{footpaperII}, and as pointed out  there,
the negative slope $\partial \Lambda/\partial T < 0$ may play an important role
in stabilizing the steady state solution.
Indeed, as with the $N_{\rm sec} = 1 $ case,
the halo $\langle T \rangle$ range of interest for galaxies with active star formation roughly
matches the temperature region where $\partial \Lambda/\partial T < 0$.
In this context, having $f_{\rm metal} \neq 0$ might be important as the metal content
can significantly steepen the cooling function gradient [Figure 7],
and thereby help stabilize the solution.


\section{Direct detection experiments}

Before concluding this study, let us 
digress here to make a 
 few remarks concerning the implications for direct detection
experiments. 
Indeed, kinetic mixing near $\epsilon \sim 10^{-9}$ was 
identified quite some time ago \cite{foot1d,foot1e} by the positive DAMA annual modulation signal \cite{damaold1,damaold2,dama3},
and provided some inspiration to think seriously about the implications of kinetic mixing of this magnitude for halo
dynamics.
The DAMA signal is yet to be confirmed by other experiments, but nevertheless constitutes impressive
evidence for the direct detection of dark matter.

The entire region of interest given in Figure 8 can be probed via direct
detection experiments. The dark nuclei can elastically scatter off target 
nuclei, leading to nuclear recoils - the classic direct detection signal favoured in
experiments. Dark hydrogen and dark helium are light compared to the typical
target nuclei used, and generate sub-keV nuclear recoils, below the threshold
of most current direct detection experiments. The CRESST-III experiment \cite{cresst}, with a threshold
below $0.1$ keV, is an important exception, and should be able to provide
a sensitive probe of the dark helium component.
If there is a heavy dark metal component then $E_R \gtrsim 1$ keV nuclear recoils are anticipated, potentially
more widely  observable in experiments.
The dark nuclei - nuclei cross section is proportional to $\epsilon^2 f_{\rm metal}$, and for
$f_{\rm metal} = 0.02$ Figure 8 indicates
a coupling strength of order $\epsilon \sqrt{f_{\rm metal}} \sim 10^{-11}$. This coupling is too small
for nuclear recoils to account for the DAMA annual modulation signal, but the large scale
xenon experiments, including XENON1T \cite{xenon1}, LUX \cite{lux}, and PandaX \cite{panda} can
probe this parameter space \cite{foot2d}.

Another important direct detection search channel  is via electron scattering. The halo temperature for 
the Milky Way is $\langle T \rangle \sim \bar m v_{\rm rot}^2$ $\sim 0.5$ keV, so halo dark electrons
can scatter off losely bound target electrons to produce  keV recoils \cite{foot2f}.
While most experiments discriminate against electron recoils, DAMA and a few others, most notably XENON100, XENON1T  
are sensitive to this interaction channel. In fact, electron recoils might be able to explain the impressive DAMA annual
modulation signal, despite the constraints reported by XENON100 \cite{xenon2,xenon3} and XENON1T \cite{xenon1}.
This possibility arises due to nontrivial plasma effects enhancing
the modulation rate, to be discussed in a moment.
Very recently, DAMA have presented new results for the period 2011-2017 with energy threshold of 1 keV \cite{damanew}.
The annual modulation amplitude appears to be sharply rising towards low recoil energies,
broadly compatible with the expectations from $T \sim 0.5$ keV electron recoils,
smeared by the large DAMA energy resolution.

This kind of plasma dark matter has a number of nontrivial features. Halo dark matter will inevitably be captured by
the Earth where it can accumulate \cite{footzz}. The captured dark matter provides an obstacle to the halo wind, and can strongly
modify halo dark matter properties (density and temperature) in the vicinity of the Earth.
The effects of this  `dark sphere of influence' can be modelled
with MHD equations, and the implications for direct detection experiments studied \cite{foot2c}.
The main conclusion is that the annual modulation signal can be greatly enhanced, even an almost maximal annual modulation
is possible.
In addition, large diurnal variation is also expected.
While these effects can help reconcile the positive DAMA annual modulation signal with the negative results reported
by XENON100 and XENON1T, accurate predictions are difficult due to the complexity of the problem. It is also
possible that the dark sphere of influence effectively suppresses direct detection signals by
shielding the Earth from halo dark matter, so the dissipative model discussed  
here cannot be definitively excluded by direct detection experiments, although it can be confirmed.


\section{Discussion and Conclusion}

Dissipative dark matter arising from a hidden sector which consists of $N_{\rm sec}$ exact copies 
of the Standard Model has been examined. The particles from each sector interact with those from 
the
other sectors by gravity and via the kinetic mixing interaction, described by the 
dimensionless parameter, $\epsilon$. 
It has been known for a long time that models of this kind 
can be consistent with large scale structure and the cosmic microwave background measurements. 
Here we have argued that such models can potentially explain various 
observations on small scales, including the observed paucity and planar distribution
of satellite galaxies, the flat velocity function of field galaxies, 
and the structure of galaxy halos.

The model features matter power suppression due to dark acoustic oscillations and dark photon
diffusion; these processes occur prior to and during dark hydrogen recombination, and so only affect
small scales.
For a given $N_{\rm sec}$, 
these scales
depend only on the kinetic mixing parameter, $\epsilon$.
To explore the
ramifications of the small scale power suppression, we have made use of the Extended Press Schechter formalism to
obtain the halo mass function. To make direct contact with observations,
we evaluated the velocity function, making use of the halo mass - rotational velocity relation suggested
by the dissipative dynamics. Matching the velocity function with the observations provided 
an estimate of the kinetic mixing parameter, shown in Figure 5.

Structure becomes exponentially suppressed for small halos, $m_{\rm halo} \lesssim 10^9\ m_\odot$.
This means that satellite galaxy's around hosts, such as the Milky Way and Andromeda, could only have formed top down.
That is, from larger scale density perturbations. 
The origin of the satellite galaxies is of course speculative, but there is a simple picture:
A galaxy scale dark matter perturbation is expected to collapse and cool ending in the formation
of a dark disk, in much the same way that the baryons evolve to produce the baryonic disk.
Such a nonlinear collapse process is not expected to be uniform, and perturbations leading to
small satellite galaxies could thereby have arisen.
If this formation mechanism is correct, then the satellite galaxies are anticipated to be 
co-rotating and orbit in the same plane as the dark disk, kinematic features consistent with observations 
of the satellites around the Milky Way \cite{satmw1,satmw2,satmw3,satmw4}, Andromeda \cite{sat2,sat2b} and Centaurus A \cite{sat3,sat3b}.

The dark disk will be heated once 
significant star formation occurs, as
the kinetic mixing interaction of strength $\epsilon \sim 10^{-10}$ transforms
Type II supernovae into powerful dark sector heat sources.
Eventually the dark disk can be heated to the point where the dark gas component completely disrupts and expands to form
a roughly spherical extended distribution.\footnote{The baryonic stars which formed during the first stage of galaxy evolution, prior to the formation
of the roughly spherical dark halo, and those which formed after, could possibly be identified with the 
thick and thin disk components of galaxies.}
The dark matter is 
then pressure supported and very hot, with temperature typically greater
than the ionization energy of dark hydrogen.
This means that
galaxy halos around galaxies with active star formation,
including spirals and dwarf irregular galaxies, take the form of a dark plasma.
Such a dark plasma halo continues to be dynamical, it can expand and contract in response to heating
and cooling processes. The system is a complicated one,
with baryons coupled to the dark matter via gravity, while the dark matter is coupled to 
the baryons via the SN sourced heating.

This dynamically evolving system can be modelled with fluid equations.
At the current epoch, galaxy halos around sufficiently isolated and unperturbed
galaxies are presumed to have evolved to a steady state configuration, governed by relatively
simple equations, Eq.(\ref{SSX}).
This may not always occur, as for some systems the halo might undergo runaway contraction or expansion.
Some important issues have not been addressed here, including the
dynamical stability of the steady state solution.
This appears to be especially nontrivial in models with 
$N_{\rm sec} > 1$.
Leaving this as an important open question, we then solved the steady state equations
(considering idealized spherically symmetric systems)
to yield the halo density, from which rotation curves can be derived.
These rotation curves have the rather interesting feature in that their shape and normalization 
are effectively dictated by 
the baryonic properties of a given galaxy; a direct consequence of the SN sourced heating.

We have examined a specific model with $N_{\rm sec} = 5$, which can serve to illustrate 
dark matter arising from multiple exact copies of the Standard model. 
It can also be loosely motivated by the $\Omega_{\rm dark} \approx 5\Omega_{b}$ cosmic
density inferred from analysis of the CMB anisotropies.
Halo heating is sourced from Type II supernovae, and the level of halo heating, 
which is set by the kinetic mixing parameter, must be sufficient to match cooling. 
[This need not require any fine tuning, the halo will evolve, 
expand or contract, etc., with SFR also dynamically evolving as it couples to the halo via gravity.]
This dynamics gives an estimate for $\epsilon$, given by the hatched region in Figure 8, 
which is consistent with the independent estimate from
matter power suppression (Figure 5).

The rotation curves which result from the steady state density solution appear to have the right 
broad properties to be consistent with the
measured rotation curves of spiral and dwarf irregular galaxies.
The approximate linear rise in the inner region $r \lesssim r_D$, and the transition between the linear 
rise and flat profile at $r/r_D \sim 1-2$ are noted features of the measured rotation curves; features 
which also arise in these steady state solutions.
With respect to variations in $r_D$ the steady state solutions are almost (but not exactly)
scale invariant, with shape depending (approximately) only on the dimensionless variable $r/r_D$.
The normalization of the halo velocity has a weak variation with respect to $r_D$, 
with $v_{\rm halo}^{\rm max} \propto (r_D)^{1/4}$.
[These results can also be understood from the analytic considerations of Sec. 6.5, 
see also \cite{footpaperII} for further discussion.]
Scale invariance is an important feature which appears to be
essential if the rotation curve measurements are to 
be explained, e.g. \cite{littlethings,stacy,sal17}.\footnote{The concept of scale invariance has been discussed earlier in the context of Modified
Newtonian Dynamics \cite{mond}.}

We have spent a lot of time studying steady state solutions.  These solutions appear to have the
capacity to provide a simple and consistent description of the current dark matter properties in galaxies
with active star formation, i.e. spirals and gas rich dwarfs. Of course, dynamical halos
influenced by heating and cooling processes could exist
in non steady state configurations. Large scale radial oscillations of the halo might
be possible, although transient in nature. Starburst galaxies are prime candidates for such oscillating systems,
with their oscillating period estimated from observations to be around 100 Myr \cite{quinn}.
In some circumstances, galaxies would undergo runaway contraction or expansion, 
depending on various factors.
Runaway contraction would see the halo condense into dark stars and become very compact.
The baryons would also become more compact due to the enhanced gravitational field.
The DDO101 and NGC3738 dwarf galaxies are candidates for such systems. In the case of runaway expansion, the plasma halo component
disperses and escapes the galaxy. The baryons, if they do not also disperse completely, will
be weakly bound. The recently observed ultra-diffuse galaxy
NGC1052-DF2 with little or no dark matter would be a candidate for such an object \cite{ultra}.

With regard to elliptical and dwarf spheroidal galaxies, these galaxy types are known
to have a very low active star formation rate.
In the absence of significant heat sources,
the dissipative dark matter would presumably have collapsed into dark stars and black holes.
As elliptical galaxies are rather common in the Universe, black holes too should be rather numerous,
which would have important implications for
gravitational wave observations \cite{GW1,GW2}. More generally, dissipative dark matter could facilitate
the early formation of supermassive black holes around the time when dark disk formation was occurring,
which might be important in view of the 
observations indicating the early formation of supermassive black holes \cite{quasar1,quasar2,quasar3}.

Clearly, the
treatment of small scale phenomena is still sketchy in many ways,
and much more work could be done to flesh out more of the details. There are many
interesting small scale issues yet to be studied, including
time-dependent phenomena 
such as rotation curves in starburst galaxies, 
evolution of galaxy dark matter from the dark disk to the expanded halo
distribution etc. 
Nevertheless, this analysis indicates that dissipative dark matter models are promising candidates potentially able to
explain
dark matter related phenomena on both
large and small scales. 
One final topic yet to be considered here is intermediate scales: dark matter
in clusters as probed by cluster collisions. 

The most well known example is the Bullet cluster \cite{bullet}.
In that system there is evidence that the dark matter from each cluster is able to pass through the other,
like the galaxy component. The baryon gas component is slowed and shock heated, and is spatially offset
from the galaxy and dark matter components.
If dark matter is dissipative, then
the most straightforward interpretation of the Bullet cluster
is that the bulk of the dissipative dark matter in each
cluster component has condensed into dark subhalos and/or dark stars, cf. \cite{zurab4}. If the subhalos are compact enough,
then these objects can pass by each other without colliding during the cluster collision.

A rough estimation for the fraction
of dark matter in the cluster that could of formed subhalos follows from the Press Schechter formalism,
\begin{eqnarray}
f_{\rm halo} = {\rm erfc} \left(
\frac{\nu}{\sqrt{2}}  \right)
\end{eqnarray}
where $\nu$ is defined in Sec.4.2.
For $m_{\rm halo} \gtrsim 10^9\ m_\odot$, and $\epsilon = 10^{-10}$, $N_{\rm sec} = 5$ we 
find that $f_{\rm halo} \approx 0.5$.
The number of halos formed below $10^9 \ m_\odot$ is insignificant
in the Press Schechter formalism
due to the matter power suppression effects arising during the linear regime. 

Of course, in the cluster environment, hydrodynamical effects, heating and cooling processes, 
dark magnetic fields etc., might well be important and significantly modify the 
Press Schechter estimate. 
Are the subhalos disrupted and destroyed or do they survive intact?
Are additional subhalos formed?
Galaxy - galaxy interactions and collisions
can strip halos of matter, and within the cluster
might well produce numerous smaller structures.
Subhalos with $m_{\rm halo} \lesssim 10^{9} m_\odot$,
if they are produced in this way, may evolve into very compact objects. 
Systems with mean halo temperature below the He II line emission peak, 
i.e. $\langle T_{\rm halo} \rangle \lesssim 10$ eV (or $v_{\rm rot}^{\rm asym} \lesssim 30$ km/s),
have stability issues and can potentially collapse into dark stars \cite{footpaperII}.
The end result will be a dark matter distribution that has both a diffuse component and a more
clumpy component (comprising dark subhalos and dark stars).
This picture can be consistent with Bullet
cluster observations provided that the diffuse component is less than around 50\% 
of the total dark baryon cluster mass in that system. 
This is presumably possible given the apparent
complexities associated with the evolution of the dissipative dark matter within clusters.
Naturally, significant variability between different clusters can arise, 
depending on the evolutionary history and other factors, and this variability
might account for the apparent diversity of cluster observations reported in several 
studies \cite{abel1,abel2,otherx}.

\section*{Appendix: An estimate of the disk pressure force}

As discussed in Section 5, baryonic and dark disks can potentially form at an early stage of
a galaxy's life.
The situation may be rather fluid as there is the possibility 
of pressure forces arising from SN generated winds and also disk heating.
The way the disks evolve is expected to depend on whether these pressure forces are important relative
to gravity.
If these pressure forces overwhelm gravity, then the
the disks are expected to be orthogonally orientated, which it has been suggested, might 
have something to do with the polar alignment of satellite planes.
In this appendix
a crude estimate of the disk pressure force due to heating of the disks is given.

Within dissipative dark matter models with kinetic mixing induced heating of dark baryons
from ordinary SN, and potentially also heating of baryons from dark SN, the conditions 
necessary for significant disk heating are readily available.
There can be concurrent star formation in both sectors
We shall assume for illustration that ordinary SN form first, so that the dark disk is heated.
(If it is the other way around, or if there is concurrent dark and ordinary SN formation,
then our conclusions are not expected to change.)
If the side of the dark disk facing the baryonic disk is heated to a temperature, $T_{\rm near}$, then
the pressure gradient creates a force per unit disk area of,
\begin{eqnarray}
{\cal F}_{\rm pressure} &= & z_{\rm disk} \langle \frac{\partial P}{\partial z} \rangle 
\nonumber \\ 
&\sim & z_{\rm disk} \langle \frac{\rho}{\bar m} \frac{\partial T}{\partial z} \rangle \
\sim  \frac{\rho T_{\rm near}}{\bar m} 
\end{eqnarray}
where $\bar m$ is the mean mass parameter, and $z_{\rm disk}$ is a measure of the disk thickness.
Also, in the above equation we have assumed the density variation can be neglected relative to the temperature variation.

For the purposes of a rough estimation, we consider a constant density disk extending to a radius, $R$,
and thickness $z_{\rm disk}$, so that $\rho = m_{\rm dark}/(\pi R^2 z_{\rm disk})$. 
The average pressure force on the dark disk is then
\begin{eqnarray}
F_{\rm pressure} &=& \pi R^2 \langle {\cal F}_{\rm pressure} \rangle \nonumber \\
&\sim & \frac{ m_{\rm dark} T_{\rm near}}{\bar m z_{\rm disk}}
\ .  \end{eqnarray} 
The gravitational force between the two disks is:
\begin{eqnarray}
F_{\rm gravity} \sim \frac{G_N m_{\rm dark} m_{\rm baryon}}{R^2}
\ .  \end{eqnarray}
Comparing $F_{\rm pressure}$ with $F_{\rm gravity}$,
\begin{eqnarray} 
\frac{F_{\rm pressure} }{F_{\rm gravity}} &\sim &
\frac{T_{\rm near} R^2}{G_N \bar m z_{\rm disk} m_{\rm baryon}} \nonumber \\
&\sim & \left( \frac{T_{\rm near}}{10 \ {\rm eV}}\right) 
\left(\frac{0.1 \ {\rm kpc}}{z_{\rm disk}}\right) \left( \frac{10^{10}\ m_\odot}{m_{\rm baryon}}\right)  
\left( \frac{R}{3\ {\rm kpc}}\right)^2 \ .
\label{95} \end{eqnarray}
This is a very crude estimate, but nevertheless it indicates that the pressure force 
can possibly overwhelm gravity.


\vskip 1.5cm
\noindent
{\large \bf Acknowledgments}

\vskip 0.4cm
\noindent
The author would like to thank C. Lagger for bringing \cite{ultra} to his attention,
and also would like to acknowledge useful correspondence with F. Hammer and M. Pawlowski.
This work was supported by the Australian Research Council.


\end{document}